\documentclass[sn-mathphys,dvipsnames,leqno,pdflatex]{sn-jnl}
\usepackage[utf8]{inputenc}
\jyear{2022}%
\normalbaroutside

\renewenvironment{table}[1][]%
{\tableorg[#1]%
\tablebodyfont%
\renewcommand\footnotetext[2][]{{\removelastskip\vskip3pt%
\let\tablebodyfont\tablefootnotefont%
\hskip0pt\if!##1!\else{\smash{$^{##1}$}}\fi##2\par}}%
}{\endtableorg}

\usepackage{amsmath}
\usepackage{amssymb}
\usepackage{mathtools}
\usepackage{hyperref}
\usepackage{cleveref}
\usepackage{algorithm}
\usepackage{algpseudocode}
\usepackage{pgfplots}
\usepackage{tabularx}

\pgfplotsset{compat=1.17}
\usepgfplotslibrary{patchplots}
\usepgfplotslibrary{groupplots}

\crefname{problem}{problem}{problems}

\newcommand{\floats}{\mathbb{F}}

\newcommand{\reals}{\mathbb{R}}
\newcommand{\integers}{\mathbb{Z}}

\newcommand{\emin}{{e_{\mathrm{min}}}}
\newcommand{\emax}{{e_{\mathrm{max}}}}
\newcommand{\qmin}{{q_{\mathrm{min}}}}
\newcommand{\qmax}{{q_{\mathrm{max}}}}

\DeclareMathSymbol{\mlq}{\mathord}{operators}{'134}
\DeclareMathSymbol{\mrq}{\mathord}{operators}{'42}

\newcommand{\MulInv}{\mathbin{\mlq{/}\mrq}}
\renewcommand{\powerset}{\mathcal{P}}

\DeclareMathOperator{\fl}{fl}
\DeclareMathOperator{\RU}{RU}
\DeclareMathOperator{\RD}{RD}
\DeclareMathOperator{\RN}{RN}
\DeclareMathOperator{\RNE}{RNE}
\DeclareMathOperator{\ufp}{ufp}
\DeclareMathOperator{\ulp}{ulp}
\DeclareMathOperator{\Feas}{Feas}
\DeclareMathOperator{\Plaus}{Plaus}
\DeclareMathOperator{\diam}{diam}
\DeclareMathOperator{\sgn}{sgn}

\DeclarePairedDelimiter\abs{\lvert}{\rvert}
\DeclarePairedDelimiter\floor{\lfloor}{\rfloor}
\DeclarePairedDelimiter\ceil{\lceil}{\rceil}

\newcommand*\from{:}

\providecommand\given{}
\newcommand\SetSymbol[1][]{%
    \mathrel{}\mathclose{}#1|\allowbreak\mathopen{}\mathrel{}}
\DeclarePairedDelimiterX\Set[1]\{\}{%
    \renewcommand\given{\SetSymbol[\delimsize]}
    #1
}

\newcommand{\ignore}[1]{}

\newcommand{\overlinestar}[1]{\overline{#1}\vphantom{#1}^*}
\newcommand{\overlineinf}[1]{\overline{#1}\vphantom{#1}_\infty}
\newcommand{\overlinephantom}[1]{\overline{#1}\vphantom{#1}}

\newcommand{\PlausRU}[1]{\phi_{#1}}

\theoremstyle{plain}
\newtheorem{theorem}{Theorem}[section]
\newtheorem{lemma}[theorem]{Lemma}
\newtheorem{conjecture}[theorem]{Conjecture}
\newtheorem{corollary}{Corollary}[theorem]

\theoremstyle{definition}
\newtheorem{definition}[theorem]{Definition}
\newtheorem{example}[theorem]{Example}
\newtheorem{problem}{Problem}

\theoremstyle{remark}
\newtheorem*{remark}{Remark}

\begin{document}
\title{Finding normal binary floating-point factors efficiently}

\author*[1]{\fnm{Mak} \sur{Andrlon}}\email{mnazecic@student.unimelb.edu.au}
\affil[1]{\orgdiv{School of Computing and Information Systems}, \orgname{The University of Melbourne}, \orgaddress{\state{Victoria} \postcode{3010}, \country{Australia}}}

\abstract{Solving the floating-point equation $x \otimes y = z$, where $x$, $y$ and $z$ belong to floating-point intervals, is a common task in automated reasoning for which no efficient algorithm is known in general. We show that it can be solved by computing a constant number of floating-point factors, and give a fast algorithm for computing successive normal floating-point factors of normal floating-point numbers in radix 2. This leads to an efficient procedure for solving the given equation, running in time of the same order as floating-point multiplication.}
\keywords{Floating-point equation solving, Interval multiplication, Decision procedures}

\maketitle

\section{Introduction}
Floating-point arithmetic has long been the prevailing system used for numerical computing, allowing efficient and precise calculation on modern hardware. Whilst it closely approximates real arithmetic, it does not always satisfy its laws. These sporadic exceptions often confuse programmers and complicate mathematical reasoning to the point that deciding even the simplest statements can take excessive amounts of time. Our interest stems from the area of formal software verification, where automated reasoning is used to prove that programs satisfy their specifications. Although modern verification systems can reason about floating-point arithmetic, it has much less support than integer or bit-vector arithmetic. This makes proving correctness far more difficult and drastically limits practical applications of formal methods. Our work seeks to fill these gaps.

Denoting floating-point multiplication by $\otimes$, we study the following problem:
\begin{problem}\label{prob:main}
Let $X, Y, Z$ be intervals over the floating-point numbers and let
\begin{equation*}
V = \Set{ (x, y, z) \given x \in X, y \in Y, z \in Z, x \otimes y = z}.
\end{equation*}
What is the smallest box\footnote{A Cartesian product of intervals.} enclosing $V$?
\end{problem}

We can give a simplified example without getting into the details of floating-point arithmetic:

\begin{example}\label{ex:fp}
For the sake of exposition, we will use decimal numbers with two decimal places, rounding downward the results of arithmetic operations. For example, since $0.50 \cdot 1.01 = 0.505$, we have $0.50 \otimes 1.01 = 0.50$, where $\otimes$ is rounded multiplication.

Let $x$ and $y$ be two-place decimal numbers, and suppose $x \otimes y = 5.00$ where $2.20 \leq x \leq 2.50$ and $1.00 \leq y \leq 2.50$. These bounds are clearly quite loose. For instance, the lower bound on $y$ is too small: if $x = 2.50$ and $y = 1.00$, then $x \otimes y = 2.50$, which is much less than the desired $5.00$.

First, we construct an equivalent condition that uses real multiplication instead of floating-point multiplication. This can be accomplished with a simple transformation. Since the numbers that round downward to $5.00$ are exactly those in $[5.00, 5.01)$, it follows that $x \otimes y = 5.00$ is equivalent to $5.00 \leq xy < 5.01$. By getting rid of the $\otimes$ operator, we can now reason using real arithmetic:
\begin{enumerate}
    \item Multiplying $x$ and $y$ and combining their bounds, we get $2.20 \leq xy \leq 6.25$.
    \item This is worse than our existing bounds on $xy$, so we cannot improve them.
    \item Dividing the bounds on $xy$ by $y$, we obtain $5.00/2.50 \leq x < 5.01/1.00$.
    \item This also does not improve our existing bounds on $x$.
    \item Dividing by $x$ instead, we obtain $5.00/2.50 \leq y < 5.01/2.20$.
    \item We can thus improve our bounds on $y$ to $2.00 \leq y < 2.27\overline{72}$.\footnote{The overline denotes repeating digits.}
\end{enumerate}
If we were allowed the full range of the reals, we would now be finished, stating that $5.00 \leq xy < 5.01$ where $2.20 \leq x \leq 2.50$ and $2.00 \leq y < 2.27\overline{72}$. We can easily show that these bounds are tight for real multiplication:
\begin{itemize}
    \item If $x = 2.20$ and $y < 2.27\overline{72}$, then $xy < 5.01$. If $y \geq 2.27\overline{72}$ instead, then $xy \geq 5.01$.
    \item If $x = 2.50$ and $y = 2.00$, then $xy = 5.00$.
\end{itemize}
Having used each of the bounds on $x$, $y$ and $xy$, we know that this set of bounds is tight on its own. However, we also require that $x$ and $y$ are decimal numbers with two decimal places. Thus, since $y < 2.27\overline{72}$, we can further conclude that $y \leq 2.27$. Since the bounds all have two decimal places now, we would ideally like to be done. If we check, though, we find that the bounds are no longer tight: if $x = 2.20$ and $y = 2.27$, then $xy = 4.994$, which is too small, and therefore $x \geq 2.21$.

We can continue tightening even further. If $x = 2.21$ instead and $y = 2.27$, then $xy = 5.0167$, which is too large. Therefore $y \leq 2.26$, and so on. From this point, division does not help us as much as it did earlier. Since the product of the bounds on $x$ and $y$ is so close to the optimal result, all of the quotients differ by less than $0.01$, and so it is equivalent to an exhaustive search.

After a few more iterations, we eventually reach the optimal solution: $2.33 \leq x \leq 2.50$ and $2.00 \leq y \leq 2.15$. This clearly satisfies our bounds, since $2.33 \cdot 2.15 = 5.0095 \in [5.00, 5.01)$. However, proving optimality is tedious without further results, requiring us to iterate the above argument for every decimal until we reach the optimum.
\end{example}

The toy example above illustrates the limits of current floating-point interval reasoning. As demonstrated, the analogous problem over the reals is straightforward, and the steps in the example suggest a general algorithm. If we were working with the much larger IEEE 754 64-bit floating-point numbers instead, we could easily be forced to iterate over millions of values: for a concrete example, take $12738103310254127 \leq x \leq 12738103379848963$, $y$ unrestricted and $z = 2^{54} - 1$ under rounding to nearest. Although $x \otimes y = z$ has no solution under these conditions, it will take 69,594,839 steps to prove this using the traditional interval-based approach. In general, the traditional algorithm struggles to prove unsatisfiability if the real relaxation of the problem is solvable, since the optimal intervals for an unsatisfiable system are empty. By contrast, the method we present in this paper arrives at the answer immediately.

We consider the following to be the primary contributions of this paper:
\begin{itemize}
    \item An algorithm to efficiently compute the next normal floating-point factor of any floating-point number in radix 2.
    \begin{itemize}
        \item This algorithm is also partially applicable in higher bases. We do not assume a particular radix in any of our results.
    \end{itemize}
    \item When the above algorithm is applicable, an efficient procedure for finding optimal interval bounds on variables of floating-point constraints of the form $x \otimes y = z$.
\end{itemize}
In addition, we believe the following secondary contributions are of interest:
\begin{itemize}
    \item A demonstration that ``round-trip'' rounding errors have useful, nontrivial piecewise polynomial extensions to the reals.
    \item A characterization of upward and downward floating-point rounding in terms of the remainder of floored division.
    \item A floating-point predicate for deciding, in any base, whether a number is floating-point factor of another floating-point number.
    \item Results on solving certain kinds of interval constraints over integer remainders.
\end{itemize}

\subsection{Related work}
Although floating-point arithmetic has existed for a long time, work on solving floating-point equations and inequalities has been sporadic and has mostly been concerned with binary floating-point arithmetic. Michel~\cite{michel2002} developed partial solutions for floating-point equations of the form $x \circ y = z$ where $\circ$ is one of rounded addition, subtraction, multiplication, or division, as well as a complete solution for rounded square root. His work contains an early description of the classical interval-based algorithm, as it applies to the floating-point numbers.\footnote{Note that the classical algorithm is optimal for any strictly increasing (or decreasing) single-variable real function.} As an early example, only normal, binary floating-point numbers and the standard rounding modes were considered. Ziv et al.~\cite{ziv2003} showed that, under mild conditions, if $z$ can take at least 2 different values, then the equation $x \otimes y = z$ has at least one solution in binary floating-point arithmetic. However, their work does not treat other bases, nor the case where $z$ is fixed. Bagnara et al.~\cite{bagnara2016} gave some extremal bounds on the $x$ and $y$ in terms of the bounds on $z$ for binary floating-point arithmetic. However, the resulting bounds are very weak, since they do not account for any of the bounds on $x$ or $y$. At present, there are no known efficient algorithms for computing the solution set of $x \otimes y = z$ in general. Indeed, in terms of similar efforts, the only basic operation which has been solved (apart from square root) is addition~\cite{gallois-wong2020,andrlon2019}.

\subsection{Outline}
In \Cref{sec:prelim}, we introduce our notation and develop the formalisms used in this paper. We also give an account of the basic properties of floating-point arithmetic that we will need, and precisely define the classical interval-based algorithm. Due to the substantial amount of notation, a glossary is supplied for reference in \Cref{sec:glossary}. 
In \Cref{sec:simp}, we relate optimal intervals, the division-based algorithm and floating-point factors, and simplify the problem by narrowing down the conditions under which the division-based algorithm does not produce an optimal result. In \Cref{sec:hardcase}, we then develop conditions for numbers being factors based on the rounding error in the product. Then, in \Cref{sec:optimizing-remainders}, we show that we can control that rounding error under certain conditions and thus directly find floating-point factors. Combining these in \Cref{sec:algorithms}, we give an algorithm to compute such factors efficiently and thus solve the problem. \Cref{sec:additional-proofs} contains some ancillary proofs used in the results of \Cref{sec:algorithms}. In \Cref{sec:conclusion}, we conclude and discuss open problems.

\section{Preliminaries}\label{sec:prelim}
Following convention, we denote the integers by $\integers$, the reals by $\reals$, and the extended reals by $\overline{\reals} = \reals \cup \Set{+\infty, -\infty}$, and the same sets excluding zero by $\integers^*$, $\reals^*$ and $\overlinestar{\reals}$. The power set of a set $X$ is denoted by $\mathcal{P}(X)$. We write $f[X]$ for the image of $X$ under a function $f$, and $f^{-1}[X]$ for the preimage of $X$ under $f$. To simplify exposition, we will denote the images of $X \times Y$ and $X \times \Set{y}$ under a binary operator $\circ$ simply by $X \circ Y$ and $X \circ y$, respectively. We will use such notation most frequently in the context of intervals. We also discuss the size of intervals in terms of their width on the extended real line. To this end, we define a notion of diameter for subsets of the extended reals:
\begin{definition}[Set diameter]
For any $X \subseteq \overline{\reals}$, we define
\begin{equation*}
\diam X = \sup{\Set{{d(x, y) \given x, y \in X}}\cup \Set{0}},
\end{equation*}
where $d \from \overline{\reals}^2 \to \overline{\reals}$ is a function such that
\begin{equation*}
d(x, y) = \begin{cases}
0 & \text{if } x = y,\\
\abs{x - y} & \text{otherwise.}
\end{cases}
\end{equation*}
\end{definition}
The diameter satisfies the following properties for all sets $X \subseteq \overline{\reals}$:
\begin{align}
\diam \emptyset = 0  && \text{(diameter of empty set)}\\
\text{for all } x \in \overline{\reals},\ \diam{\Set{x}} = 0 && \text{(diameter of singleton)}\\
\text{if}\ \inf X < \sup X,\ \text{then}\ \diam X = \sup X - \inf X\\
\text{for all } a \in \reals^*,\ \diam aX = \abs{a} \diam X && \text{(absolute homogeneity)}\\
\text{for all } b \in \reals,\ \diam (X + b) = \diam X && \text{(translation invariance)}
\end{align}
Importantly, the diameter of an interval is equal to the distance between its endpoints:
\begin{lemma}
For all $a, b \in \overline{\reals}$, if $a < b$, then
\begin{equation*}
\diam{[a, b]} = \diam{(a, b]} = \diam{[a, b)} = \diam{(a, b)} = b - a.
\end{equation*}
\end{lemma}
Additionally, if two intervals overlap, then the larger one contains at least one of the endpoints of the smaller one:
\begin{lemma}\label{lem:overlapping-interval-endpoints}
Let $a, b \in \reals$ and let $I \subseteq \overline{\reals}$ be an interval. If $[a, b] \cap I$ is nonempty and $\diam I > b - a$, then either $a \in I$ or $b \in I$.
\end{lemma}
Since the subject of this work is solving equations similar to $xy = z$, we will often need to use division. However, division alone does not provide what we need. For instance, if $y = z = 0$, then any finite and nonzero $x$ satisfies, even though $z/y= 0/0$ is undefined. In order to succinctly describe these solution sets, we introduce a special ``division'' operator:
\begin{definition}
For any $Y, Z \subseteq \overline{\reals}$, we define
\begin{equation*}
    Z \MulInv Y = \Set{x \in \overline{\reals} \given \exists y \in Y (xy \in Z)}.
\end{equation*}
\end{definition}
The following properties then hold for all $Y, Z \subseteq \overline{\reals}$ and $a \in \reals^*$:
\begin{align}
    a Z \MulInv Y &= Z \MulInv (Y/a),\\
    Z \MulInv (Y \cap \reals^*) &= Z / (Y \cap \reals^*),\\
    Z \MulInv 0 &= \begin{cases}
        \reals,& \text{if } 0 \in Z,\\
        \emptyset,& \text{otherwise,}
    \end{cases}\\
    Z \MulInv +\infty &= \begin{cases}
        (0, +\infty],& \text{if } {+\infty} \in Z \text{ and } {-\infty} \notin Z,\\
        [-\infty, 0),& \text{if } {+\infty} \notin Z \text{ and } {-\infty} \in Z,\\
        \overlinestar{\reals},& \text{if } {+\infty} \in Z \text{ and } {-\infty} \in Z,\\
        \emptyset, & \text{otherwise,}
    \end{cases}\\
    \shortintertext{Furthermore, for all $Y', Z' \subseteq \overline{\reals}$, we also have the following properties:}
    Z \MulInv (Y \cup Y') &= (Z \MulInv Y) \cup (Z \MulInv Y'),\\
    (Z \cup Z') \MulInv Y &= (Z \MulInv Y) \cup (Z' \MulInv Y).
\end{align}
\begin{remark}
Note that, since $Y = (Y \cap \reals^*) \cup (Y \setminus \reals^*)$, we have
\begin{equation*}
    Z \MulInv Y = (Z/(Y \cap \reals^*)) \cup (Z \MulInv (Y \setminus \reals^*)).
\end{equation*}
It is also worth noting that $Y \setminus \reals^* = Y \cap \Set{0,-\infty,+\infty}$. This allows us to compute $Z \MulInv Y$ as the union of an ordinary division and at most three special cases as described above.
\end{remark}

We also define the box operator as follows:
\begin{definition}[Box closure]
A \emph{box} is any Cartesian product of (extended) real intervals. For any set $X \subseteq \overline{\reals}^n$, we define its box closure $\Box X$ as the unique smallest $n$-dimensional box containing $X$ as a subset.\footnote{Existence and uniqueness follow since every nonempty subset of $\overline{\reals}$ is bounded and thus has both a supremum and infimum.}
\end{definition}
For any sets $X, Y \subseteq \overline{\reals}^n$ we have:
\begin{align}
    X \subseteq \Box X &&\rlap{\text{(extensivity)}}\\
    \text{if } X \subseteq Y, \text{then } \Box X \subseteq \Box Y, &&\rlap{\text{(monotonicity)}}\\
    \Box \Box X = \Box X &&\rlap{\text{(idempotence)}}\\
    \intertext{Furthermore, for all subsets $A_1, \ldots, A_n$ of $\overline{\reals}$, where the product is the Cartesian product, we have the following property:}
    \Box \prod_{i=1}^{n} A_i = \prod_{i=1}^{n} \Box A_i&&\rlap{\text{(distributivity)}}
\end{align}
That is, the box of a product of sets is exactly the product of the boxes of those sets.
\begin{remark}
A 1-dimensional box is an interval, a 2-dimensional box is a rectangle, a 3-dimensional box is a cuboid, and so on. Also note that a box does not necessarily contain its boundary: an open interval is still an interval, and therefore a box. However, if $X$ is a finite subset of $\overline{\reals}$, then $\Box X = [\min X, \max X]$. In general, if $X$ is finite (or otherwise contains its boundary), then $\Box X$ is a Cartesian product of closed intervals.
\end{remark}

\subsection{Floating-point arithmetic}
Much work has been done on specifying floating-point arithmetic in various formal systems~\cite{brain2015,boldo2011,boldo2017,harrison2006,harrison1999,yu2013,rummer2010,miner1995,loiseleur1997,carreno1995,carreno1995a,barrett1989,boldo2006,harrison1995,jacobsen2015,jacobi2005,miner1995a,daumas2001,melquiond2012,pan1990}. However, to the best of the author's knowledge, there is no well-established and precise informal language of floating-point theory. For the sake of completeness and rigor, the rest of this section will be devoted to defining our terms. Readers interested in a deeper introduction may wish to consult a reference such as Muller et al.~\cite{muller2018} or Goldberg~\cite{goldberg1991}.

\subsubsection{Floating-point numbers, briefly}
A finite floating-point number is usually written in the style of scientific notation as $\pm d_1.d_2d_3\cdots{}d_p \times \beta^e$, where each digit $d_i$ lies in $\Set{0, \dots, \beta - 1}$. The $\pm d_1.d_2d_3\cdots{}d_p$ part is the \emph{significand}, the number of digits $p$ is the \emph{precision}, $\beta$ is the \emph{base} (commonly 2 or 10), and $e$ is the \emph{exponent}. For example, with precision $p = 3$ and base $\beta = 2$, we can write the decimal number 0.75 as either $1.10 \times 2^{-1}$ or $0.11 \times 2^0$. Of the two representations, the former is preferred, as it is \emph{normalized}---that is, it has no leading zeros---and is therefore unique.

With a higher precision we can represent more numbers. For instance, we need $p \geq 4$ to represent the binary number $1.101$ exactly. Regardless of precision, however, there are still limits imposed by the choice of base. Just as $\frac{1}{3}$ lacks a finite decimal expansion, for any given base there are numbers that cannot be represented. Notably, the decimal $0.1$ is $0.000\overline{1100}$ in binary, and thus it has no representation in any binary floating-point system. As an example, if we take $\beta = 2$ and $p = 3$, then $1.10 \times 2^{-4}$ and $1.11 \times 2^{-4}$ are consecutive floating-point numbers, but they correspond to the decimal numbers $0.09375$ and $0.109375$, respectively.

Consequently, the sum, product or ratio of two precision-$p$ base-$\beta$ floating-point numbers is rarely itself a precision-$p$ base-$\beta$ floating-point number and thus computing with them typically involves accepting some degree of rounding error. In ordinary usage, we minimize rounding error by \emph{rounding to nearest}; that is, by choosing the floating-point number closest to the real number we are rounding. Other rounding modes such as rounding upward, downward, or towards zero (also known as truncation) are typically only used in situations where the direction of the rounding is more important than the accuracy of a single operation. For example, the simplest way to avoid underapproximating the set of true results in interval arithmetic is to always round upper bounds upward and lower bounds downward.

\subsubsection{Definitions and terminology}
The arithmetic described in this work is compatible with and generalizes that of the IEEE 754 standard, with two exceptions: we omit signed zeros\footnote{Although positive and negative zero correspond to the same real value, they are distinct, since $1/{+0} = +\infty$ and $1/{-0} = -\infty$. However, the $\MulInv$ operator allows us to sidestep the issue of division by zero.} ``NaN'' values, because they are essentially irrelevant to the problem but add clutter. \Cref{sec:glossary} contains a summary of the definitions introduced here. We begin with a construction of the floating-point numbers.
\begin{definition}[Floating-point numbers]\label{def:floats}
A \emph{floating-point format} is a quadruple of integers $(\beta, p, \emin, \emax)$ such that $\beta \geq 2$, $p \geq 1$, and $\emin \leq \emax$, where $\beta$ is the \emph{base}, $p$ is the \emph{precision}, $\emin$ is the \emph{minimum exponent}, and $\emax$ is the \emph{maximum exponent}. For convenience, we also define the \emph{minimum quantum exponent}\footnote{In IEEE 754 parlance, the ``quantum'' of a floating-point number is the value of a unit in the last digit of its significand, often referred to as the ``unit in last place''. This is necessarily an integer power of $\beta$, and the quantum exponent is the exponent of that power.} $\qmin = \emin - p + 1$ and the \emph{maximum quantum exponent} $\qmax = \emax - p + 1$. Given these quantities, we define the set $\floats^*$ of finite nonzero floating-point numbers as follows:
\begin{equation*}
\floats^* = \Set{M \cdot \beta^{e - p + 1} \given M, e \in \integers, 0 < \abs{M} < \beta^p, \emin \leq e \leq \emax}.
\end{equation*}
From here, we define the set $\floats$ of finite floating-point numbers and the set $\overline{\floats}$ of floating-point numbers:
\begin{align*}
\floats &= \floats^* \cup \Set{0},\\
\overline{\floats} &= \floats \cup \Set{-\infty, +\infty}.
\end{align*}
\end{definition}

To avoid repetition, from this point forward we will assume that the sets of floating-point numbers correspond to some floating-point format $(\beta, p, \emin, \emax)$.

From the above definition, we immediately obtain the following:
\begin{lemma}
$\floats^*$, $\floats$ and $\overline{\floats}$ are each closed under negation.
\end{lemma}
Furthermore, the largest finite floating-point number and the smallest positive floating-point number have simple expressions:
\begin{gather}
\max \floats^* = \max \floats = (\beta^p - 1)\beta^\qmax\\
\min \overline{\floats} \cap (0, +\infty] = \beta^\qmin
\end{gather}

We further categorize the elements of $\floats^*$ as follows:
\begin{definition}[Normal and subnormal numbers]
A nonzero finite floating-point number $x \in \floats^*$ is \emph{normal} if $\abs{x} \geq \beta^\emin$, or \emph{subnormal} otherwise.
\end{definition}
The normal numbers are so called because they each have a normalized representation in scientific notation, restricting the exponent to $\Set{\emin, \dots, \emax}$. By contrast, a subnormal number can only be written without leading zeros by choosing an exponent below $\emin$. Note that we do not consider zero to be normal or subnormal.

A number written in scientific notation has an exponent and significand. Similarly, each finite floating-point number has an associated exponent and significand determined by the floating-point format. We will now define these concepts and generalize them to the finite reals.
\begin{definition}[Exponent and significand]
The functions $E \from \reals \to \integers$ and $Q \from \reals \to \integers$ are defined as follows:
\begin{align*}
E(x) &= \begin{cases}
\floor{\log_\beta \abs{x}}, & \abs{x} \geq \beta^\emin,\\
\emin, & \abs{x} < \beta^\emin.
\end{cases}\\
Q(x) &= E(x) - p + 1.
\end{align*}
Given $x \in \reals$, the integers $E(x)$ and $Q(x)$ are the \emph{exponent} and \emph{quantum exponent} of $x$, respectively. The \emph{significand} of $x$ is the number $x\beta^{-E(x)}$. If $x\beta^{-Q(x)}$ is an integer, then it is the \emph{integral significand}~\cite{muller2018} of $x$.
\end{definition}
To justify these names, we establish that these indeed compute the exponent of a number written as a product of a significand and a power of $\beta$, respecting the minimum exponent:
\begin{lemma}[Exponent law]\label{lem:exponent-law}
Let $m \in \reals$ and $e \in \integers$ such that $\abs{m} < \beta$ and $e \geq \emin$.
If either $\abs{m} \geq 1$ or $e = \emin$, then $E(m \beta^e) = e$. Otherwise, $E(m \beta^e) < e$.
\end{lemma}
\begin{proof}
First, note that if $\abs{m \beta^e} \geq \beta^\emin$, then
\begin{align*}
E(m \beta^e)
&= \floor{\log_\beta \abs{m \beta^e}}\\
&= \floor{\log_\beta \abs{m} + e}\\
&= \floor{\log_\beta \abs{m}} + e.
\end{align*}
We now proceed by cases:
\begin{itemize}
    \item Suppose $\abs{m} \geq 1$. Then $\abs{m \beta^e} \geq \beta^\emin$ and $\floor{\log_\beta \abs{m}} = 0$, and thus $E(m \beta^e) = e$.
    \item Suppose $e = \emin$ and $\abs{m} < 1$. Then $\abs{m \beta^e} < \beta^\emin$ and so we have $E(m \beta^e) = \emin = e$ immediately by the definition of $E$.
    \item Suppose $e > \emin$ and $\abs{m} < 1$. If $\abs{m \beta^e} \geq \beta^\emin$, then $\floor{\log_\beta \abs{m}} < 0$ and hence $E(m \beta^e) = \floor{\log_\beta \abs{m}} + e < e$. If $\abs{m \beta^e} < \beta^\emin$ instead, then $E(m \beta^e) = \emin < e$.
\end{itemize}
Since the result holds in all cases, we are finished.
\end{proof}
Expanding on the above, the function $E$ also satisfies the following properties for all $x, y \in \reals$:
\begin{align}
E(x) = E(-x) && \text{(evenness)}\\
\text{if } \abs{x} \leq \abs{y}, \text{ then } E(x) \leq E(y) && \text{(piecewise monotonicity)}\\
\text{if } x \in \floats, \text{ then } \emin \leq E(x) \leq \emax && \text{(exponent bound)}\\
\abs{x \beta^{-E(x)}} < \beta && \text{(significand bound)}\\
1 \leq \abs{x \beta^{-E(x)}} \text{ if and only if } \beta^\emin \leq \abs{x}   && \text{(normality)}\\
\intertext{From the above properties, the definition of $Q$ immediately implies the following:}
Q(x) = Q(-x) && \text{(evenness)}\\
\text{if } \abs{x} \leq \abs{y}, \text{ then } Q(x) \leq Q(y) && \text{(piecewise monotonicity)}\\
\text{if } x \in \floats, \text{ then } \qmin \leq Q(x) \leq \qmax && \text{(quantum exponent bound)}\\
\abs{x \beta^{-Q(x)}} < \beta^p && \text{(integral significand bound)}\\
\beta^{p - 1} \leq \abs{x \beta^{-Q(x)}}  \text{ iff } \beta^\emin \leq \abs{x} && \text{(normality)}\\
\intertext{Additionally, the following fact establishes an important connection between integral significands and the finite floating-point numbers:}
\text{if } x \in \floats, \text{ then } x \beta^{-Q(x)} \in \integers && \text{(integrality)}
\end{align}
In other words, every finite floating-point number has an integral significand.

\begin{definition}[Unit in first/last place~\cite{rump2008}]
The functions $\ufp, \ulp : \reals \to \reals$ are defined as follows:
\begin{align*}
\ufp(x) &= \begin{cases}
\beta^{\floor{\log_\beta{\abs{x}}}}, & x \neq 0,\\
0, & x = 0,
\end{cases}\\
\ulp(x) &= \ufp(x)\beta^{1-p}.
\end{align*}
For any $x \in \reals$, we refer to $\ufp(x)$ as the \emph{unit in first place} of $x$, and to $\ulp(x)$ as the \emph{unit in last place} of $x$.\footnote{Note that these definitions are independent of exponent range. In particular, readers should be aware that they do not treat subnormal numbers specially, unlike the $Q$ and $E$ functions. In effect, they assume unbounded exponents.}
\end{definition}
For all $x, y \in \reals$ and $n \in \integers$, the following statements hold:
\begin{align}
\ufp(x) \leq \abs{x} && \text{(lower bound)}\\
\text{if } x \neq 0, \text{ then } {\ufp(x) \leq \abs{x} < \beta \ufp(x)} && \text{(bounds)}\\
\ufp(x) = \abs{x} \text{ iff } x = 0 \text{ or } \abs{x} = \beta^k \text{ for some } k \in \integers && \text{(exactness)}\\
\ufp(x \beta^n) = \ufp(x) \beta^n && \text{(scaling)}\\
\ufp(x) \ufp(y) = \ufp(x \ufp(y)) && \text{(product law)}\\
\text{if } y \neq 0, \text{ then } {\ufp(x)/\ufp(y)} = \ufp(x/\ufp(y)) && \text{(quotient law)}
\intertext{There is also a close relationship between $\ufp$ and $E$. In particular, the base-$\beta$ logarithm of the unit in first place is the exponent (if exponents were unbounded):}
\text{if } \abs{x} \geq \beta^\emin, \text{ then } {\ufp(x) = \beta^{E(x)}}\\
\text{if } \abs{x} < \beta^\emin, \text{ then } {\ufp(x) < \beta^{E(x)}}
\end{align}
The following result allows us to further simplify computing the unit in first place of a quotient by separating out the significands:
\begin{lemma}\label{lem:ufp-of-ratio}
For all $x, y \in \reals^*$, we have
\begin{gather*}
    \ufp(x/y) = \ufp(m_x/m_y) \frac{\ufp(x)}{\ufp(y)},\\
        \ufp(m_x/m_y) = \begin{cases}
    1/\beta & \text{if } \abs{m_x} < \abs{m_y},\\
    1 & \text{otherwise,}
    \end{cases}
\end{gather*}
where $m_x = x/\ufp(x)$ and $m_y = y/\ufp(y)$.
\end{lemma}
In a similar vein, we have the following lemma on the exponents of quotients:
\begin{lemma}\label{lem:z-below-x-exponent}
Let $x, z \in \reals^*$ and let $m_x = x\beta^{-E(x)}$ and $m_z = z\beta^{-E(z)}$. If $\beta^\emin \leq \abs{z/x}$, then $E(z/x) = E(z) - E(x) + \log_\beta \ufp(m_z/m_x)$.
\end{lemma}
\begin{proof}
Suppose $\beta^\emin \leq \abs{z/x}$. Then,
\begin{align*}
\beta^{E(z/x)}
&= \ufp(z/x)\\
&= \ufp\left(\frac{m_z \beta^{E(z)}}{m_x \beta^{E(x)}}\right)\\
&= \ufp(m_z/m_x)\beta^{E(z) - E(x)}.
\end{align*}
Taking logarithms, we obtain the desired result.
\end{proof}

Having defined the basic structure of $\overline{\floats}$, we now turn to the definition of the computational operations involving the floating-point numbers.
\begin{definition}[Rounding]
A rounding function is any function $f \from \overline{\reals} \to \overline{\floats}$. The downward rounding function $\RD \from \overline{\reals} \to \overline{\floats}$ and the upward rounding function $\RU \from \overline{\reals} \to \overline{\floats}$ are defined as follows:
\begin{align*}
\RD(x) &= \max\,\Set{y \in \overline{\floats} \given x \leq y},\\
\RU(x) &= \min\,\Set{y \in \overline{\floats} \given x \geq y}.
\end{align*}
A rounding function $f$ is \emph{faithful} if and only if $f(x) \in \Set{\RD(x), \RU(x)}$ for all $x \in \overline{\reals}$.
\end{definition}

It is easy to show that both $\RD$ and $\RU$ are nondecreasing\footnote{Also known as monotonic, (weakly) increasing, (weakly) order-preserving, or isotone.} functions. In addition, they satisfy the following properties for all $x \in \overline{\reals}$:
\begin{align}
\RD(x) \leq x \leq \RU(x) && \text{(lower/upper bound)}\\
\RU(-x) = -\RD(x) && \text{(duality)}\\
x \in \overline{\floats} \text{ if and only if } \RU(x) = \RD(x)   && \text{(exactness)}
\end{align}
In addition, the following gap property holds for all $x \in \reals$:
\begin{equation}\label{eq:gap-property}
\RU(x) - \RD(x) =
\begin{cases}
\infty & \text{if } \abs{x} > \max \floats,\\
0 & \text{if } x \in \floats,\\
\beta^{Q(x)} & \text{otherwise.}\\
\end{cases}
\end{equation}
We also give a partial definition of \emph{rounding to nearest}, which is the default rounding mode used in practice:
\begin{definition}[Rounding to nearest]
A rounding function $f$ \emph{rounds to nearest} if $\abs{f(x) - x} = \min_{y \in \overline{\floats}}\,\abs{y - x}$ for all $x \in [\min \floats, \max \floats]$.
\end{definition}
We denote by $\RN$ an arbitrary nondecreasing and faithful rounding to nearest. Note that this is not a complete definition! In particular, it does not give a tie-breaking rule for when $x$ is the exact midpoint between two floating-point numbers. It also does not fully specify how to round values outside $[\min \floats, \max \floats]$. We leave these unspecified since IEEE 754 specifies two different round-to-nearest modes with different rules in these cases.\footnote{The two tie-breaking rules used in IEEE 754 are ``ties to even'' (choose the result with an even integral significand)  and ``ties away from zero'' (choose the larger magnitude result). The former is the default mode and must be supported. The latter mode need only be available for decimal systems. Additionally, a ``ties to zero'' (truncation) rule is used for the ``augmented'' addition, subtraction and multiplication operations added in IEEE 754-2019~\cite{ieee754_2019}, which return the rounding error alongside the principal result.} In particular, even though $+\infty$ and $-\infty$ are infinitely far apart from any finite number, according to IEEE 754, they may still result from rounding finite values to nearest. More specifically, it states that the rounding to nearest of any $x$ such that $\abs{x} \geq \beta^\emax(\beta - \beta^{1-p}/2)$ is the infinity with the same sign as $x$. Note that this bound is the exact midpoint of $\max \floats$ and $\beta^{\emax + 1}$.

We shall denote by $\fl$ an arbitrary nondecreasing faithful rounding function. This condition is satisfied by most practical rounding modes, including $\RD$, $\RU$, and rounding to nearest as defined above. The following properties hold for all $x \in \overline{\reals}$:
\begin{align}
x \in \overline{\floats} \text{ if and only if } \fl(x) = x && \text{(exactness)}\\
\fl(\fl(x)) = \fl(x) && \text{(idempotence)}
\end{align}

Throughout this paper, we will frequently use the preimages of floating-point intervals under rounding. We will often rely on the following lemma:
\begin{lemma}
Let $X \subseteq \overline{\floats}$. If $X$ is a nonempty floating-point interval, then $\fl^{-1}[X]$ is an extended real interval such that $\fl^{-1}[X] \supseteq [\min X, \max X]$.
\end{lemma}
\begin{proof}
Suppose $X$ is a nonempty floating-point interval, and suppose to the contrary that $\fl^{-1}[X]$ is not an interval. Then there are some $a, b \in \fl^{-1}[X]$ and $y \in \overline{\reals}$ such that $a < y < b$ and $y \notin \fl^{-1}[X]$. Since $\fl$ is nondecreasing, it follows that $\fl(a) \leq \fl(y) \leq \fl(b)$. However, since $\fl(a), \fl(b) \in X$ where $X$ is a floating-point interval, we thus have $\fl(y) \in X$. Contradiction! Hence $\fl^{-1}[X]$ is an interval.

Now, since $\fl$ is the identity over the floating-point numbers, $\fl(x) = x$ for all $x \in X$ and hence $\fl^{-1}[X] \supseteq X$. Since $\overline{\floats}$ is finite, $X$ is also finite, and hence $[\min X, \max X]$ is the smallest interval containing $X$ as a subset. Therefore, since $\fl^{-1}[X]$ is an interval, it follows that $\fl^{-1}[X] \supseteq [\min X, \max X]$.
\end{proof}

To ensure the generality of our results, we define floating-point multiplication in terms of $\fl$ rather than fixing a specific rounding function:\footnote{Although more precise results can sometimes be found by considering a specific rounding function, the limitations of the method described in this paper stem from the unpredictability of upward- and downward-rounded quotients, which are independent of $\fl$.}
\begin{definition}[Multiplication]
The floating-point multiplication operator $\otimes$ is defined by $x \otimes y  = \fl(xy)$ for all $x, y \in \overline{\floats}$ such that the product $xy$ is defined.
\end{definition}
Note that the extended real product is defined if and only if the factors do not include both zero and an infinity. In particular, the product of an extended real with a nonzero (finite) real number is always defined. As a consequence, however, $\overline{\reals}$ is not closed under multiplication, and so neither is $\overline{\floats}$. In fact, neither $\floats^*$ nor $\floats$ are necessarily closed under floating-point multiplication, since it is possible for a sufficiently large product to round to infinity or a sufficiently small nonzero product to round to zero.

However, since $\fl$ is nondecreasing, we can still carry over some important ordering properties from the reals.
\begin{lemma}\label{lem:mul-order-property}
Let $a \in \floats^*$ and $x, y \in \overline{\floats}$ such that $x \leq y$. If $a > 0$, then $a \otimes x \leq a \otimes y$. If $a < 0$, then $a \otimes y \leq a \otimes x$.
\end{lemma}
Since $\overline{\floats}$ is a subset of $\overline{\reals}$, it is totally ordered by the usual order relation. As such, there is a simple definition of an interval over $\overline{\floats}$:
\begin{definition}[Floating-point interval]
A \emph{floating-point interval} is any set $X \cap \overline{\floats}$ where $X$ is an extended real interval.
\end{definition}
With the exception of $[-\infty, +\infty] \cap \overline{\floats}$, due to the gaps between floating-point numbers, every floating-point interval has many representations in terms of extended real intervals. However, since $\overline{\floats}$ is finite, each of its nonempty subsets has a minimum and a maximum, making it trivial to find the smallest interval containing it. For all $a, b \in \overline{\reals}$, we have the following:
\begin{align}
\text{if } a, b \in \overline{\floats} \text{ and } a \leq b, \text{ then } a, b \in [a, b] \cap \overline{\floats} && \text{(tightness)}\\
[a, b] \cap \overline{\floats} = [\RU(a), \RD(b)] \cap \overline{\floats} && \text{(rounding)}
\end{align}
That is, if the bounds are in $\overline{\floats}$, they are necessarily the tightest possible bounds, and intersecting an interval with $\overline{\floats}$ corresponds to rounding its bounds ``inward''.

\subsection{The classical iterative procedure}

As shown in \Cref{ex:fp}, we can use real division to correctly improve the bounds on the factors of a floating-point product. Although it did not find optimal bounds immediately, the division-based approach shown in the example is still essential to solving the problem efficiently, so let us describe it more rigorously.

In the example at the start, we reasoned in terms of inequalities. However, a general method will be easier to describe in terms of sets. Let $X, Y, Z \subseteq \overline{\floats}$ be floating-point intervals and let $x \in X$.
First, note that by the definition of floating-point multiplication, for all $y \in Y$, we have $x \otimes y \in Z$ if and only if $xy \in \fl^{-1}[Z]$. This greatly simplifies things by allowing us to characterize $x$ using extended real division in most cases. More precisely, using our augmented ``division'', we have $x \otimes y \in Z$ for some $y \in Y$ if and only if $x \in \fl^{-1}[Z] \MulInv Y$. We relate this to our problem statement as follows:
\begin{definition}[Solution sets and optimality]
A \emph{problem instance} is any triple $(X, Y, Z)$ where $X, Y, Z \subseteq \overline{\floats}$. The \emph{solution set} of $(X, Y, Z)$ is the set of triples
\begin{equation*}
    V = \Set{(x, y, z) \given x \in X, y\in Y, z \in Z, x \otimes y = z}.
\end{equation*}
A problem instance is \emph{satisfiable} if and only if its solution set is nonempty. The triple $(X, Y, Z)$ has \emph{optimal bounds} if and only if $\Box V = \Box X \times \Box Y \times \Box Z$.
\end{definition}
\begin{lemma}\label{lem:solution-set-box-decomposition}
Let $X, Y, Z \subseteq \overline{\floats}$ and let $V$ be the solution set of $(X, Y, Z)$. Then $\Box V = \Box X' \times \Box Y' \times \Box Z'$, where
\begin{align*}
    X' &= X \cap (\fl^{-1}[Z] \MulInv Y),\\
    Y' &= Y \cap (\fl^{-1}[Z] \MulInv X),\\
    Z' &= Z \cap (X \otimes Y).
\end{align*}
\end{lemma}
\begin{proof}
Let $V_1$, $V_2$ and $V_3$ be the sets containing, for each triple in $V$, its first, second, or third element, respectively. Then,
\begin{equation*}
    \Box V = \Box V_1 \times \Box V_2 \times \Box V_3.
\end{equation*}
Now, by the definition of $\MulInv$, it is easy to check that $V_1 = X'$, $V_2 = Y'$, and $V_3 = Z'$, and the result follows immediately.
\end{proof}

Although useful, it is impractical to directly use the above result to solve the problem. Since $Y$ is a \emph{floating-point} interval, it has many gaps, and so it is very expensive to compute the set of quotients. These gaps were invisible in \Cref{ex:fp} because we were working with inequalities, rather than the exact sets. Translated in terms of sets, what we did corresponds to enlarging the set of denominators $Y$ to the extended real interval $\Box Y$. By enlarging $X$ and $Z$ similarly, we obtain a precise definition of our algorithm:
\begin{definition}[Classical iteration]\label{def:classical-iteration}
We define the function $F \from \powerset(\overline{\reals})^3 \to \powerset(\overline{\reals})^3$ by $F(X, Y, Z) = (X', Y', Z')$ where
\begin{align*}
    X' &= X \cap (\fl^{-1}[Z] \MulInv \Box Y),\\
    Y' &= Y \cap (\fl^{-1}[Z] \MulInv \Box X),\\
    Z' &= Z \cap \fl[\Box X \Box Y].
\end{align*}
\end{definition}

Since $\Box Y$ is a superset of $Y$, it follows that $\fl^{-1}[Z] \MulInv \Box Y$ is a superset of $\fl^{-1}[Z] \MulInv Y$, and so this substitution is sound, though it may not produce an optimal result. Importantly, however, a quotient of intervals is easy to compute and represent, since it must itself be either an interval or a union of two disjoint intervals.
\begin{lemma}[Soundness]
Let $X, Y, Z \subseteq \overline{\floats}$ and let $x \in X$, $y \in Y$, $z \in Z$. If $x \otimes y = z$, then $x \in X', y \in Y', z \in Z'$ where $(X', Y', Z') = F(X, Y, Z)$.
\end{lemma}
\begin{proof}
Suppose $x \otimes y = z$. Then, since $x \otimes y = \fl(xy) \in \fl[\Box X \Box Y]$ and $z \in Z$, we have $z \in Z'$ by definition. Since $z \in Z$, we also have $xy \in \fl^{-1}[Z]$ by definition. Thus, since $y \in \Box Y$, we have $x \in \fl^{-1}[Z] \MulInv \Box Y$ by definition, and hence $x \in X'$. Since $x \in \Box X$, we also similarly have $y \in Y'$, as desired.
\end{proof}

It is easy to prove that iterating $F$ converges in finite time, since the sets involved are finite, and we only shrink them:
\begin{lemma}[Termination]
For all $X_1, Y_1, Z_1 \subseteq \overline{\floats}$, the sequences of sets given by
\begin{equation*}
    (X_n, Y_n, Z_n) = F(X_{n-1}, Y_{n-1}, Z_{n-1}),\; n>1
\end{equation*}
are eventually constant.
\end{lemma}
\begin{proof}
Since $\overline{\floats}$ is finite, $X_1$, $Y_1$ and $Z_1$ are finite. Therefore, since the definition of $F$ implies that the sequences $X_n$, $Y_n$ and $Z_n$ are monotonically decreasing by the subset relation, they must be eventually constant.
\end{proof}

Furthermore, individual applications of $F$ can be computed quickly:
\begin{lemma}[Time complexity]\label{lem:f-time-complexity}
If $X$, $Y$, and $Z$ are floating-point intervals or are each a disjoint union of an all-nonnegative and an all-nonpositive floating-point interval, then $F(X, Y, Z)$ can be computed using $O(1)$ arithmetic operations and computations of the preimage of a floating-point interval under $\fl$.
\end{lemma}
\begin{remark}
Although the number of arithmetic operations needed does not increase with the problem parameters, the individual cost of computing them does. In the sign-significand-exponent encoding, floating-point numbers (in any base) take up $O(p + \log(\emax - \emin + 1))$ bits. If $\fl$ belongs to one of usual families of rounding functions (e.g.\ $\RD$, $\RU$, or $\RN$), both it and its preimage can be computed in time linear in the length of its input. Thus, floating-point addition can be computed in time linear in the length of the addends and multiplying two floating-point numbers takes time $O(M(p) + \log(\emax - \emin + 1))$, where $M(p)$ is the time complexity of multiplying two $p$-digit integers. Correctly rounding division is more difficult, but it can be computed using Newton-Raphson division in same amount of time asymptotically as multiplication. Thus, for most choices of $\fl$, computing $F$ takes time $O(M(p) + \log(\emax - \emin + 1))$.\footnote{Note that to obtain this complexity result rigorously, we must define a family of rounding functions suitably indexed by floating-point format, as well as a matching indexed variant of $F$.} Note that we do not take the base, precision or exponent range as constants in our analysis, as doing so tends to collapse the time complexity of floating-point algorithms to $O(1)$, precluding any meaningful comparison. We also do not assume a specific multiplication algorithm, as there are many to choose from with different characteristics.
\end{remark}
However, given that what we are computing is a relaxation of the original problem, it is unclear how quickly the iteration converges, or even that it necessarily converges to the optimum. Certainly, it can be quite slow in some cases, as we saw in \Cref{ex:fp}.
Although we do not have a concrete answer, we offer the following conjecture on the rate of convergence:
\begin{conjecture}\label{conj:f-convergence-rate}
Let $X_1, Y_1, Z_1 \subseteq \overline{\floats}$ be floating-point intervals, and define the sequences of sets
\begin{equation*}
    (X_n, Y_n, Z_n) = F(X_{n-1}, Y_{n-1}, Z_{n-1}),\; n>1.
\end{equation*}
Then, for all $m > 2$, the sets $X_{m} \setminus X_{m+1}$, $Y_{m} \setminus Y_{m+1}$ and $Z_{m} \setminus Z_{m+1}$ contain at most 4 members each.
\end{conjecture}
That is, we conjecture that the rate of convergence is constant. The intuition behind this is thus: after two iterations, we will have propagated the bounds from one factor to the other, and back again. After that, however, the bounds are within one floating-point number of the optimum of the real relaxation, so there is no significant progress to be gained from interval reasoning. Instead, each step only removes at most one floating-point number from the end of each interval in the sets. (Each set after the first iteration may be a disjoint union of two intervals, if the initial intervals contain both positive and negative numbers.) This is clearly rather slow, so we make a further conjecture: that, in the worst case, iterating $F$ takes a number of steps to converge exponential in the precision:
\begin{conjecture}\label{conj:f-worst-case-complexity}
Denote by $\overlinephantom{\floats}_n$ the floating-point numbers with format $(\beta, n, \emin, \emax)$ for each precision $n \geq 2$ and let $\fl_n$ be an associated nondecreasing, faithful rounding function. Let $F_n$ be $F$ parameterized over $\fl_n$. Let $\mu_n(X, Y, Z) = m$ be the least positive integer such that $F^{m+1}_n(X, Y, Z) = F^m_n(X, Y, Z)$, where $F^k_n$ denotes $F_n$ composed with itself $k$ times, where $X, Y, Z \subseteq \overlinephantom{\floats}_n$. If $f \from \Set{2, 3, \ldots} \to \integers$ is a function such that
\begin{equation*}
f(n) = \max{\Set{\mu_n(X, Y, Z) \given X, Y, Z \text{ are floating-point intervals over } \overlinephantom{\floats}_n}},
\end{equation*}
then $f(n) = \Theta(\beta^{cn})$ for some positive real $c$.
\end{conjecture}
\begin{remark}
Proving an upper bound is trivial, as it is of the same order of steps as brute force search. The lower bound is tricky. In order to prove it, we essentially need to prove that bad (i.e.\ exponentially slow) instances exist in every precision. It would perhaps be easier to show that $f$ is not dominated by any function of lesser order. For this, we merely need an infinite number of precisions with bad instances. A promising candidate is the set of all instances of the form $(X, \overline{\floats}, \Set{\beta^p - 1})$, where $X$ is the widest floating-point interval centered on $\sqrt{\beta^{p-1}(\beta^p - 1)}$ such that the instance is unsatisfiable. (The unsatisfiable example we gave earlier for 64-bit IEEE floating-point numbers is one such instance.)
\end{remark}

Leaving aside the rate of convergence, we would also like to know what exactly we are converging to. We can show that whenever $\Box X$ and $\Box Y$ do not contain zero, convergence implies that the bounds on $X$ and $Y$ are optimal. In such cases, we obtain explicit solutions to the equation $x \otimes y = z$ where $x$ or $y$ is an endpoint of $X$ or $Y$, respectively.
\begin{lemma}[Conditional partial optimality]\label{lem:conditional-optimality}
Let $X, Y, Z \subseteq \overline{\floats}$. If $Z$ is a floating-point interval, and neither $\Box X$ nor $\Box Y$ contains zero, and $F(X, Y, Z) = (X, Y, Z)$, then one of the following statements holds:
\begin{enumerate}
    \item $X$, $Y$ and $Z$ are empty.
    \item The elements of $X$ have the same sign as the elements of $Y$, the elements of $Z$ are nonnegative, and $\min X \otimes \max Y \in Z$ and $\max X \otimes \min Y \in Z$.
    \item The elements of $X$ have the opposite sign of the elements of $Y$, the elements of $Z$ are nonpositive, and $\min X \otimes \min Y \in Z$ and $\max X \otimes \max Y \in Z$.
\end{enumerate}
\end{lemma}
\begin{proof}
Suppose the conditions hold. Then, if any of $X$, $Y$, or $Z$ are empty, the definition of $F$ trivially implies that they are all empty, so the result holds. Suppose instead that $X$, $Y$ and $Z$ are nonempty. Since $\Box X$ is an interval not containing zero, it is a subset of either $[-\infty, 0)$ or $(0, +\infty]$, and thus the elements of $X$ all have the same sign. Similarly, we also find that every element of $Y$ has the same sign. We now proceed by cases:

Suppose that the elements of $X$ have the same sign as the elements of $Y$. Then $\fl(xy) \geq 0$ for all $x \in \Box X$ and $y \in \Box Y$ where the product is defined. Therefore, by the definition of $F$, the elements of $Z$ are all nonnegative.
\begin{enumerate}
    \item Let $x \in \Box X$ and $y \in \Box Y$ such that $\min X \cdot y \in \fl^{-1}[Z]$ and $x \cdot \max Y \in \fl^{-1}[Z]$. (These exist by the definition of $\MulInv$.)
    \begin{enumerate}
        \item Suppose $\min X > 0$ and $\max Y > 0$. Then, since $\min X \leq x$ and $y \leq \max Y$, multiplying and rounding gives
        \begin{equation*}
            \min Z \leq \fl(\min X \cdot y) \leq \min X \otimes \max Y \leq \fl(x \cdot \max Y) \leq \max Z,
        \end{equation*}
        and since $Z$ is a floating-point interval, we thus have $\min X \otimes \max Y \in Z$.
        \item Suppose $\min X < 0$ and $\max Y < 0$ instead. Then, similarly, we obtain
        \begin{equation*}
            \fl(x \cdot \max Y) \leq \min X \otimes \max Y \leq \fl(\min X \cdot y),
        \end{equation*}
        and thus $\min X \otimes \max Y \in Z$.
    \end{enumerate}
    \item Let $x \in \Box X$ and $y \in \Box Y$ instead such that $\max X \cdot y \in \fl^{-1}[Z]$ and $x \cdot \min Y \in \fl^{-1}[Z]$.
    \begin{enumerate}
        \item Suppose $\max X > 0$ and $\min Y > 0$. Then, since $x \leq \max X$ and $\min Y \leq y$,
        \begin{equation*}
            \fl(x \cdot \min Y) \leq \max X \otimes \min Y \leq \fl(\max X \cdot y),
        \end{equation*}
        and thus $\max X \otimes \min Y \in Z$.
        \item Suppose $\max X < 0$ and $\min Y < 0$ instead. Then, since $x \leq \max X$ and $\min Y \leq y$,
        \begin{equation*}
            \fl(\max X \cdot y) \leq \max X \otimes \min Y \leq \fl(x \cdot \min Y),
        \end{equation*}
        and thus $\max X \otimes \min Y \in Z$.
    \end{enumerate}
\end{enumerate}

Suppose instead that the elements of $X$ have the opposite sign of the elements of $Y$. Then $\fl(xy) \leq 0$ for all $x \in \Box X$ and $y \in \Box Y$ and thus the elements of $Z$ are all nonpositive.
\begin{enumerate}
    \item Let $x \in \Box X$ and $y \in \Box Y$ such that $\min X \cdot y \in \fl^{-1}[Z]$ and $x \cdot \min Y \in \fl^{-1}[Z]$.
    \begin{enumerate}
        \item Suppose $\min X > 0$ and $\min Y < 0$. Then, since $\min X \leq x$ and $\min Y \leq y$, multiplying and rounding gives
        \begin{equation*}
            \min Z \leq \fl(x \cdot \min Y) \leq \min X \otimes \min Y \leq \fl(\min X \cdot y) \leq \max Z,
        \end{equation*}
        and therefore $\min X \otimes \min Y \in Z$.
        \item Suppose $\min X < 0$ and $\min Y > 0$ instead. Then, similarly, we obtain
        \begin{equation*}
            \fl(\min X \cdot y) \leq \min X \otimes \min Y \leq \fl(x \cdot \min Y),
        \end{equation*}
        and thus $\min X \otimes \min Y \in Z$.
    \end{enumerate}
    \item Let $x \in \Box X$ and $y \in \Box Y$ instead such that $\max X \cdot y \in \fl^{-1}[Z]$ and $x \cdot \max Y \in \fl^{-1}[Z]$.
    \begin{enumerate}
        \item Suppose $\max X > 0$ and $\max Y < 0$. Then, since $x \leq \max X$ and $y \leq \max Y$,
        \begin{equation*}
            \fl(\max X \cdot y) \leq \min X \otimes \min Y \leq \fl(x \cdot \max Y),
        \end{equation*}
        and so $\max X \otimes \max Y \in Z$.
        \item Suppose $\max X < 0$ and $\max Y > 0$ instead. Then, similarly,
        \begin{equation*}
            \fl(x \cdot \max Y) \leq \min X \otimes \min Y \leq \fl(\max X \cdot y),
        \end{equation*}
        and thus $\max X \otimes \max Y \in Z$.
    \end{enumerate}
\end{enumerate}
Since we have exhausted all cases, the proof is complete.
\end{proof}
\begin{remark}
We can use this result in the general case by splitting $X$ and $Y$ into positive and negative parts, though doing this does mean that may need to solve multiple smaller instances to solve the overall instance. Alternatively, we could reformulate $F$ to split $X$ and $Y$ appropriately.
\end{remark}
\begin{remark}
Note that if there are no solutions, this implies that iterating $F$ eventually proves it by converging to empty sets.
\end{remark}

Now, although the above result is very informative, it does not tell us about the bounds on $Z$ when $F$ converges. We now show that, unfortunately, iterating $F$ does not always produce optimal bounds:
\begin{example}\label{ex:classical-is-suboptimal}
Consider $\beta = 10$, $p = 2$ and $\emin \leq 0 \leq \emax$. (That is, 1-place decimal floating-point numbers.) Let $X = [1.2, 1.4] \cap \floats$, $Y = [1.2, 1.3] \cap \floats$ and $Z = [1.5, 1.8] \cap \floats$ and suppose $\fl$ is rounding to nearest, ties to even. We first show that $F(X, Y, Z) = (X, Y, Z)$. In the following, note that $\fl^{-1}[Z] = (1.45, 1.85]$:
\begin{align*}
    \fl^{-1}[Z] \MulInv \Box Y
    &= (1.45, 1.85] / [1.2, 1.3]\\
    &= (1.1\overline{153846}, 1.541\overline{6}]\\
    &\supseteq X,\\
    \fl^{-1}[Z] \MulInv \Box X
    &= (1.45, 1.85] / [1.2, 1.4]\\
    &= (1.03\overline{571428}, 1.541\overline{6}]\\
    &\supseteq Y,\\
    \fl[\Box X \Box Y]
    &= \fl[[1.2, 1.4] \cdot [1.2, 1.3]]\\
    &= \fl[[1.44, 1.82]]\\
    &= [1.4, 1.8] \cap \floats\\
    &\supseteq Z.
\end{align*}
From the definition of $F$, the above confirms that $F(X, Y, Z) = (X, Y, Z)$ indeed. It is also easy to verify that the bounds on $X$ and $Y$ are optimal, as per \Cref{lem:conditional-optimality}:
\begin{gather*}
\min X \otimes \max Y = \fl(1.2 \cdot 1.3) = \fl(1.56) = 1.6 \in Z,\\
\max X \otimes \min Y = \fl(1.4 \cdot 1.2) = \fl(1.68) = 1.7 \in Z.
\end{gather*}
However, if we enumerate all the products of the members of $X$ and $Y$, we find that none of them round to $\min Z = 1.5$:
\begin{table}[H]
    \centering
    \setlength\aboverulesep{0ex}
    \setlength\belowrulesep{0ex}
    \begin{tabular}{c|c c c c}
    \toprule
    $\cdot$ & 1.2 & 1.3 & 1.4\\
    \midrule
    1.2 & 1.44 & 1.56 & 1.68 \\ 
    1.3 & 1.56 & 1.69 & 1.82 \\ 
    \bottomrule
    \end{tabular}
    \begin{tabular}{c|c c c c}
    \toprule
    $\otimes$ & 1.2 & 1.3 & 1.4\\
    \midrule
    1.2 & 1.4 & 1.6 & 1.7 \\ 
    1.3 & 1.6 & 1.7 & 1.8 \\ 
    \bottomrule
    \end{tabular}
    \caption{Real and rounded products of the elements of $X$ and $Y$.}
    \label{tab:xy-products}
\end{table}
From \Cref{tab:xy-products}, we see that $X \otimes Y = \Set{1.4, 1.6, 1.7, 1.8}$, and so $Z \cap (X \otimes Y) = [1.6, 1.8] \cap \floats$ is a proper subset of $Z$ that is in fact a floating-point interval. Therefore $\Box X \times \Box Y \times \Box Z$ is not the smallest box enclosing the solution set.
\end{example}

\Cref{ex:classical-is-suboptimal} demonstrates that the classical algorithm does not necessarily produce optimal results. However, there is a simple strategy we can use to remedy this. To find optimal bounds on $Z$---that is, find the least and greatest values $z \in Z$ such that $x \otimes y = z$ is satisfied by some $x \in X$ and $y \in Y$---we can use binary search. Now, testing satisfiability is simple if we can find optimal bounds on any part: if a problem instance is unsatisfiable, then the optimal bounds are all empty sets. The key, then, is to find an efficient optimization method. In this paper, we develop such a procedure for the bounds on factors.

\section{Simplifying the problem}\label{sec:simp}

\subsection{Sufficient and necessary conditions}
In order to simplify the problem, we will need to precisely determine the conditions under which the division-based algorithm produces optimal results. In the second half of \Cref{ex:fp}, we encountered certain combinations of values for which the equation $x \otimes y = z$ has no solutions over the floating-point numbers. Unlike with the reals, there is no guarantee that we can find a value for $x$ satisfying the equality for any given $y$ and $z$. The existence of a solution is necessary for any optimal bound, however, and so the following definition will be useful:
\begin{definition}[Floating-point factors]
A floating-point number $x$ is a floating-point factor of a floating-point number $z$ if and only if $x \otimes y = z$ for some $y \in \overline{\floats}$. Given a set $Z \subseteq \overline{\floats}$, we say that $x$ is \emph{feasible} for $Z$ if and only if $x$ is a floating-point factor of some $z \in Z$. The set of all floating-point factors of the members of $Z$ is denoted $\Feas(Z)$.
\end{definition}
For the sake of brevity, we will typically simply write ``factor'' when it is clear from context that we are discussing floating-point factors. It is worth noting that sets of factors satisfy the following closure property:
\begin{lemma}\label{lem:feas-closed-under-negation}
For any $Z \subseteq \overline{\floats}$, the set $\Feas(Z)$ is closed under negation.
\end{lemma}
\begin{proof}
Let $Z \subseteq \overline{\floats}$. If $\Feas(Z)$ is empty, it is trivially closed under negation. Suppose $\Feas(Z)$ is nonempty instead, and let $x \in \Feas(Z)$. Then there is some $y \in \overline{\floats}$ such that $x \otimes y \in Z$. Therefore,
\begin{align*}
-x \otimes -y
&= \fl(-x \cdot -y)\\
&= \fl(xy)\\
&= x \otimes y \in Z,
\end{align*}
and so $-x \in \Feas(Z)$.
\end{proof}
\begin{corollary}
For any $x \in \overline{\floats}$ and $Z \subseteq \overline{\floats}$, we have
\begin{equation*}
\max{\Set{y \in \Feas(Z) \given y \leq x}} = -\min{\Set{y \in \Feas(Z) \given y \geq -x}}.
\end{equation*}
\end{corollary}
\begin{remark}
The above corollary implies that we only need one procedure to compute both the minimum and maximum of $[a, b] \cap \Feas(Z)$ for any $a, b \in \overline{\floats}$.
\end{remark}

The following two lemmas are crucial.  First, we prove that the real quotients from the division-based algorithm can still provide us with useful bounds.
\begin{lemma}\label{lem:interval-div-witnesses}
Let $x \in \overline{\floats}$, and let $Y, Z \subseteq \overline{\floats}$ where $Y$ is a floating-point interval. If $\fl(xy) \in Z$ for some $y \in \Box Y$, then
\begin{enumerate}
    \item $x \otimes s \geq \min Z$ for some $s \in Y$, and
    \item $x \otimes t \leq \max Z$ for some $t \in Y$.
\end{enumerate}
\end{lemma}
\begin{proof}
Suppose $\fl(xy) \in Z$ for some $y \in \Box Y$. Since $\min Y$ and $\max Y$ are floating-point numbers and $\min Y \leq y \leq \max Y$, by the monotonicity of rounding, we have $\min Y \leq \RD(y) \leq \max Y$ and $\min Y \leq \RU(y) \leq \max Y$. Since $Y$ is a floating-point interval, it thus follows that $\RD(y) \in Y$ and $\RU(y) \in Y$. We shall first deal with the special cases of $x$ being zero or infinite.

Suppose $x = 0$. Since $xy$ is defined by assumption, $y$ must be finite, and hence $xy = 0$. Thus $\fl(xy) = 0$. Since $y$ is finite, at least one of either $\RD(y)$ or $\RU(y)$ is finite, and thus either $x \RD(y) = 0$ or $x \RU(y) = 0$. Therefore either $x \otimes \RD(y) = 0$ or $x \otimes \RU(y) = 0$. Since $\fl(xy) = 0$ and $\min Z \leq \fl(xy) \leq \max Z$, the result follows.

Suppose $x = +\infty$ or $x = -\infty$ instead. Since $xy$ is defined by assumption, $y$ must be nonzero and hence either $xy = +\infty$ or $xy = -\infty$. Therefore $\fl(xy) = xy$. Since $y$ is nonzero, at least one of $\RU(y)$ or $\RD(y)$ is nonzero. If $\RD(y)$ is nonzero, then it has the same sign as $y$ and hence $x y = x \RD(y)$. Similarly, if $\RU(y)$ is nonzero, then it has the same sign as $y$ and hence $xy = x \RU(y)$. Therefore either $x \otimes \RD(y) = \fl(xy)$ or $x \otimes \RU(y) = \fl(xy)$, and hence the result.

We now handle the remaining case of $x$ being finite and nonzero. Suppose $x \in \floats^*$ instead. Then multiplication by $x$ is always defined. Note that $\min Y \leq y \leq \max Y$ and $\min Z \leq \fl(xy) \leq \max Z$. We shall consider each sign of $x$ separately:
\begin{itemize}
    \item Suppose $x$ is positive. Multiplying by $x$, it follows that
    \begin{equation*}
    x \min Y \leq xy \leq x \max Y,
    \end{equation*}
    and by the monotonicity of rounding, we have
    \begin{equation*}
    x \otimes \min Y \leq \fl(xy) \leq x \otimes \max Y.
    \end{equation*}
    Combining these bounds with our previous bounds on $\fl(xy)$, we obtain
    \begin{align*}
    x \otimes \min Y \leq &\fl(xy) \leq \max Z,\\
    \min Z \leq &\fl(xy) \leq x \otimes \max Y,
    \end{align*}
    and hence the result holds.
    \item Suppose $x$ is negative. Then, similarly, we have
    \begin{equation*}
    x \otimes \min Y \geq \fl(xy) \geq x \otimes \max Y.
    \end{equation*}
    Therefore, since $\fl(xy) \in Z$, it follows that
    \begin{align*}
    x \otimes \min Y \geq &\fl(xy) \geq \min Z,\\
    \max Z \geq &\fl(xy) \geq x \otimes \max Y,
    \end{align*}
    and thus we obtain the result.
\end{itemize}
Since we have proved the result for all possible cases, we are finished.
\end{proof}
Next, we show that the solution set is exactly the set of feasible quotients:
\begin{lemma}\label{lem:contains-feasible-factor-equivalence}
Let $x \in \overline{\floats}$, and let $Y, Z \subseteq \overline{\floats}$ be floating-point intervals. Then there exists $y \in Y$ such that $x \otimes y \in Z$ if and only if
\begin{enumerate}
    \item $x$ is feasible for $Z$, and
    \item $\fl(xa) \in Z$ for some $a \in \Box Y$.
\end{enumerate}
\end{lemma}
\begin{proof}
If $x \otimes y \in Z$ for some $y \in Y$, then $x \in \Feas(Z)$ by definition and $\fl(xy) = x \otimes y \in Z$ where $y \in \Box Y$ trivially. Suppose instead that $x$ is feasible for $Z$, and that $\fl(xa) \in Z$ for some $a \in \Box Y$. Then, by definition, there is some $w \in \overline{\floats}$ such that $x \otimes w \in Z$. Further, by \Cref{lem:interval-div-witnesses}, we have $x \otimes s \geq \min Z$ and $x \otimes t \leq \max Z$ for some $s, t \in Y$. If $w \in Y$, then the result follows immediately. Since $Y$ is a floating-point interval, if instead $w \notin Y$, then either $w < \min Y$ or $w > \max Y$. For the following, note that $\min Z \leq x \otimes w \leq \max Z$. We shall divide the problem into four cases, depending on the sign of $x$ and the region $w$ lies in:
\begin{enumerate}
    \item Suppose $w < \min Y$. Then $w < s$ and $w < t$.
    \begin{enumerate}
        \item Suppose $x \geq 0$. Multiplying, we obtain $x \otimes w \leq x \otimes t$, and thus $\min Z \leq x \otimes t \leq \max Z$. Since $Z$ is a floating-point interval, it follows that $x \otimes t \in Z$.
        \item Suppose $x < 0$. Then, we can multiply $w < s$ by $x$ to obtain $x \otimes w \geq x \otimes s$, and thus $\max Z \geq x \otimes s \geq \min Z$. Since $Z$ is a floating-point interval, we have $x \otimes s \in Z$.
    \end{enumerate}
    \item Suppose $w > \max Y$. Then $w > s$ and $w > t$.
    \begin{enumerate}
        \item Suppose $x \geq 0$. Then $x \otimes w \geq x \otimes s$, and thus $x \otimes s \in Z$ by the reasoning of case (1)(b).
        \item Suppose $x < 0$. Then $x \otimes w \leq x \otimes t$, and hence $x \otimes t \in Z$ by case (1)(a).
    \end{enumerate}
\end{enumerate}
Since we have either $x \otimes s \in Z$ or $x \otimes t \in Z$ in all cases, the result follows.
\end{proof}
\begin{remark}
This is not a solution on its own. Note that $a$ is not necessarily a floating-point number!
\end{remark}

This brings us to our first major result. The following theorem breaks the problem of computing optimal bounds into two simpler problems: computing bounds using the division-based algorithm, and then shrinking those bounds to the nearest feasible values.
\begin{theorem}\label{thm:classical-intersect-feasible}
Let $X, Y, Z \subseteq \overline{\floats}$ be nonempty floating-point intervals, and let
\begin{align*}
\hat{X} &= X \cap (\fl^{-1}[Z] \MulInv \Box Y),\\
X' &= \Set{x \in X \given \exists y \in Y\,(x \otimes y \in Z)}.
\end{align*}
Then $X' = \hat{X} \cap \Feas(Z)$.
\end{theorem}
\begin{proof}
By \Cref{lem:contains-feasible-factor-equivalence}, for all $x \in \overline{\floats}$, we have $x \in X'$ if and only if $x \in X$, $x \in \Feas(Z)$ and $x \in \fl^{-1}[Z] \MulInv \Box Y$, and the result follows immediately.
\end{proof}
\begin{remark}
This result implies that the division-based algorithm produces optimal bounds if and only if it produces feasible bounds. That is, $\min \hat{X} = \min X'$ if and only if $\min \hat{X}$ is feasible for $Z$. Therefore, the difficult case is when the bounds produced by the division-based algorithm are infeasible.
\end{remark}

\Cref{thm:classical-intersect-feasible} shows that we can solve \Cref{prob:main} by using the classical division-based algorithm to compute the set of quotients and then intersecting it with the set of factors. As such, the remainder of this paper will focus on solving this new problem:
\begin{problem}
Let $x \in \overline{\floats}$ and let $Z \subseteq \overline{\floats}$ be a floating point interval. What is the least (greatest) number feasible for $Z$ that is no less (greater) than $x$?
\end{problem}

\subsection{Narrowing down infeasibility}
Due to \Cref{thm:classical-intersect-feasible}, we would like to devise a method to find the factors nearest any given non-factor. To that end, we now identify some more practical conditions for feasibility. The following lemma gives a simple and direct test for feasibility, and also shows that floating-point multiplication and division are closely related:
\begin{lemma}\label{lem:equiv-for-feasible}
Let $x \in \floats^*$ and let $Z \subseteq \overline{\floats}$ be a floating-point interval. Let $z \in Z$. Then $x$ is feasible for $Z$ if and only if either $x \otimes \RD(z/x) \in Z$ or $x \otimes \RU(z/x) \in Z$.
\end{lemma}
\begin{proof}
If either $x \otimes \RD(z/x) \in Z$ or $x \otimes \RU(z/x) \in Z$, then $x$ is trivially feasible for $Z$. For the other half, suppose $x$ is feasible for $Z$. Then $x \otimes y \in Z$ for some $y \in \overline{\floats}$. Now, let $Y = \fl^{-1}[Z]/x$. Then $y \in Y$ and $z/x \in Y$, and since $Z$ is a floating-point interval, it follows that $Y$ is an interval. Therefore, if $y \leq z/x$, then $y \leq \RD(z/x) \leq z/x$, and so $\RD(z/x) \in Y$. If $y > z/x$ instead, then $y \geq \RU(z/x) \geq z/x$, and thus $\RU(z/x) \in Y$. Since either $\RD(z/x) \in Y$ or $\RU(z/x) \in Y$, we have either $x \otimes \RD(z/x) \in Z$ or $x \otimes \RU(z/x) \in Z$, as desired.
\end{proof}
\begin{corollary}
Let $x \in \floats^*$ and $z \in \overline{\floats}$. Then $x$ is a floating-point factor of $z$ if and only if either $x \otimes \RD(z/x) = z$ or $x \otimes \RU(z/x) = z$.
\end{corollary}
Additionally, it is worth mentioning that the above serves as a manual proof of a general statement of the automatically derived ``invertibility condition'' for multiplication of Brain et al.~\cite{brain2019}

Although it is clear that if the set of products is large, there will be many solutions, it is unclear how large exactly it needs to be. The following results give a more precise idea of how its diameter relates to the set of solutions. In the following lemma, we describe a lower bound on the diameter sufficient to ensure feasibility in all cases.
\begin{lemma}\label{lem:wide-z-implies-feasible}
Let $Z \subseteq \overline{\floats}$ be a floating-point interval, and let $x \in \floats^*$ and $z \in Z$. If $\abs{z/x} \leq \max \floats$ and $\diam \fl^{-1}[Z] > \abs{x}\beta^{Q(z/x)}$, then $x$ is feasible for $Z$.
\end{lemma}
\begin{proof}
If $z/x \in \overline{\floats}$, then $x \otimes (z/x) = z$, and hence $x$ is feasible for $Z$. Suppose $z/x \notin \overline{\floats}$ instead, and suppose $\abs{z/x} \leq \max \floats$ and $\diam \fl^{-1}[Z] > \abs{x}\beta^{Q(z/x)}$. First, note that $\RD(z/x) < z/x < \RU(z/x)$, and hence $z$ lies strictly between $x\RD(z/x)$ and $x\RU(z/x)$. More precisely, let $I = x \cdot [\RD(z/x), \RU(z/x)]$. Then $z \in I$, and hence $I \cap \fl^{-1}[Z]$ is nonempty. Since $Z$ is a floating-point interval, $\fl^{-1}[Z]$ is an interval. We shall now show that $\fl^{-1}[Z]$ is strictly wider than $I$:
\begin{align*}
\diam I
&= \abs{x} (\RU(z/x) - \RD(z/x))\\
&= \abs{x} \beta^{Q(z/x)}\\
&< \diam \fl^{-1}[Z].
\end{align*}
Therefore, since $I$ and $\fl^{-1}[Z]$ are overlapping intervals, by \Cref{lem:overlapping-interval-endpoints}, it follows that either $\min I \in \fl^{-1}[Z]$ or $\max I \in \fl^{-1}[Z]$. Hence either $x \otimes \RU(z/x) \in Z$ or $x \otimes \RD(z/x) \in Z$, and so $x$ is feasible for $Z$.
\end{proof}
The next lemma builds on the previous result to give a width independent of the denominator under which virtually all values are feasible.
\begin{lemma}\label{lem:very-wide-z-implies-feasible}
Let $Z \subseteq \overline{\floats}$ be a floating-point interval, and let $x \in \floats^*$ and $z \in Z$. If $\beta^\emin \leq \abs{z/x} \leq \max \floats$ and $\diam \fl^{-1}[Z] \geq \beta^{Q(z) + 1}$, then $x$ is feasible for $Z$.
\end{lemma}
\begin{proof}
Suppose $\beta^\emin \leq \abs{z/x} \leq \max \floats$ and $\diam \fl^{-1}[Z] \geq \beta^{Q(z) + 1}$. Then,
\begin{align*}
\abs{x} \beta^{Q(z/x)}
&= \abs{x} \beta^{E(z/x) - p + 1}\\
&\leq \abs{x} \abs{z/x} \beta^{1-p}\\
&= \abs{z} \beta^{1 - p}\\
&< \beta^{E(z) + 1} \beta^{1 - p}\\
&= \beta^{Q(z) + 1}\\
&\leq \diam \fl^{-1}[Z].
\end{align*}
Therefore, by \Cref{lem:wide-z-implies-feasible}, $x$ is feasible for $Z$.
\end{proof}
\begin{remark}
For any ordinary rounding function (e.g.\ $\RD$, $\RU$, or $\RN$), the requirement that $\diam \fl^{-1}[Z] \geq \beta^{Q(z) + 1}$ roughly corresponds to $Z$ containing at least $\beta$ floating-point numbers with the same exponent as $z$ (cf.\ Ziv et al.~\cite{ziv2003}).
\end{remark}

The next lemma provides an even stronger bound, under the condition that the significand of the numerator is no greater in magnitude than the significand of the denominator:
\begin{lemma}\label{lem:conditional-wide-z-implies-feasible}
Let $Z \subseteq \overline{\floats}$ be a floating-point interval, and let $x \in \floats^*$ and $z \in Z$. Let $m_x = x \beta^{-E(x)}$ and $m_z = z \beta^{-E(z)}$. If $\diam \fl^{-1}[Z] \geq \beta^{Q(z)}$, $\abs{m_z} \leq \abs{m_x}$, and $\beta^\emin \leq \abs{z/x} \leq \max \floats$, then $x$ is feasible for $Z$.
\end{lemma}
\begin{proof}
Suppose $\beta^\emin \leq \abs{z/x} \leq \max \floats$. If $\abs{m_z} = \abs{m_x}$, then $\abs{z/x} = \beta^{E(z) - E(x)}$ and hence $z/x \in \floats$, so $x$ is trivially feasible for $Z$. Suppose $\abs{m_z} < \abs{m_x}$ instead, and also suppose $\diam \fl^{-1}[Z] \geq \beta^{Q(z)}$. Then $\log_\beta \ufp(m_z/m_x) \leq -1$ and so by \Cref{lem:z-below-x-exponent}, it follows that $E(z/x) \leq E(z) - E(x) - 1$. Thus,
\begin{align*}
\abs{x}\beta^{Q(z/x)}
&= \abs{x}\beta^{E(z/x) - p + 1}\\
&\leq \abs{x}\beta^{E(z) - E(x) - p}\\
&= \abs{m_x}\beta^{Q(z) - 1}\\
&< \beta^{Q(z)}\\
&\leq \diam \fl^{-1}[Z].
\end{align*}
Therefore, by \Cref{lem:wide-z-implies-feasible}, $x$ is feasible for $Z$.
\end{proof}
\begin{remark}
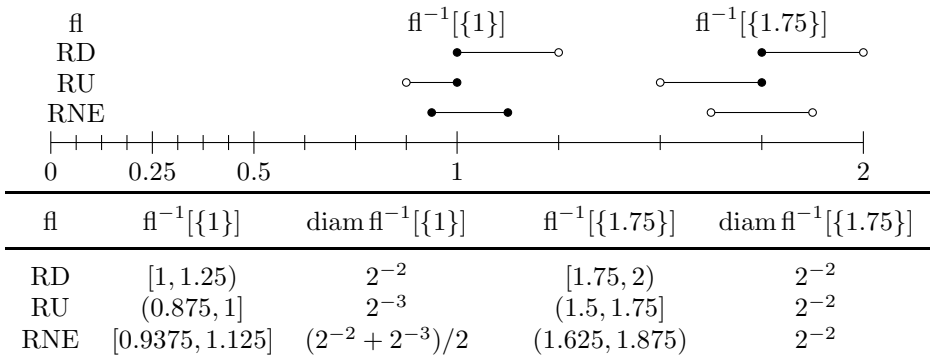
\begin{figure}\centering%
\begin{tikzpicture}[x=0.45\linewidth,outer xsep=0,inner xsep=0,every node/.style={inner xsep=0,outer xsep=0}]]
\draw (0, 0) -- (2, 0);
\foreach \x in {0, 0.25, 0.5, 1, 2} {
    \draw (\x, 0.15) -- (\x, -0.15) node[below]{$\x$};
}
\foreach \x in {1/16, 2/16, 3/16, 5/16, 6/16, 7/16, 5/8, 6/8, 7/8, 5/4, 6/4, 7/4} {
    \draw (\x, 0.1) -- (\x, -0.1);
}

\begin{scope}[yshift=4mm]

\begin{scope}
\draw[fill=black] (15/16, 0) circle (0.5mm) -- (9/8, 0) circle (0.5mm);
\draw (1.625, 0) -- (1.875, 0);
\draw[fill=white] (1.625, 0) circle (0.5mm);
\draw[fill=white] (1.875, 0) circle (0.5mm);
\node at (0.0625, 0){$\RNE$};
\end{scope}

\begin{scope}[yshift=8mm]
\draw (1, 0) -- (5/4, 0);
\draw[fill=black] (1, 0) circle (0.5mm);
\draw[fill=white] (5/4, 0) circle (0.5mm);

\draw (1.75, 0) -- (2, 0);
\draw[fill=black] (1.75, 0) circle (0.5mm);
\draw[fill=white] (2, 0) circle (0.5mm);

\node at (0.0625, 0){$\RD$};
\end{scope}

\begin{scope}[yshift=4mm]
\draw (7/8, 0) -- (1, 0);
\draw[fill=white] (7/8, 0) circle (0.5mm);
\draw[fill=black] (1, 0) circle (0.5mm);

\draw (1.5, 0) -- (1.75, 0);
\draw[fill=white] (1.5, 0) circle (0.5mm);
\draw[fill=black] (1.75, 0) circle (0.5mm);

\node at (0.0625, 0){$\RU$};
\end{scope}

\begin{scope}[yshift=12mm]
\node at (0.0625, 0){$\fl$};
\node at (1, 0){$\fl^{-1}[\Set{1}]$};
\node at (1.75, 0){$\fl^{-1}[\Set{1.75}]$};
\end{scope}
\end{scope}
\end{tikzpicture}
\begin{tabular}{c c c c c c}
\toprule
$\fl$ & $\fl^{-1}[\Set{1}]$ & $\diam \fl^{-1}[\Set{1}]$ && $\fl^{-1}[\Set{1.75}]$ & $\diam \fl^{-1}[\Set{1.75}]$\\
\midrule
$\RD$ & $[1, 1.25)$ & $2^{-2}$ && $[1.75, 2)$ & $2^{-2}$ \\
$\RU$ & $(0.875, 1]$ & $2^{-3}$ && $(1.5, 1.75]$ & $2^{-2}$ \\
$\RNE$ & $[0.9375, 1.125]$ & $(2^{-2} + 2^{-3})/2$ && $(1.625, 1.875)$ & $2^{-2}$ \\
\bottomrule
\end{tabular}
\caption{Preimages of $1$ and $1.75$ under floating-point rounding, with $(\beta, p, \emin, \emax) = (2, 3, -2, 1)$. $\RNE$ is rounding to nearest, breaking ties by choosing numbers with even integral significands.}\label{fig:preimage-diagram}
\end{figure}%
Any ``ordinary'' rounding function almost always satisfies the condition on the diameter on the preimage. Specifically, we always have $\diam \fl^{-1}[\Set{z}] = \beta^{Q(z)}$ unless $\abs{z}$ is normal and a power of $\beta$. This can be seen in \Cref{fig:preimage-diagram}; note the different diameters of the preimages of 1 under rounding.
\end{remark}
Note that \Cref{lem:wide-z-implies-feasible,lem:conditional-wide-z-implies-feasible} are both easier to apply when using standard rounding modes, or more specifically, when the preimage under rounding of any floating-point number is not too narrow. Although the usual rounding functions are both faithful and monotonic, and these are indeed very useful properties, they are not enough on their own:
\begin{example} We shall construct an ill-behaved rounding function which is still both faithful and monotonic. Let $f \from \overline{\reals} \to \overline{\floats}$ be a function such that
\begin{equation*}
f(x) =
\begin{cases}
\RD(x) & \text{if } x \leq \max \floats,\\
+\infty & \text{otherwise}.
\end{cases}
\end{equation*}
Since $\RD$ is nondecreasing and faithful, it is easy to see that $f$ is also nondecreasing and faithful. However, the only real that it rounds to $\max \floats$ is $\max \floats$ itself. That is, we have $f^{-1}[\Set{\max \floats}] = \Set{\max \floats}$, and so the preimage under rounding has zero diameter. Further, given $x, y \in \overline{\floats}$, we have $x \otimes y = \max \floats$ if and only if $xy = \max \floats$ exactly. This means that finding the floating-point factors of $\max \floats$ requires integer factorization. Specifically, since $\max \floats = (\beta^p - 1)\beta^\qmax$, the floating-point factors of $\max \floats$ are exactly the floating-point numbers with a nonnegative exponent (i.e.\ integers) whose integral significands divide $\beta^p - 1$.
\end{example}

As demonstrated above, faithfulness and monotonicity are not enough to ensure that a rounding function has a tractable preimage. We therefore restrict our attention to a better-behaved class of rounding functions. In particular, we are interested in rounding functions that allocate to each floating-point number at least half of the gap around it:
\begin{definition}[Equitable rounding]\label{def:equitable}
A rounding function $f$ is equitable if it is faithful, monotonic, and $\diam f^{-1}[\Set{x}] \geq \beta^{Q(x) - k}$ where $k = 1$ if $\abs{x\beta^{-E(x)}} = 1$ or $k = 0$ otherwise for all $x \in \floats^*$.
\end{definition}
\begin{remark}
Note that we do not require the preimages under rounding of $0$, $+\infty$, and $-\infty$ to have a minimum diameter. This is because they already have every nonzero finite floating-point number as a floating-point factor, and so we do not need their preimages at all.
\end{remark}
Given this definition, the following is then straightforward (albeit tedious) to prove:
\begin{lemma}
$\RD$ and $\RU$ are equitable rounding functions.
\end{lemma}
However, $\RN$ is not necessarily equitable, since it is unclear what rounding to nearest means for numbers outside $[\min \floats, \max \floats]$ and thus our definition does not fully specify the preimages of $\min \floats$ and $\max \floats$ under $\RN$.
For instance, we could choose $\RN$ such that $\RN^{-1}[\Set{x}] = ((x^- + x)/2, x]$ when $x = \max \floats$. This is clearly not equitable, since $\diam \RN^{-1}[\Set{x}] = \beta^\qmax/2$. However, this inconvenience is not as important as it may seem. As mentioned earlier, IEEE 754-style rounding to nearest specifies that a number rounds to an infinity if and only if its magnitude is no less than $(\max \floats + \beta^{\emax + 1})/2$. This in turn means that the preimages of $\min \floats$ and $\max \floats$ under rounding have a diameter of $\beta^\qmax$, as required for equitability. In particular, the preimages of $\min \floats$ and $\max \floats$ under any equitable rounding to nearest are at least as large as those under IEEE 754-style rounding:
\begin{lemma}
$\RN$ is an equitable rounding function if and only if $\RN^{-1}[\Set{+\infty}] \subseteq I$ and $\RN^{-1}[\Set{-\infty}] \subseteq -I$ where $I = [(\max \floats + \beta^{\emax + 1})/2, +\infty]$.
\end{lemma}
Therefore, since IEEE 754-style systems are virtually the only ones in use, we do not consider equitability to be an especially onerous requirement. 

Now, under the assumption of equitable rounding, we can substantially refine the conclusion of \Cref{lem:conditional-wide-z-implies-feasible}.
\begin{lemma}\label{lem:equitable-z-below-x}
Let $x \in \floats^*$ and $z \in \overline{\floats}$, and let $m_x = x \beta^{-E(x)}$ and $m_z = z \beta^{-E(z)}$. If $\fl$ is equitable and $\beta^\emin \leq \abs{z/x} \leq \max \floats$ and $\abs{m_z} \leq \abs{m_x}$, then either $\abs{m_z} = 1$ or $x$ is a floating-point factor of $z$.
\end{lemma}
\begin{proof}
Suppose $\fl$ is equitable and $\beta^\emin \leq \abs{z/x} \leq \max \floats$ and $\abs{m_z} \leq \abs{m_x}$. Suppose also that $\abs{m_z} \neq 1$. Then, by the definition of equitable rounding, we have $\diam \fl^{-1}[\Set{z}] \geq \beta^{Q(z)}$. Therefore by \Cref{lem:conditional-wide-z-implies-feasible}, it follows that $x$ is feasible for $\Set{z}$, and hence the result.
\end{proof}
\begin{remark}
Note that here we show that $x$ is a factor of $z$, whereas \Cref{lem:wide-z-implies-feasible,lem:conditional-wide-z-implies-feasible} only prove that it is feasible for some superset $Z$.
\end{remark}
Combining the previous results, we obtain a much clearer picture of what it means when a number is infeasible:
\begin{theorem}\label{thm:infeasible-summary}
Let $Z \subseteq \overline{\floats}$ be a floating-point interval, and let $x \in \floats^*$ and $z \in Z$. Let $m_x = x\beta^{-E(x)}$ and $m_z = z\beta^{-E(z)}$. If $\fl$ is equitable and $\abs{z/x} \leq \max \floats$ and $x$ is not feasible for $Z$, then either
\begin{enumerate}
    \item $\abs{z/x} < \beta^\emin$, or
    \item $1 = \abs{m_z} < \abs{m_x}$ and $\diam \fl^{-1}[Z] < \beta^{Q(z)}$, or
    \item $\abs{m_x} < \abs{m_z}$ and $\diam \fl^{-1}[Z] < \beta^{Q(z) + 1}$.
\end{enumerate}
\end{theorem}
\begin{proof}
Follows immediately from \Cref{lem:wide-z-implies-feasible,lem:conditional-wide-z-implies-feasible,lem:equitable-z-below-x}.
\end{proof}
\begin{remark}
When rounding the quotient produces a subnormal number, as in case (1), the loss of significant digits can drastically magnify the error and make $x$ infeasible unless $Z$ is exceptionally wide. This makes subnormal quotients difficult to handle in general, since the difference between $z$ and $x \otimes \fl(z/x)$ may be as large as $x$ itself.

Since we assume $\fl$ is equitable, case (2) is the only instance where $x$ is infeasible and $\abs{m_z} \leq \abs{m_x}$. Although it is a very constrained special case, it shall take some work to dispense with, as we will see later. The most common cause of infeasibility is case (3). In this instance, we have $\abs{m_x} \leq \abs{m_z}$, but $Z$ is too narrow to provide a direct solution, likely due to containing fewer than $\beta$ numbers. Note that this means that having more than one number in $Z$ usually suffices for feasibility when $\beta = 2$.
\end{remark}
\Cref{thm:infeasible-summary} gives a straightforward classification of cases where infeasibility can occur. Although we are not able to give a full treatment of the subnormal quotient case in this paper, we will present an efficient solution for the other two cases. In the next section, we develop criteria for feasibility that are more amenable to a computational solution.

\section{Analyzing the error}\label{sec:hardcase}
\begin{figure}[b]
\centering
\begin{tikzpicture}
\pgfplotsset{set layers}
\begin{axis}[
  scale only axis,
  axis x line*=bottom,
  axis y line*=left,
  width=0.89\textwidth,
  height=0.8\textwidth,
  ymin=-0.2, ymax=0.2,
  xmin=0, xmax=280,
  ytick={-0.1,0,0.1},
  yticklabels={$2.4$,$2.5$,$2.6$},
  minor ytick={-0.05,0.05},
  grid=major,
  xtick={0,10,100,190,280},
  xticklabels={0,0.1,1,10,100},
  legend entries={$x\RU(2.5/x)$,$x\RD(2.5/x)$},
  legend pos=north west,
  mark size=0.8pt,
  xlabel=$x$,
]
    \addplot+[
      ycomb,
      opacity=0.5,
      line width=0.5pt,
      every mark/.append style={solid,opacity=1}
    ] table[y index=2]{combined-rounding.tsv};
    \addplot+[
      ycomb,
      opacity=0.5,
      line width=0.5pt,
      every mark/.append style={solid,opacity=1}
    ] table[y index=3]{combined-rounding.tsv};
\end{axis}
\begin{axis}[
  scale only axis,
  width=0.89\textwidth,
  height=0.8\textwidth,
  ymin=-0.2, ymax=0.2,
  xmin=0, xmax=280,
  axis y line*=right,
  axis x line*=top,
  ytick=\empty,
  minor ytick={-0.05,0.05},
  xtick={10,100,190,280},
  xticklabels={$\beta^{-1}$,$\beta^0$,$\beta^1$,$\beta^2$},
]
\end{axis}
\end{tikzpicture}
\caption{Graphs of exact products of floating-point divisors $x$ and rounded quotients $\RD(z/x)$ and $\RU(z/x)$ where $(\beta, p, \emin, \emax) = (10, 2, -1, 1)$ and $z = 2.5$. The x-axis shows all positive finite floating-point numbers, spaced equally.}
\label{fig:exact-products-error}
\end{figure}
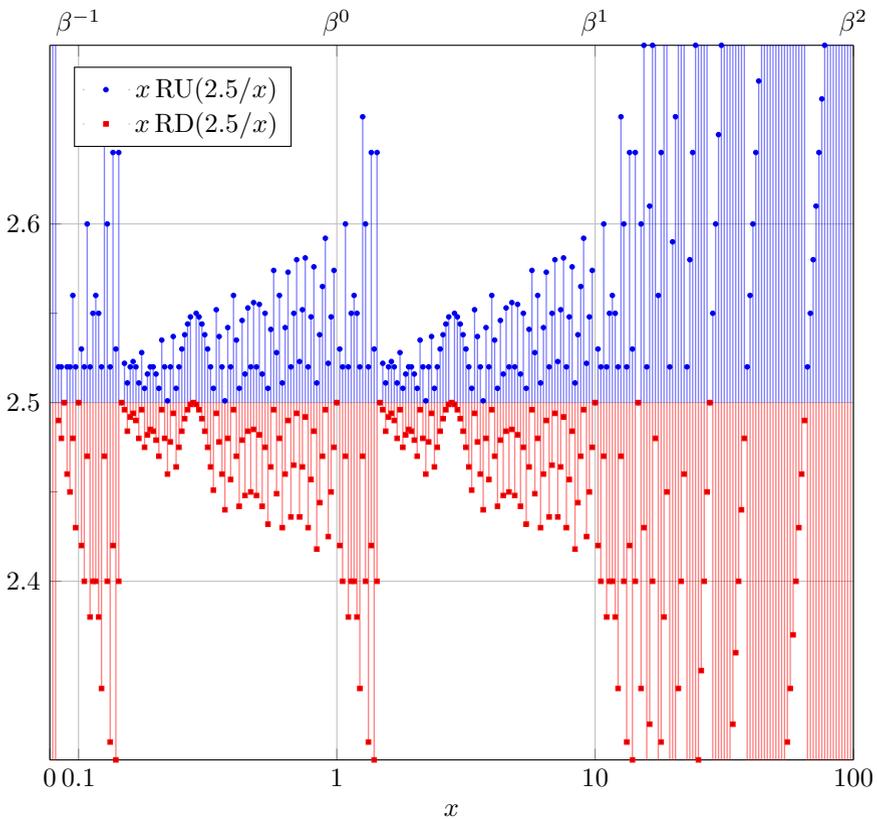
When computing a real number, we are often forced to approximate it due to the limitations of the number system. For systems as complicated as the floating-point numbers, the error in this approximation---the \emph{rounding error}---can seem unpredictable. The notion is simple: given a rounding function $f$, the rounding error of some $x \in \reals$ with respect to $f$ is just $f(x) - x$. However, it is difficult to tell from this what values the rounding error can take and how it varies with $x$. In our case, we are more precisely interested in a ``round-trip'' error. Roughly speaking, according to \Cref{lem:equiv-for-feasible}, a floating-point number is feasible for a given set if and only if we can undo division by using multiplication. In other words, it is feasible if and only if the error in the ``round-trip'' is small enough.

For any given floating-point numerator $z$ and denominator $x$, any error in attempting to round-trip multiplication using division stems from rounding the quotient $z/x$. This is easy to show: since $\fl$ is assumed to be faithful, if $z/x$ is also a floating-point number, then $x \otimes \fl(z/x) = x$ exactly. Consequently, any instance where $x \otimes \fl(z/x) \neq x$ implies that $z/x$ is not exactly representable. This is the common case, since the quotient of two floating-point numbers rarely fits into the available number of digits.\footnote{More precisely, if $\beta$ is prime (such as when $\beta = 2$), then the quotient either fits or has no finite base-$\beta$ representation. For a proof, see \Cref{lem:quotients-in-prime-bases}.} However, the product is also rounded, so an inexact quotient may still be close enough. For instance, if we round downward, it is sufficient that the product lies in $[z, z^+)$, where $z^+$ is the next floating-point number after $z$. A number of such examples can be seen in \Cref{fig:exact-products-error}, as well as a few instances where the product has no error (note that the graph is centered on $z$). Other patterns are visible as well, such as periodic behavior and a reduction in error whenever the significand of $x$ is greater than the significand of $z$.

In the remainder of this section, we shall develop ways of manipulating this round-trip error, with the goal of using integer optimization to find floating-point factors. The key insight comes from \Cref{lem:equiv-for-feasible}, which tells us that feasibility can be decided using only $\fl$ and upward- or downward-rounded quotients. Further, since we assume $\fl$ is faithful, we also know that multiplication is also either rounded upward or downward. Therefore, we only really need to be concerned with rounding errors relative to $\RU$ and $\RD$ to solve the problem. However, the definitions of $\RU$ and $\RD$ are very cumbersome to manipulate directly. In order to make progress, we must first find a more practical description of rounding. We will do this by expressing rounding error in terms of the remainder of floored division. Furthermore, for the sake of simplicity, we will consider rounding over floating-point numbers with unbounded exponents.\footnote{Due to the difficulties with subnormal numbers mentioned in the remark to \Cref{thm:infeasible-summary}, the specific approach presented here does not benefit from restricting exponents, and doing so would drastically complicate the presentation.} We will also show how this relates to the original case.

\subsection{Further preliminaries}
\subsubsection{The mod operator}
We shall use a generalized notion of the remainder of a division. Given a numerator $n$, denominator $d$, and quotient $q$, a remainder $r$ satisfies $n = qd + r$ and $\abs{r} < \abs{d}$. These quantities are traditionally integers, but we extend the operation to the real numbers. Given $n$ and $d$, we shall determine the remainder $r$ by choosing the quotient $q = \floor{n/d}$. Thus, we define the binary operator $\bmod$ as follows:
\begin{definition}[Remainder of floored division]
For any numerator $x \in \reals$ and denominator $y \in \reals^*$, we define $(x \bmod y) = x - y\floor{x/y}$.
\end{definition}
\begin{remark}
To avoid excessive parenthesization, the mod operator defined here is nonassociative and has the lowest precedence. That is, within the parentheses, everything before mod is the first argument, and everything after mod is the second argument.
\end{remark}
Since this is an extension of the usual notion of the remainder to \emph{real} numerators and denominators, not all the same properties will hold exactly. However, there is still a close connection with modular arithmetic, which the next two lemmas will illustrate. Firstly, the remainder is periodic in the numerator, with a period equal to the denominator. That is, we can add or subtract integer multiples of the denominator to the numerator without affecting the result:

\begin{lemma}
For any $x \in \reals$, $y \in \reals^*$, and $n \in \integers$, $(x + ny \bmod y) = (x \bmod y)$.
\end{lemma}
\begin{proof} Follows directly by a straightforward computation.
\end{proof}
Secondly, if we restrict ourselves wholly to the integers and fix the denominator, equality of remainders is equivalent to congruence:
\begin{lemma}
For all $a, b \in \integers$, $n \in \integers^*$, we have $(a \bmod n) = (b \bmod n)$ if and only if $a \equiv b \pmod{\abs{n}}$.
\end{lemma}
\begin{proof}
If $(a \bmod n) = (b \bmod n)$, then $a - n \floor{a/n} = b - n \floor{b/n}$, and hence $a \equiv b \pmod{\abs{n}}$. Suppose instead that $a \equiv b \pmod{\abs{n}}$. Then there exist $p, q, r \in \integers$ such that $a = pn + r$ and $b = qn + r$. As $(pn + r \bmod n) = (qn + r \bmod n) = (r \bmod n)$, the result follows.
\end{proof}
As might be expected from being so closely related to congruence, the $\bmod$ operator has many useful properties, though we shall only need a few. The following properties hold for all $x \in \reals$ and $a, y \in \reals^*$:
\begin{align}
\abs{(x \bmod y)} < \abs{y} && \text{(upper bound)}\\
(ax \bmod ay) = a(x \bmod y)&&\text{(distributive law)}\\
\text{if } x, y \in \integers, \text{ then } (x \bmod y) \in \integers&&\text{(integrality)}\\
\text{if } (x \bmod y) \neq 0, \text{ then } (-x \bmod y) = y - (x \bmod y) &&\text{(complement)}\\
\shortintertext{It is also worth noting that the remainder (when nonzero) has the same sign as the denominator:}
y(x \bmod y) \geq 0
\shortintertext{Finally, the following results simplify some common cases of the numerator:}
(0 \bmod y) = 0\\
(x \bmod y) = 0 \text{ if and only if } x = ny \text{ for some } n \in \integers\\
(x \bmod y) = x \text{ if and only if } 0 \leq x/y < 1\\
(x \bmod y) = x \text{ if and only if } xy \geq 0 \text { and } \abs{x} < \abs{y}
\end{align}
The proofs of these statements are simple and follow easily from the definition of $\bmod$.

\subsubsection{Floating-point numbers with unbounded exponents}
We define these sets almost identically to the bounded case. For compatibility with their bounded counterparts, we inherit the base $\beta$ and precision $p$:
\begin{definition}[Unbounded floating-point numbers]
We define the following sets:
\begin{align*}
    \floats_\infty^* &= \Set{M \beta^q \given M, q \in \integers, 0 < \abs{M} < \beta^p},\\
    \floats_\infty &= \floats_\infty^* \cup \Set{0},\\
    \overlineinf{\floats} &= \floats_\infty \cup \Set{-\infty, +\infty}.
\end{align*}
\end{definition}
\begin{remark}
Note that the unbounded sets are strict supersets of their bounded counterparts. That is, we have $\floats^* \subset \floats_\infty^*$, $\floats \subset \floats_\infty$, and $\overline{\floats} \subset \overlineinf{\floats}$.
\end{remark}
The unbounded floating-point numbers satisfy the following properties for all $x \in \overlineinf{\floats}$ and $n \in \integers$:
\begin{align}
    \text{if } \abs{n} \leq \beta^p, \text{then } n &\in \floats_\infty\\
    \text{if } \beta^\emin \leq \abs{x} \leq \max \floats, \text{ then } x &\in \floats^*\\
    x\beta^n &\in \overlineinf{\floats} && \text{(scaling by powers of $\beta$)}\\
    \text{if } x \in \floats_\infty^*, \text{ then } x/{\ulp(x)} &\in \integers && \text{(integral significand)}
\end{align}
Proving the existence of a unique upward or downward rounding takes a little more work in the unbounded case, since the sets are infinite, so their subsets might not have a minimum or maximum. For instance, the set $(0, 1] \cap \overlineinf{\floats}$ has no minimum, since it contains every $\beta^{-n}$ where $n \geq 0$. However, we only have finitely bounded infinite sets around zero:
\begin{lemma}\label{lem:closed-interval-intersect-floats-is-finite}
Let $a, b \in \reals$. If $0 \notin [a, b]$, then $[a, b] \cap \overlineinf{\floats}$ is a finite set.
\end{lemma}
\begin{proof}
Let $S = [a, b] \cap \overlineinf{\floats}$. If $S$ is empty, then the result follows trivially. Suppose $S$ is nonempty instead and suppose that $0 \notin [a, b]$. Then $a \leq b$, and $a$ and $b$ must be nonzero and have the same sign. We shall consider the case of each sign separately.

Suppose $a$ and $b$ are positive. Let $e_a = \floor{\log_\beta a}$ and $e_b = \floor{\log_\beta b}$, and let
\begin{equation*}
   T = \Set{M \beta^{e-p+1} \given M, e \in \integers, 0 < M < \beta^p, e_a \leq e \leq e_b + p - 1}. 
\end{equation*}
Clearly, $T$ is finite. Now, let $s \in S$. Since $s$ is positive, by definition we have $s = M \beta^q$ for some $M, q \in \integers$ where $0 < M < \beta^p$. Thus, to show that $s \in T$, it suffices to show that $e_a - p + 1 \leq q \leq e_b$. Therefore,
\begin{align*}
    \beta^q \leq M \beta^q = s \leq b < \beta^{e_b + 1}
\end{align*}
so $q < e_b + 1$ and hence $q \leq e_b$. Similarly,
\begin{align*}
    \beta^{e_a} \leq a \leq s = M\beta^{q} < \beta^{q + p}
\end{align*}
so $e_a - p < q$ and hence $e_a - p + 1 \leq q$. Therefore $s \in T$, and hence $S \subseteq T$, and since $T$ is finite, so is $S$.

Suppose instead $a$ and $b$ are negative. Then $-b$ and $-a$ are positive, so $0 \notin [-b, -a]$, and thus $[-b, -a] \cap \overlineinf{\floats}$ is finite. Since $\overlineinf{\floats}$ is closed under negation, we have $[-b, -a] \cap \overlineinf{\floats} = -([a, b] \cap \overlineinf{\floats}) = -S$, and hence $S$ is also finite.
\end{proof}
\begin{corollary}
For any $x \in \overline{\reals}$, the sets $\Set{y \in \overlineinf{\floats} \given y \leq x }$ and $\Set{y \in \overlineinf{\floats} \given y \geq x }$ have a minimum and a maximum.
\end{corollary}
We can now define downward and upward rounding over the unbounded floating-point numbers:
\begin{definition}[Unbounded rounding]
We define the functions $\RD_\infty, \RU_\infty \from \overline{\reals} \to \overlineinf{\floats}$ as follows:
\begin{align*}
\RD_\infty(x) &= \max{\Set{y \in \overlineinf{\floats} \given y \leq x }},\\
\RU_\infty(x) &= \min{\Set{y \in \overlineinf{\floats} \given y \geq x }}.
\end{align*}
\end{definition}
As expected, these functions are equal to their bounded counterparts over the range of the normal numbers:
\begin{lemma}
For all $x \in \overline{\reals}$, if $\beta^\emin \leq \abs{x} \leq \max \floats$, then $\RD(x) = \RD_\infty(x)$ and $\RU(x) = \RU_\infty(x)$.
\end{lemma}

We also have a similar gap property to \Cref{eq:gap-property}:
\begin{lemma}
For all $x \in \overline{\reals}$, we have either $\RD_\infty(x) = \RU_\infty(x) = x$ or $\RU_\infty(x) - \RD_\infty(x) = \ulp(x)$.
\end{lemma}
Note that the gap is always finite, since having unbounded exponents means that every finite real rounds to another finite real. Only $+\infty$ and $-\infty$ round to themselves.

\subsubsection{Plausibility}
In order to effectively translate our problem into the realm of unbounded floating-point numbers, we will also need a counterpart to our earlier concept of feasibility. Note that we keep the same rounding function: even though $\fl$ is not surjective with respect to the unbounded floating-point numbers, it is compatible since it already maps every extended real to a bounded floating-point number.
\begin{definition}[Plausibility]
Given $Z \subseteq \overlineinf{\floats}$ and $x \in \overlineinf{\floats}$, we say that $x$ is \emph{plausible} for $Z$ if and only if $\fl(xy) \in Z$ for some $y \in \overlineinf{\floats}$. We denote the set of all unbounded floating-point numbers plausible for $Z$ by $\Plaus(Z)$.
\end{definition}
\begin{remark}
If $Z \subseteq \overline{\floats}$, then we immediately have $Z \subseteq \Feas(Z) \subseteq \Plaus(Z)$.
\end{remark}
Similarly to the sets of factors, we have the following closure property:
\begin{lemma}\label{lem:plausible-closure}
For all $Z \subseteq \overlineinf{\floats}$, the set $\Plaus(Z)$ is closed under negation and scaling by powers of $\beta$.
\end{lemma}
Due to exponents being unrestricted, we have the following:
\begin{lemma}
For all $Z \subseteq \overlineinf{\floats}$, we have $1 \in \Plaus(Z)$ if and only if $Z$ is nonempty.
\end{lemma}
Perhaps most importantly, we have an analogue to \Cref{lem:equiv-for-feasible} (the proof is identical, \emph{mutatis mutandis}):
\begin{lemma}\label{lem:equiv-for-plausible}
Let $Z \subseteq \overline{\floats}$ be a floating-point interval, and let $x \in \floats^*_\infty$ and $z \in Z$. Then $x \in \Plaus(Z)$ if and only if either $\fl(x \RD_\infty(z/x)) \in Z$ or $\fl(x \RU_\infty(z/x)) \in Z$.
\end{lemma}
We also have an analogue to \Cref{lem:conditional-wide-z-implies-feasible}:
\begin{lemma}\label{lem:conditional-wide-z-implies-plausible}
Let $Z \subseteq \overline{\floats}$ be a floating-point interval, and let $x \in \floats^*_\infty$ and $z \in Z$. Let $m_x = x/ulp(x)$ and $m_z = z/\ulp(z)$. If $\diam \fl^{-1}[Z] \geq \ulp(z)$ and $\abs{m_z} \leq \abs{m_x}$, then $x$ is plausible for $Z$.
\end{lemma}

As expected, feasibility and plausibility agree when the quotients are normal:
\begin{lemma}\label{lem:normal-implies-plausible-equiv-feasible}
Let $Z \subseteq \overline{\floats}$ be a floating-point interval, and let $x \in \floats^*$ and $z \in Z$. If $\beta^\emin \leq \abs{z/x} \leq \max \floats$, then $x \in \Feas(Z)$ if and only if $x \in \Plaus(Z)$.
\end{lemma}
\begin{proof}
If $x \in \Feas(Z)$, then $x \in \Plaus(Z)$ trivially. Suppose $x \in \Plaus(Z)$ and $\beta^\emin \leq \abs{z/x} \leq \max \floats$. Then $\RD(z/x) = \RD_\infty(z/x)$ and $\RU(z/x) = \RU_\infty(z/x)$, so by \Cref{lem:equiv-for-plausible},
we have either
\begin{align*}
    x \otimes \RD(z/x) &= \fl(x \RD_\infty(z/x)) \in Z,\\
    \shortintertext{or}
    x \otimes \RU(z/x) &= \fl(x \RU_\infty(z/x) \in Z,
\end{align*}
and therefore $x \in \Feas(Z)$ by definition.
\end{proof}

Since exponents are unbounded, there is always a next and a previous plausible number. We will find it useful to have a function giving the smallest plausible number greater than some lower bound. (Note that $\overline{\floats}$ being finite means that we cannot have a similar total function for the feasible numbers.) First, we prove its existence:
\begin{lemma}
Let $Z \subseteq \overline{\floats}$ be a floating-point interval, and let $x \in \reals^*$. Let $A = \Set{y \in \Plaus(Z) \given y \geq x}$. Then $A$ is nonempty and has a minimum if and only if $Z$ is nonempty.
\end{lemma}
\begin{proof}
If $Z$ is empty, then $\Plaus(Z)$ is trivially empty and hence so is $A$. Suppose instead that $Z$ is nonempty and let $z \in Z$ and let $t = z/s$ where
\begin{equation*}
    s =
\begin{cases}
\beta \ufp(x) & \text{if } x > 0,\\
-\ufp(x) & \text{otherwise.}
\end{cases}
\end{equation*}
Clearly, $x \leq s$. Now, since $\abs{s}$ is a power of $\beta$ and $z \in \overlineinf{\floats}$, we have $s, t \in \overlineinf{\floats}$. Therefore $\fl(st) = \fl(z) = z$, and hence $s, t \in \Plaus(Z)$ by definition and so $s \in A$.

Now, let $L = [x, s] \cap \Plaus(Z)$ and $L' = [x, s] \cap \overlineinf{\floats}$. Since $x$ and $s$ have the same sign by definition, by \Cref{lem:closed-interval-intersect-floats-is-finite}, it follows that $L'$ is finite. Hence, since $s \in L \subseteq L'$, it follows that $L$ is also finite, and since it is nonempty, it has a minimum. Now, let $R = (s, +\infty] \cap \Plaus(Z)$. Since $\min L \leq s$, it follows that $\min L < r$ for all $r \in R$. Thus, since $A = L \cup R$, we have $\min L \leq a$ for all $a \in A$, and hence $\min A = \min L$.
\end{proof}
With this result, we can now define our ``round up to plausible'' function:
\begin{definition}
For each nonempty $Z \subseteq \overline{\floats}$, we define the function $\PlausRU{Z} \from \reals^* \to \overlineinf{\floats}$ by
\begin{equation*}
    \PlausRU{Z}(x) = \min{\Set{y \in \Plaus(Z) \given y \geq x}}.
\end{equation*}
\end{definition}
It is easy to see from the definition that these functions are nondecreasing:
\begin{lemma}
For all $Z \subseteq \overline{\floats}$ and $x, y \in \reals^*$, if $x \leq y$, then $\PlausRU{Z}(x) \leq \PlausRU{Z}(y)$.
\end{lemma}
Thus, since plausible numbers are a superset of feasible numbers, we have the following:
\begin{lemma}\label{lem:plausible-is-lower-bound}
Let $Z \subseteq \overline{\floats}$, $x \in \reals^*$ and $y \in \Feas(Z)$. Then, if $x \leq y$, then $\PlausRU{Z}(x) \leq y$. 
\end{lemma}
That is, each of these functions gives a lower bound on the next feasible number.
\subsection{Relating feasibility, plausibility, and rounding errors}
We begin by giving expressions for rounding in terms of remainders. Note that the following result directly gives a rounding error term.
\begin{lemma}\label{lem:rounding-in-terms-of-mod}
For all $x \in \reals^*$, we have
\begin{align*}
\RD_\infty(x) &= x - (x \bmod \ulp(x)),\\
\RU_\infty(x) &= x + (-x \bmod \ulp(x)).
\end{align*}
\end{lemma}
\begin{proof}
Let $x' = x - (x \bmod \ulp(x))$, $x'' = x + (-x \bmod \ulp(x))$ and $M = x/\ulp(x)$. Then, rearranging using the definition of $\bmod$, we obtain $x' = \ulp(x) \floor{M}$ and $x'' = \ulp(x) \ceil{M}$ and thus $x' \leq x \leq x''$. Since $\beta^{p-1} \leq \abs{M} < \beta^p$, it follows that $\abs{\floor{M}} \leq \beta^p$ and $\abs{\ceil{M}} \leq \beta^p$. Thus $\floor{M}, \ceil{M} \in \floats_\infty$, and since $\ulp(x)$ is a power of $\beta$, we have $x', x'' \in \floats_\infty$. Thus, by the definitions of $\RD_\infty$ and $\RU_\infty$,
\begin{equation*}
x' \leq \RD_\infty(x) \leq x \leq \RU_\infty(x) \leq x''.   
\end{equation*}
Now, if $M \in \integers$, then $x' = x''$ and hence the result follows immediately. Suppose $M \notin \integers$ instead. Then $x \notin \floats_\infty$, and so $\RU_\infty(x) - \RD_\infty(x) = \ulp(x) = x'' - x'$. Therefore,
\begin{equation*}
    x' \leq \RD_\infty(x) \leq x \leq \RD_\infty(x) + \ulp(x) \leq x' + \ulp(x),
\end{equation*}
and hence $x' = \RD_\infty(x)$ and $x'' = \RU_\infty(x)$, as desired.
\end{proof}
The above result allows us to manipulate the rounding error using the various laws obeyed by the mod operator. The following result gives us an expression for the round-trip error between any two consecutive plausible numbers.
\begin{lemma}\label{lem:round-trip-up-to-next-plausible}
Let $Z \subseteq \overline{\floats}$ and $z \in Z$. Let $x \in \reals^*$. If $z \in \floats^*$, then for all $t \in [x, \PlausRU{Z}(x)]$, we have $\ufp(x) \leq \abs{t} \leq \beta \ufp(x)$ and
\begin{align*}
    t \RD_\infty(z/t) &= z - (z \bmod t \ulp(z/x)),\\
    t \RU_\infty(z/t) &= z + (-z \bmod t \ulp(z/x)).
\end{align*}
\end{lemma}
\begin{proof}
Suppose $z \in \floats^*$ and let $t \in [x, \PlausRU{Z}(x)]$. Let $m_x = x/\ufp(x)$, $m_z = z/\ufp(z)$, and $m_t = t/\ufp(t)$. By \Cref{lem:rounding-in-terms-of-mod} and the distributivity of multiplication over $\bmod$,
\begin{align*}
    t \RD_\infty(z/t) &= z - (z \bmod t \ulp(z/t)),\\
    t \RU_\infty(z/t) &= z + (-z \bmod t \ulp(z/t)),
\end{align*}
so for the equalities it suffices to show that $\ulp(z/t) = \ulp(Z/x)$, or equivalently, that $\ufp(z/t) = \ufp(z/x)$. Note that if $\ufp(x) = \ufp(t)$ and $\ufp(m_x) = \ufp(m_t)$, then by \Cref{lem:ufp-of-ratio}, we have
\begin{align*}
    \ufp(z/t)
    &= \ufp(m_z/m_t)\ufp(z)/\ufp(t)\\
    &= \ufp(m_z/m_x)\ufp(z)/\ufp(x)\\
    &= \ufp(z/x).
\end{align*}
Also note that if $\ufp(t) = \ufp(x)$, then $\ufp(x) \leq \abs{t} < \beta \ufp(t)$ immediately.

We proceed by cases. In the following, note that by \Cref{lem:plausible-closure}, every power of $\beta$ and every scaling of $z$ by a power of $\beta$ is plausible for $Z$.

\begin{itemize}
    \item Suppose $\abs{m_z} < m_x$. Then by \Cref{lem:ufp-of-ratio}, we have $\ufp(m_z/m_x) = 1/\beta$. Then, since $x < \beta \ufp(x) \in \Plaus(Z)$, we have $t \leq \beta \ufp(x)$.
    \begin{itemize}
        \item Suppose $t < \beta \ufp(x)$. Then, since $\ufp(x) \leq x \leq t$, we have $\ufp(t) = \ufp(x)$ and hence $\abs{m_z} < m_x \leq m_t$. Therefore, by \Cref{lem:ufp-of-ratio}, we have $\ufp(m_z/m_t) = 1/\beta$ as well, as desired.
        \item Suppose $t = \beta \ufp(x)$ instead. Then,
        \begin{align*}
            \ufp(z/t)
            &= \ufp\left(\frac{z}{\beta \ufp(x)}\right)\\
            &= \frac{1}{\beta} \cdot \frac{\ufp(z)}{\ufp(x)}\\
            &= \ufp(m_z/m_x) \ufp(z)/\ufp(x)\\
            &= \ufp(z/x).
        \end{align*}
    \end{itemize}
    \item Suppose $0 < m_x \leq \abs{m_z}$ instead. Then, since $\abs{m_z} \ufp(x) \in \Plaus(Z)$, we have
    \begin{equation*}
        x \leq t \leq \abs{m_z} \ufp(x) < \beta \ufp(x).
    \end{equation*}
    Thus $\ufp(t) = \ufp(x)$ and hence
    \begin{equation*}
        m_x \leq m_t = t/\ufp(x) \leq \abs{m_z}.
    \end{equation*}
    Thus by \Cref{lem:ufp-of-ratio}, we also have $\ufp(m_z/m_x) = \ufp(m_z/m_t) = 1$, as desired.
    \item Suppose $-\abs{m_z} \leq m_x < 0$ instead. Then, since $-\ufp(x) \in \Plaus(Z)$, we have
    \begin{equation*}
        x \leq t \leq -\ufp(x),
    \end{equation*}
    so $\ufp(t) = \ufp(x)$ and hence
    \begin{equation*}
        -\abs{m_z} \leq m_x \leq m_t \leq -1.
    \end{equation*}
    Therefore, by \Cref{lem:ufp-of-ratio}, we have $\ufp(m_z/m_x) = \ufp(m_z/m_t) = 1$, as desired.
    \item Suppose $m_x < -\abs{m_z}$ instead. Then $\ufp(m_z/m_x) = 1/\beta$ by \Cref{lem:ufp-of-ratio}, and since $-\abs{m_z} \ufp(x) \in \Plaus(Z)$, we have
    \begin{equation*}
        x \leq t \leq -\abs{m_z} \ufp(x),
    \end{equation*}
    and thus $\ufp(x) = \ufp(t)$.
    \begin{itemize}
        \item Suppose $t < -\abs{m_z} \ufp(x)$. Then \begin{equation*}
            \ufp(m_z/m_x) = \ufp(m_z/m_t) = 1/\beta,
        \end{equation*}
        and hence $\ufp(z/t) = \ufp(z/x)$, as desired.
        \item Suppose $t = -\abs{m_z} \ufp(x)$ instead. Then,
        \begin{align*}
            t \ulp(z/x)
            &= -\abs{m_z} \ufp(x) \ulp(z/x)\\
            &= -\abs{m_z} \ufp(x) \ufp(m_z/m_x) \frac{\ufp(z)}{\ufp(x)} \beta^{1-p} \\
            &= -\abs{m_z}\ufp(z)\beta^{1-p}\\
            &= -\frac{\abs{z}}{\beta^{p-1}},
        \end{align*}
        hence $z$ is an integer multiple of $t \ulp(z/x)$. Thus
        \begin{equation*}
            (z \bmod t \ulp(z/x))  = (-z \bmod t \ulp(z/x)) = 0,
        \end{equation*}
        and so the rounding error is zero.
        Similarly, since
        \begin{equation*}
            z/t = \frac{z}{-\abs{m_z} \ufp(x)} = -\sgn(z)\frac{\ufp(z)}{\ufp(x)},
        \end{equation*}
        it is a (possibly negated) power of $\beta$, and thus in $\floats^*_\infty$. Therefore,
        \begin{equation*}
            t \RD_\infty(z/t) = t \RU_\infty(z/t) = t \cdot \frac{z}{t} = z
        \end{equation*}
        exactly, and hence the result follows.
    \end{itemize}
\end{itemize}
We have thus proven the result in all cases, so we are finished.
\end{proof}
The following lemma constrains the feasibility of tiny quotients. Essentially, if any especially small quotients are feasible, so are any larger ones, up to a limit. This result will be useful in determining whether there are any floating-point factors around zero. Although we cannot easily find subnormal factors, we can often eliminate them when they do not exist.
\begin{lemma}\label{lem:tiny-feasible-implies-larger-feasible}
Let $Z \subseteq \overline{\floats}$ be a floating-point interval, $x, y \in \floats^*$, and $z \in Z$. If $y \in \Feas(Z)$ and $\abs{y} \leq \abs{x} \leq \abs{z}/\max \floats$, then $x \in \Feas(Z)$.
\end{lemma}
\begin{proof}
If $\abs{x} = \abs{z}/\max \floats$ exactly, then $\abs{x} \otimes \max \floats = \abs{z}$ and hence $x \in \Feas(Z)$ by \Cref{lem:feas-closed-under-negation}. Suppose $y \in \Feas(Z)$ and $\abs{y} \leq \abs{x} < \abs{z/\max \floats}$ instead. Then $\abs{z/x} > \max \floats$ and hence $z$ is nonzero. We first prove the result for positive $x$ and $y$, before then generalizing to all cases. In the following, note that by \Cref{lem:equiv-for-feasible}, we have either $y \otimes \RU(z/y) \in Z$ or $y \otimes \RD(z/y) \in Z$.

Suppose $z > 0$. Since $\abs{z/x} > \max \floats$ and $0 < y \leq x$ by assumption, we have $\RD(z/y) = \RD(z/x) = \max \floats$ and $\RU(z/y) = +\infty$. Therefore, if $y \otimes \RU(z/y) = +\infty \in Z$, then $x \otimes +\infty = +\infty \in Z$ also, and hence $x$ is feasible for $Z$. If $y \otimes \RD(z/y) \in Z$ instead, then, since $y \leq x$, multiplying by $\RD(z/y)$ gives,
\begin{equation*}
\min Z
\leq y \otimes \RD(z/y)
\leq x \otimes \RD(z/y)
= x \otimes \RD(z/x)
\leq z
\leq \max Z,
\end{equation*}
and since $Z$ is a floating-point interval, it follows that $x \otimes \RD(z/x) \in Z$. Therefore $x$ is feasible for $Z$.

Suppose $z < 0$ instead. Then, we similarly have $\RD(z/y) = -\infty$ and $\RU(z/y) = \RU(z/x) = \min \floats$. Hence, if $y \otimes \RD(z/y) = -\infty \in Z$, then $x \otimes -\infty = -\infty \in Z$, and so $x$ is feasible for $Z$. If $y \otimes \RU(z/y) \in Z$ instead, then since $y \leq x$, multiplying by $\RU(z/y)$ gives
\begin{equation*}
\max Z \geq y \otimes \RU(z/y) \geq x \otimes \RU(z/y) = x \otimes \RU(z/x) \geq z \geq \min Z,
\end{equation*}
and since $Z$ is a floating-point interval, it follows that $x \otimes \RU(z/x) \in Z$. Therefore $x$ is feasible for $Z$.

We have thus shown the result for positive $x$ and $y$. Now, suppose instead that $y$ is negative ($x$ can have any sign). Then $\abs{y}$ is feasible for $Z$ by \Cref{lem:feas-closed-under-negation}, and thus $\abs{x}$ is feasible for $Z$ by the earlier result. Therefore $\abs{x}$ is feasible for $Z$ regardless of the sign of $y$, and hence by \Cref{lem:feas-closed-under-negation}, so is $x$.
\end{proof}
The following theorem allows us to reduce the question of finding feasible numbers to that of finding plausible numbers, when applicable:
\begin{theorem}\label{thm:reduction-to-plausible}
Let $Z \subseteq \overline{\floats}$ be a floating-point interval, and let $x \in \floats^*$ and $z \in Z$. Let
\begin{align*}
    A &= \Set{y \in \Feas(Z) \given y \geq x},\\
    B &= \Set{y \in \Plaus(Z) \given y \geq x},\\
    C &= \Set{y \in \Plaus(Z) \given y \geq \abs{x}}.
\end{align*}
If $x$ is normal and $\beta^\emin \leq \abs{z/x} \leq \max \floats$, then either $A$ is empty or $\min A \in \Set{\min B, \min C}$.
\end{theorem}
\begin{proof}
Suppose $x$ is normal, $\beta^\emin \leq \abs{z/x} \leq \max \floats$, and $A$ is nonempty. Let $a = \min A$, $b = \min B$, and $c = \min C$. Since $\Feas(Z) \subseteq \Plaus(Z)$, we have $a \in B$ and thus $x \leq b \leq a$.

We first show that if $c \leq \max \floats$, then $c \in \Feas(Z)$. Let $t = z/\beta^{E(z/x)}$. By \Cref{lem:plausible-closure}, since $\abs{z} \in \Plaus(Z)$, we have $\abs{t} \in \Plaus(Z)$. Now, since $\abs{z/x} \geq \beta^\emin$, we have $\abs{z/x} \geq \beta^{E(z/x)}$. Taking reciprocals and multiplying by $\abs{z}$, we obtain $\abs{x} \leq \abs{z}/\beta^{E(z/x)} = \abs{t}$ and therefore $\abs{t} \in C$. Thus we obtain $\abs{x} \leq c \leq \abs{t}$, and taking reciprocals and multiplying by $\abs{z}$ again, it follows that
\begin{align*}
    \max \floats \geq \abs{z/x} \geq \abs{z/c} &\geq \abs{z/t} = \beta^{E(z/x)} \geq \beta^\emin.
\end{align*}
Therefore, by \Cref{lem:normal-implies-plausible-equiv-feasible}, since $\beta^\emin \leq \abs{x} \leq c$, if $c \leq \max \floats$, then $c \in \floats^*$ and thus $c \in \Feas(Z)$.

Next, we dispose of two trivial cases:
\begin{itemize}
    \item Suppose $0 \in \Feas(Z)$. Then $0 \otimes y \in Z$ for some $y \in \floats$ and hence $0 \in Z$. Thus $0 \otimes x = 0 \in Z$, and so $x \in \Feas(Z)$. Therefore $a \leq x$, and hence $x = a = b$.
    \item Suppose $+\infty \in \Feas(Z)$. Then $+\infty \otimes y \in Z$ for some nonzero $y \in \overline{\floats}$. If $y > 0$, then $+\infty \otimes y = +\infty = +\infty \otimes x$, and so $x \in \Feas(Z)$. If $y < 0$, then $+\infty \otimes y = -\infty = -\infty \otimes x$, and so $x \in \Feas(Z)$ again. Thus $a \leq x$ and so $x = a = b$.
\end{itemize}
We can now proceed to the proof of the main statement. Suppose $0 \notin \Feas(Z)$ and $+\infty \notin \Feas(Z)$ instead. Then $a < +\infty$, and so $a \leq \max \floats$. We shall divide the problem into two cases, depending on the sign of $x$.

Suppose $x > 0$. Then $B = C$, and hence $b = c \leq a \leq \max \floats$. Therefore $c \in \Feas(Z)$ by our earlier result, so $a \leq c$ also, and hence $a = b = c$, as desired.

Suppose $x < 0$ instead. Then, since $x$ is normal, we have $x \leq -\beta^\emin$. Since $-\beta^\emin \in \Plaus(Z)$ by \Cref{lem:plausible-closure}, we have $-\beta^\emin \in B$ and thus $x \leq b \leq -\beta^\emin$. Therefore $\beta^\emin \leq \abs{b} \leq \abs{x} \leq \max \floats$ and hence $b \in \floats^*$ and $\beta^\emin \leq \abs{z/x} \leq \abs{z/b}$. Thus, if $\abs{z/b} \leq \max \floats$, then $b \in \Feas(Z)$ by \Cref{lem:normal-implies-plausible-equiv-feasible} and hence $a = b$. Suppose $\abs{z/b} > \max \floats$ instead. Then $\abs{b} < \abs{z}/\max \floats$. We further divide this into two subcases, depending on the sign of $a$. Note that $a \neq 0$ since $0 \notin \Feas(Z)$ by assumption.
\begin{itemize}
    \item Suppose $a$ is negative. Then $b \leq a < 0$, and thus $0 < \abs{a} \leq \abs{b} < \abs{z}/\max \floats$. Since $a \in \Feas(Z)$ by definition, \Cref{lem:tiny-feasible-implies-larger-feasible} implies that $b \in \Feas(Z)$ also and thus $a = b$.
    \item Suppose $a$ is positive. Then $[x, 0] \cap \Feas(Z)$ is empty, and since $\Feas(Z)$ is closed under negation, it follows that $a > \abs{x}$. Thus, since $a \in \Plaus(Z)$, we have $a \in C$ and hence $c \leq a \leq \max \floats$. Therefore $c \in \Feas(Z)$ by our earlier result, so $a \leq c$ and therefore $a = c$, as desired.
\end{itemize}
Therefore the result holds in all cases, and we are done.
\end{proof}

We now give the main result of this section. The following theorem reduces the problem to an optimization problem over the integral significands of plausible numbers.

\begin{theorem}\label{thm:plaus-ru-in-terms-of-integer-mod}
Let $Z \subseteq \overline{\floats}$ be a floating-point interval and $z \in Z \cap \floats^*$. Then, for all $x \in \floats^*_\infty$, we have $\PlausRU{Z}(x) = M \ulp(x)$ where
\begin{align*}
    M &= \min{\Set{ N \in A \cup B \given N \geq M_x }},\\
    A &= \Set{N \in \integers^* \given -(M_z \beta^{p-k} \bmod N) \in I \beta^{p-k}},\\
    B &= \Set{N \in \integers^* \given (-M_z \beta^{p-k} \bmod N) \in I \beta^{p-k}},\\
    I &= \frac{\fl^{-1}[Z] - z}{\ulp(z)},\\
    k &= \begin{cases}
        0 & \text{if } \abs{M_x} > \abs{M_z},\\
        1 & \text{otherwise,}
    \end{cases}\\
    M_x &= \frac{x}{\ulp(x)},\\
    M_z &= \frac{z}{\ulp(z)}.
\end{align*}
\end{theorem}
\begin{proof}
Let $y = \PlausRU{Z}(x)$ and $M_y = y/\ulp(x)$. Since $y$ is plausible by definition, we have either $\fl(y \RD_\infty(z/y)) \in Z$ or $\fl(y \RU_\infty(z/y)) \in Z$ by \Cref{lem:equiv-for-plausible}. Therefore, by \Cref{lem:round-trip-up-to-next-plausible}, we have either
\begin{align*}
z - (z  \bmod y \ulp(z/x)) &\in \fl^{-1}[Z],
\shortintertext{or}
z + (-z  \bmod y \ulp(z/x)) &\in \fl^{-1}[Z].
\end{align*}
Since $\ulp(M_z/M_x) = \beta^{k - p}$ by \Cref{lem:ufp-of-ratio}, subtracting $z$ and then dividing by $\ulp(z)\cdot\ulp(M_z/M_x)$ gives us either
\begin{align*}
-(M_z \beta^{p-k}  \bmod M_y) &\in I \beta^{p-k},
\shortintertext{or}
(-M_z \beta^{p-k}  \bmod M_y) &\in I \beta^{p-k}.
\end{align*}
By \Cref{lem:round-trip-up-to-next-plausible}, we have $\ufp(x) \leq \abs{y} \leq \beta \ufp(x)$ and thus $\beta^{p-1} \leq \abs{M_y} \leq \beta^p$. If $\abs{M_y} < \beta^p$, then $\ufp(y) = \ufp(x)$, and hence $M_y$ is an integer. If $\abs{M_y} = \beta^p$ instead, then $M_y$ is again an integer, and thus $M_y \in A \cup B$ in all cases. Hence, since $x \leq y$ and thus $M_x \leq M_y$ by definition, we obtain $M_x \leq M \leq M_y$. Hence $\beta^{p-1} \leq \abs{M} \leq \beta^p$ and $x \leq M \ulp(x) \leq y$, and therefore $M \ulp(x) \in \floats^*_\infty$. Then, by \Cref{lem:round-trip-up-to-next-plausible}, we have
\begin{align*}
    t \RD_\infty(z/t) &= z - (z \bmod t \ulp(z/x)),\\
    t \RU_\infty(z/t) &= z + (-z \bmod t \ulp(z/x)),
\end{align*}
where $t = M \ulp(x)$.

Next we prove that $t \in \Plaus(Z)$. Substituting the definition of $I$ into the definitions of $A$ and $B$ and rearranging, we obtain
\begin{align*}
    A &= \Set{N \in \integers^* \given z - (z \bmod N \ulp(z) \beta^{k-p}) \in \fl^{-1}[Z]},\\
    B &= \Set{N \in \integers^* \given z + (-z \bmod N \ulp(z) \beta^{k-p}) \in \fl^{-1}[Z]}.
\end{align*}
Now, since $\beta^{k-p} = \ulp(M_z/M_x)$, we have
\begin{align*}
    M \ulp(z) \beta^{k-p}
    &= M \ulp(z) \ulp(M_z/M_x)\\
    &= M \ulp(z/M_x)\\
    &= M \ulp(x) \ulp(z/x)\\
    &= t \ulp(z/x).
\end{align*}
Hence, since $M \in A \cup B$, we have either
\begin{align*}
    z - (z \bmod t \ulp(z/x)) &= t \RD_\infty(z/t) \in \fl^{-1}[Z],\\
\shortintertext{or}
    z + (-z \bmod t \ulp(z/x)) &= t \RU_\infty(z/t) \in \fl^{-1}[Z].
\end{align*}
Since $t \in \floats^*_\infty$, we thus have $t \in \Plaus(Z)$. Therefore $y \leq t$ by definition, and hence $y = t = M \ulp(x)$, as desired.
\end{proof}
By transforming the problem of finding plausible numbers into an integer optimization problem, we make it much more straightforward to analyze. Even so, the optimization problem presented in \Cref{thm:plaus-ru-in-terms-of-integer-mod} is unusual, since what is being optimized is a remainder. However, as we saw earlier in \Cref{fig:exact-products-error}, there are patterns in the round-trip error. In the next section, we will make use of them to bound the error and thus solve the optimization problem efficiently.

\section{Minimizing remainders of a constant divided by a variable}\label{sec:optimizing-remainders}
In the previous section, we reduced the problem of finding factors to an optimization problem over remainders where the dividend is constant but the divisor is not. The underlying problem is not specific to the floating-point numbers, and we can state it generally as follows:
\begin{problem}
Given a rational $x$ and an interval $I$ containing zero, what are the integers $n$ such that $(x \bmod n) \in I$? Specifically, given a fixed integer $m$, what is the least (or greatest) such $n$ where $n \geq m$ (or $n \leq m$)?
\end{problem}
Naively, this problem can be solved by exhaustive search on the divisors in at most $\abs{x} + 1$ trials, but this is clearly impractical. We do not have a provably efficient solution to the general problem, and it may well be the case that one does not exist at all.\footnote{For instance, if $x$ is a positive integer and $I = [0, 0]$, it is equivalent to factorization.} However, the specific case we are interested in can be solved in a constant number of arithmetic operations.

\begin{figure}
\centering
\begin{tikzpicture}
  \begin{axis}[
    width=\textwidth,
    height=\textwidth,
    xlabel={$n$},
    xmin=0, xmax=100,
    xtickmin=1,
    ymin=-100, ymax=100,
    samples at={1,...,100},
    mark size=0.8pt,
    legend entries={${-1000} \bmod n$,$\floor{-1000/n}$},
    legend pos=north west
  ] 
    \addplot+[
        ycomb,
        opacity=0.5,
        line width=0.5pt,
        every mark/.append style={solid,opacity=1}
    ] {-1000-x*floor(-1000/x)};
    \addplot+[
        ycomb,
        opacity=0.5,
        line width=0.5pt,
        every mark/.append style={solid,opacity=1}
    ] {floor(-1000/x)};
    \draw ({rel axis cs:0,0}|-{axis cs:0,0}) -- ({rel axis cs:1,0}|-{axis cs:0,0});
  \end{axis}
\end{tikzpicture}
\caption{Graphs of $(-1000 \bmod n)$ and $\floor{-1000/n}$, respectively above and below the x-axis. The upper graph follows a parabola whenever the lower graph behaves linearly (e.g.\ around $n = 30$). Similarly, linear segments in the upper graph correspond to constant segments in the lower (e.g.\ near $n= 80$).}
\label{fig:a-mod-n}
\end{figure}
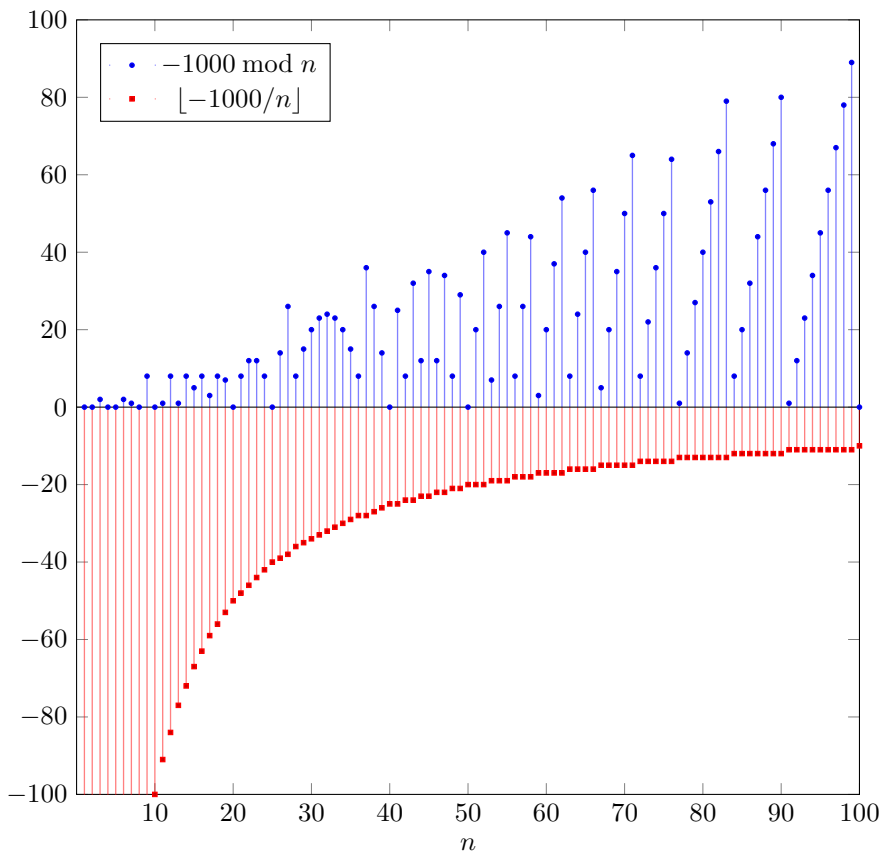
To better grasp the problem, let us first observe how the remainder varies with the divisor. Plotting $(-1000 \bmod n)$ in \Cref{fig:a-mod-n}, we see parabolas near $n = 30$, as well as between $n = 20$ and $n = 25$. After $n = 40$, a pattern of ``stepped'' parabolas emerges, with an increasing number of ``steps'' later on. Finally, the parabolas start disappearing entirely around $n = 60$ and give way to a linear pattern. As one might expect from rounding being expressible in terms of remainder, these patterns are also visible in \Cref{fig:exact-products-error}.

This behavior may seem an unusual coincidence at first, but it occurs for any choice of numerator. Algebraically, the reason is remarkably simple. By the definition of the $\bmod$ operator, we have
\begin{equation*}
(-1000 \bmod n) = -1000 - n \floor{-1000/n}.   
\end{equation*}
Thus, if $\floor{-1000/n}$ is constant over a certain interval, then $(-1000 \bmod n)$ is linear in $n$ over that interval. If $\floor{-1000/n}$ changes linearly, then $(-1000 \bmod n)$ is instead quadratic in $n$ over that interval. This relationship can be clearly seen in \Cref{fig:a-mod-n}, comparing the plots above and below the x-axis. The ``stepped parabolas'' correspond to values of $n$ where the slope of $-1000/n$ is close to linear, but less than 1, and so $\floor{-1000/n}$ does not change at every step.

More formally, consider integers $a, n, i$ such that $n$ and $n + i$ are nonzero, and let $q = \floor{\frac{a}{n}}$:
\begin{description}
    \item[Constant quotient] If $\floor{\frac{a}{n+i}} = q$, then
    \begin{align*}
    (a \bmod n + i)
    &= a - (n+i)q\\
    &= (a \bmod n) - iq.
    \end{align*}
    \item[Linear quotient] If there is some integer $d$ such that $\floor{\frac{a}{n + i}} = q + di$, then
    \begin{align*}
    (a \bmod n + i)
    &= a - (n+i)(q + di)\\
    &= (a \bmod n) - (di^2 + (q+nd)i).
    \end{align*}
\end{description}
Note that the above still holds if we permit the variables to take arbitrary real values. In this way we can also find parameters corresponding to each parabola inside the ``stepped'' segments.\footnote{Specifically, it corresponds to the second case with $d=1/w$, where $w$ is the number of integers $m$ satisfying $k = \floor{a/m}$. We can see this below the x-axis in \Cref{fig:a-mod-n}, where $w$ is the ``width'' of a level (e.g.\ around $n=80$, we have $w=7$).}

In our work, we will focus on linear and quadratic segments. These are especially useful, since we already have the tools to work with them. The following observations are key:
\begin{enumerate}
    \item Given a linear or quadratic segment, we can directly check it for a satisfying divisor by solving a linear or quadratic equation, respectively.
    \item The graph of remainders can always be partitioned into such segments.\footnote{For instance, we can trivially assign each pair of consecutive divisors its own segment.}
    \item Therefore, we can search for a solution one segment at a time, rather than one divisor at a time.
\end{enumerate}
With a good (that is, small) partition, we can shrink the search space significantly. However, this is still an exhaustive search, so proving the absence of a solution requires checking every segment. Furthermore, it is not clear that it is necessarily possible to construct a partition that is asymptotically smaller than magnitude of the dividend.

For the specific case we are interested in, however, the number of segments does not matter: for a suitable partition, we can show that the remainder at either the start or the end of any segment always satisfies the bounds. More specifically, according to \Cref{thm:plaus-ru-in-terms-of-integer-mod}, we are interested in remainders of the form $(-ab \bmod n)$ or $(ab \bmod n)$ for nonzero integers $a, b, n$. We now study the solution space under these special conditions.

\subsection{Conditions for a solution via extrapolation}

\begin{figure}
\centering
\begin{tikzpicture}
\begin{groupplot}[group style={group size=1 by 2}]
    \nextgroupplot[
        xlabel={$x$},
        ylabel={${-1000} \bmod x$},
        xmin=0,
        xmax=100,
        xtickmin=1,
        ymin=0, ymax=100,
        mark size=0.6pt,
        domain=1:100
    ]
    \pgfplotsextra{%
     \clip (axis cs:0,0) -- (axis cs:100,0) -- (axis cs:100,100) -- cycle;}
    \addplot+[
        only marks,
        samples at={1,...,100}
    ] {-1000-x*floor(-1000/x)};
    \addplot+[
        patch,
        patch type=line,
        no marks,
        draw=black,
    ] table{mod-lin-points.tsv};
    \draw[draw=black,fill=black] (axis cs:0,0) rectangle (axis cs:5,5);
    \nextgroupplot[
        xlabel={$x$},
        ylabel={${-x^2 - 1000} \bmod x$},
        xmin=0,
        xmax=100,
        xtickmin=1,
        ymin=0, ymax=100,
        mark size=0.6pt,
        domain=1:100
    ]
    \pgfplotsextra{%
     \clip (axis cs:0,0) -- (axis cs:100,0) -- (axis cs:100,100) -- cycle;}
    \addplot+[
        only marks,
        samples at={1,...,100}
    ] {-1000-x*floor(-1000/x)};
    \addplot+[
        patch,
        patch type=bezier spline,
        no marks,
        draw=black,
    ] table{mod-bezier-points.tsv};
    \draw[draw=black,fill=black] (axis cs:0,0) rectangle (axis cs:9,9);
\end{groupplot}
\end{tikzpicture}
\caption{Graphs of $(-1000 \bmod x)$ and $(-x^2 - 1000 \bmod x)$. The dots indicate points where $x$ is an integer.}
\label{fig:real-extension-of-mod}
\end{figure}
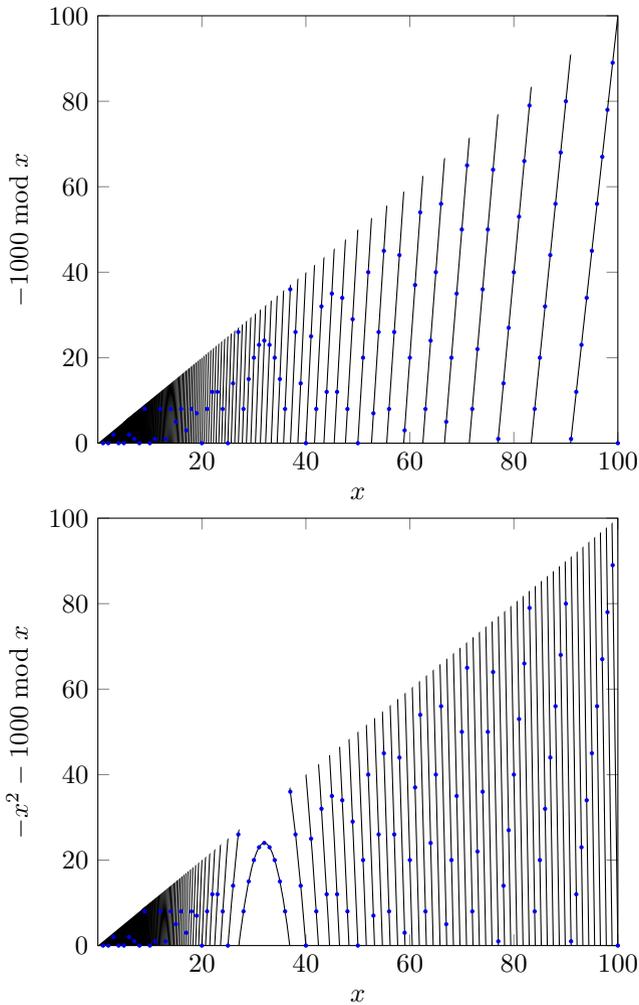
Given nonzero integers $a$ and $b$, we construct an appropriate partition using a function $f \from \reals^* \to \reals$ such that $f(x) = (-x^2 - \abs{ab} \bmod x)$. Clearly, we have $f(n) = (-\abs{ab} \bmod n)$ for all nonzero integers $n$. Since the numerator is a quadratic itself, the interval between each pair of consecutive roots of $f$ is a quadratic segment (\Cref{fig:real-extension-of-mod}, bottom). Now, note the following:
\begin{enumerate}
    \item A segment contains solutions if and only if the value of least magnitude within it satisfies the bounds.
    \item In each segment, the value of least magnitude is either at the start or the end of the segment.
    \item The start and the end of each segment lie between two consecutive roots of $f$, and are obtained by rounding those roots.
\end{enumerate}
Therefore, it suffices to prove that if $f(x) = 0$, then either $(-ab \bmod \floor{x})$ or $(-ab \bmod \ceil{x})$ lies within the bounds. Note that a linear mapping does not suffice: although it fits better when $x$ is large (this can be seen in \Cref{fig:real-extension-of-mod}), it does not provide sufficiently strong bounds when $x$ is small. The next result further explains the behavior of real extensions of remainders. In particular, we see that a quadratic extension makes quadratic segments have constant quotients.
\begin{lemma}\label{lem:nonzero-mod-implies-constant-quotient}
Let $I$ be an interval not containing zero and let $f$ be a real-valued function continuous over $I$. If $(f(x) \bmod x) \neq 0$ for all $x \in I$, then $\floor{f(x)/x} = \floor{f(y)/y}$ for all $x, y \in I$.
\end{lemma}
\begin{proof}
We proceed by contraposition. Suppose $\floor{f(x)/x} \neq \floor{f(y)/y}$ for some $x, y \in I$.
Without loss of generality, we assume that $\floor{f(x)/x} < \floor{f(y)/y}$.
Let $q \from I \to \reals$ be a function such that $q(t) = f(t)/t$. Then, since $f$ is continuous over $I$, so is $q$.
If $\floor{q(y)} \leq q(x)$, then $\floor{q(y)} \leq \floor{q(x)}$ by monotonicity, which is a contradiction. Therefore $q(x) < \floor{q(y)} \leq q(y)$, and hence by the intermediate value theorem there is some $w \in (x, y] \subseteq I$ such that $q(w) = \floor{q(y)}$. Thus $f(w)$ is an integer multiple of $w$, and so $(f(w) \bmod w) = 0$, as desired.
\end{proof}
\begin{corollary}
Let $I$ be an interval not containing zero and let $f$ be a real-valued function continuous over $I$. If $(f(x) \bmod x) \neq 0$ for all $x \in I$, then $(f(x) \bmod x) = f(x) - xq$ for all $x \in I$, where $q$ is an integer such that $q = \floor{f(y)/y)}$ for all $y \in I$.
\end{corollary}
The following lemma allows us to express the remainders of interest as quadratics.
\begin{lemma}\label{lem:quadratic-restriction-correspondence}
Let $a, b \in \integers$ and let $f, g, h \from \reals^* \to \reals$ be functions such that
\begin{align*}
    f(t) &= (-t^2 - \abs{ab} \bmod t),\\
    g(t) &= (-ab \bmod t),\\
    h(t) &= -(ab \bmod t).
\end{align*}
Let $I$ be an interval not containing zero. Let $q = \floor{(-x^2 - \abs{ab})/x}$ for some $x \in I$. If $f(t) \neq 0$ for all $t \in I$, then for all $n \in I \cap \integers$,
\begin{align*}
    f(n) &= -n^2 - qn - \abs{ab},\\
    g(n) &=
    \begin{cases}
    f(n) & \text{if } ab \geq 0,\\
    n - f(n) & \text{otherwise.}
    \end{cases}\\
    h(n) &= g(n) - n.
\end{align*}
\end{lemma}
\begin{proof}
Suppose $f(t) \neq 0$ for all $t \in I$ and let $n \in I \cap \integers$. Then, by \Cref{lem:nonzero-mod-implies-constant-quotient} we immediately have $f(n) = -n^2 - qn - \abs{ab}$. Since $n$ is an integer, we also have $f(n) = (-\abs{ab} \bmod n)$. Thus, since $f(n)$ is nonzero, it follows that $n$ does not divide $ab$. Therefore $g(n)$ and $h(n)$ are nonzero, and hence
\begin{equation*}
g(n) - n = -(n - (-ab \bmod n)) = -(ab \bmod n) = h(n).
\end{equation*}
If $ab \geq 0$, then $f(n) = (-ab \bmod n) = g(n)$. If $ab < 0$, then $f(n) = (ab \bmod n) = -h(n) = n - g(n)$, and thus $g(n) = n - f(n)$ as desired.
\end{proof}

We now proceed to demonstrate the conditions under which the roots of the quadratic mapping provide a solution. We first derive a simpler expression for the quadratic passing through a given root.
\begin{lemma}\label{lem:root-delta-simplification}
Let $a, b, \delta \in \reals$ and let $x \in \reals^*$. Then $-(x + \delta)^2 + (x + \delta)k - ab = - \delta^2 -(x - \frac{ab}{x})\delta$ where $k = \frac{x^2 + ab}{x}$.
\end{lemma}
\begin{remark}
Note that $(-x^2 - ab \bmod x) = 0$ if and only if $\frac{-x^2-ab}{x}$ is an integer.
\end{remark}
We now bound the magnitude of that quadratic in terms of the difference between the factors in the numerator.
\begin{lemma}\label{lem:delta-near-root-general-bound}
Let $a, b, x \in \reals$. If $1 < a + 1 \leq \abs{x} \leq b - 1$, then $\abs{x - ab/x} < b - a - 1$.
\end{lemma}
\begin{proof}
Let $f \from \reals^* \to \reals$ be a function such that $f(t) = t - \frac{ab}{t}$ and suppose $1 < a + 1 \leq \abs{x} \leq b - 1$. Since $a$ and $b$ are positive, we have $ab > 0$ and so $f$ is strictly increasing over $(-\infty, 0)$ and over $(0, +\infty)$. Hence $f(a+1) \leq f(\abs{x}) \leq f(b - 1)$. Now, since $b - 1$ and $a$ are positive, we have
\begin{align*}
(b - a - 1) - f(b-1)
&= (b - a - 1) - \left(b - 1 - \frac{ab}{b-1}\right)\\
&= \frac{ab}{b-1} - a\\
&= \frac{a}{b-1}\\
&> 0,
\end{align*}
and hence $b - a - 1 > f(b-1)$. Similarly, since $b$ and $a+1$ are positive,
\begin{align*}
-(b - a - 1) - f(a+1)
&= -(b - a -1) - \left(a + 1 - \frac{ab}{a+1}\right)\\
&= \frac{ab}{a+1} - b\\
&= -\frac{b}{a+1}\\
&< 0,
\end{align*}
and so $-(b - a - 1) < f(a+1)$. Therefore, $\abs{f(\abs{x})} < b - a - 1$. If $x > 0$, then $\abs{x} = x$ and the result follows immediately. If $x < 0$ instead, we have
\begin{equation*}
\abs{f(\abs{x})} = \abs{f(-x)} = \abs{-f(x)} = \abs{f(x)},
\end{equation*}
and so the result follows.
\end{proof}
\begin{remark}
We assume that $a + 1 \leq x \leq b - 1$ in order to obtain the necessary bound. When $a$ and $b$ are integers, if $x = a$ or $x = b$, then $(-ab \bmod x) = 0$, which lies within the bounding interval by assumption.
\end{remark}

The following lemma is an auxiliary result.
\begin{lemma}\label{lem:neg-delta-product-nonintegral}
Let $x \in \reals$ and $\delta \in (0, 1)$. If $\delta x$ and $(\delta - 1)(x+1)$ are integers, then they are both nonzero and have opposite signs.
\end{lemma}
\begin{proof}
Suppose $\delta x$ and $(\delta - 1)(x+1)$ are integers. If $x = 0$, then $(\delta - 1)(x+1) = \delta - 1 \notin \integers$, which is a contradiction. If $0 < \abs{x} \leq 1$, then $\delta x \notin \integers$, which is also a contradiction. Therefore $\abs{x} > 1$, and hence $x + 1$ is nonzero and has the same sign as $x$. Therefore, since $\delta - 1$ is negative, it follows that $\delta x$ and $(\delta - 1)(x+1)$ are nonzero and have opposite signs.
\end{proof}

We now show that rounding a root of the quadratic extension gives a solution for a sufficiently wide interval. However, we assume that the factors in the numerator are within a factor of two of each other. Otherwise, the bound from \Cref{lem:delta-near-root-general-bound} grows too quickly to be useful. Luckily, this assumption always holds when $\beta = 2$, which is the most common case by far.
\begin{lemma}\label{lem:rounded-root-solves-equitable}
Let $a, b \in \integers$ and let $x \in \reals^*$. Let $I$ be an interval containing zero. If $0 < a \leq \abs{x} \leq b$, $(-x^2 - ab \bmod x) = 0$, $b \leq 2a$ and $\diam I \geq a$, then for some $\hat{x}_1, \hat{x}_2 \in \Set{\floor{x}, \ceil{x}}$,
\begin{itemize}
    \item either $(-ab \bmod \hat{x}_1) \in I$ or $-(ab \bmod \hat{x}_1) \in I$, and
    \item either $(ab \bmod \hat{x}_2) \in I$ or $-(-ab \bmod \hat{x}_2) \in I$.
\end{itemize}
\end{lemma}
\begin{proof}
Suppose $0 < a \leq \abs{x} \leq b$, $(-x^2 - ab \bmod x) = 0$, $b \leq 2a$ and $\diam I \geq a$. We shall first dispose of some trivial cases. If $x \in \integers$, then $x = \floor{x} = \ceil{x}$, so
\begin{equation*}
(-ab \bmod \floor{x}) = (-ab \bmod x) = (-x^2 - ab \bmod x) = 0 \in I,
\end{equation*}
and hence the result. Since $0 \in I$, if $a < \abs{x} < a + 1$ instead, then either $\abs{\floor{x}} = a$ or $\abs{\ceil{x}} = a$, and hence the result holds trivially. Similarly, if $b - 1 < \abs{x} < b$, then either $\abs{\floor{x}} = b$ or $\abs{\ceil{x}} = b$, and the result follows trivially again.

Suppose $x \notin \integers$ and $a + 1 < \abs{x} < b - 1$ instead, and let $f \from \reals \to \reals$ be a function such that $f(t) = -t^2 + kt - ab$ where $k = \frac{x^2 + ab}{x}$. Then, since $(-x^2 - ab \bmod x) = 0$, it follows that $k$ is an integer and thus $f(t)$ is an integer whenever $t$ is an integer. Now, let $\delta = \ceil{x} - x$ and  $y = \frac{ab}{x} - x - \delta$. Then, by \Cref{lem:root-delta-simplification},
\begin{equation*}
\begin{split}
f(\ceil{x})
&= -\ceil{x}^2 + k\ceil{x} - ab\\
&= -(x + \delta)^2 + k(x + \delta) - ab\\
&= -\delta^2 - \left(x-\frac{ab}{x}\right)\delta \\
&= \delta\left(\frac{ab}{x} - x - \delta\right)\\
&= \delta y.
\end{split}
\end{equation*}
Since $x$ is not an integer, it follows that $\ceil{x} - \floor{x} = 1$ and hence $\floor{x} = x + \delta - 1$. Thus, by \Cref{lem:root-delta-simplification} again,
\begin{equation*}
\begin{split}
f(\floor{x})
&= -\floor{x}^2 + k\floor{x} - ab\\
&= -(x + (\delta - 1))^2 +k(x + (\delta - 1)) - ab\\
&= -(\delta - 1)^2 - \left(x - \frac{ab}{x}\right)(\delta-1)\\
&= (\delta-1)\left(\frac{ab}{x} - x - (\delta - 1)\right)\\
&= (\delta-1) (y + 1).
\end{split}
\end{equation*}
Now, since $b \leq 2a$, by \Cref{lem:delta-near-root-general-bound}, we have $\abs{ab/x - x} < b - a - 1 \leq a - 1$. Hence, applying the triangle inequality,
\begin{align*}
    \abs{f(\ceil{x})}
    &= \abs{\delta y}\\
    &< \abs{ab/x - x - \delta}\\
    &\leq \abs{ab/x - x} + \delta\\
    &< \abs{ab/x - x} + 1\\
    &< a,
\end{align*}
and
\begin{align*}
    \abs{f(\floor{x})}
    &= \abs{(\delta - 1)(y + 1)}\\
    &< \abs{ab/x - x - (\delta - 1)}\\
    &\leq \abs{ab/x - x} + (1 - \delta)\\
    &< \abs{ab/x - x} + 1\\
    &< a,
\end{align*}
and
\begin{align*}
\abs{f(\ceil{x}) - f(\floor{x})}
&= \abs{\delta y - (\delta - 1)(y+1)}\\
&= \abs{y + 1 - \delta}\\
&= \abs{ab/x - x + 1 - 2\delta}\\
&\leq \abs{ab/x - x} + \abs{1 - 2\delta}\\
&< \abs{ab/x - x} + 1\\
&< a.
\end{align*}
Hence the distance between each of $f(\floor{x})$, $f(\ceil{x})$ and $0$ is strictly less than $\diam I \geq a$.
Let $c = \min R$ and $d = \max R$ where $R = \Set{f(\floor{x}), f(\ceil{x})}$. Then $\diam{[c, d]} = d - c < b - a$. Since $f(\floor{x})$ and $f(\ceil{x})$ are both integers and $\delta \in (0, 1)$, it follows by \Cref{lem:neg-delta-product-nonintegral} that $\delta y$ and $(\delta - 1) (y + 1)$ are nonzero and have opposite signs and hence $c < 0 < d$. Thus both $[c, d]$ and $I$ contain zero, and since $\diam{[c, d]} < a \leq \diam I$, we have either $c \in I$ or $d \in I$ by \Cref{lem:overlapping-interval-endpoints}, and therefore either $f(\floor{x}) \in I$ or $f(\ceil{x}) \in I$. Similarly, since $[-d, -c]$ also contains zero and has the same diameter as $[c, d]$, we also have either $-f(\floor{x}) \in I$ or $-f(\ceil{x}) \in I$.

Now, in the following, note that we have both $\abs{f(\floor{x})} < a < \abs{\floor{x}}$ and $\abs{f(\ceil{x})} < a < \abs{\ceil{x}}$, which drastically simplifies working with mod.
\begin{itemize}
    \item Suppose $f(\ceil{x})$ and $x$ have the same sign. Then $-f(\floor{x})$ also has the same sign, and $-f(\ceil{x})$ and $f(\floor{x})$ have the opposite sign. Therefore,
    \begin{align*}
    f(\ceil{x}) &= (f(\ceil{x}) \bmod \ceil{x}) = (-ab \bmod \ceil{x}),\\
    -f(\floor{x}) &= (-f(\floor{x}) \bmod \floor{x}) = (ab \bmod \floor{x}),
    \end{align*}
    and thus
    \begin{align*}
    -f(\ceil{x}) &= -(-ab \bmod \ceil{x}),\\
    f(\floor{x}) &= -(ab \bmod \floor{x}).
    \end{align*}
    Hence either $(-ab \bmod \ceil{x}) \in I$ or $-(ab \bmod \floor{x}) \in I$, and also either $(ab \bmod \floor{x}) \in I$ or $-(-ab \bmod \ceil{x}) \in I$.
    \item Suppose instead $f(\ceil{x})$ and $x$ have opposite signs. Then $f(\floor{x})$ and $-f(\ceil{x})$  have the same sign as $x$, and hence $-f(\floor{x})$ has the opposite sign of $x$. Therefore,
    \begin{align*}
    f(\floor{x}) = (f(\floor{x}) \bmod \floor{x}) = (-ab \bmod \floor{x}),\\
    -f(\ceil{x}) = (-f(\ceil{x}) \bmod \ceil{x}) = (ab \bmod \ceil{x}),
    \end{align*}
    and thus
    \begin{align*}
    -f(\floor{x}) = -(-ab \bmod \floor{x}),\\
    f(\ceil{x}) = -(ab \bmod \ceil{x}).
    \end{align*}
    Hence either $(-ab \bmod \floor{x}) \in I$ or $-(ab \bmod \ceil{x}) \in I$, and also either $(ab \bmod \ceil{x}) \in I$ or $-(-ab \bmod \floor{x}) \in I$.
\end{itemize}
Therefore, for some $\hat{x}_1, \hat{x}_2 \in \Set{\floor{x}, \ceil{x}}$, we have either $(-ab \bmod \hat{x}_1) \in I$ or $-(ab \bmod \hat{x}_1 \in I)$, and also either $(ab \bmod \hat{x}_2) \in I$ or $-(-ab \bmod \hat{x}_2) \in I$.
\end{proof}

Combining the above result with \Cref{thm:plaus-ru-in-terms-of-integer-mod}, we can finally show that, for equitable rounding functions, roots of the quadratic extension give a bound on plausible integral significands. With the help of \Cref{lem:quadratic-restriction-correspondence}, this will let us compute plausible numbers efficiently in base 2.
\begin{theorem}\label{thm:root-to-solution}
Let $Z \subseteq \overline{\floats}$ be a floating-point interval, and let $z \in Z \cap \floats^*$ and $x \in \floats^*_\infty$. Let $f \from \reals^* \to \reals$ be a function such that $f(t) = (-t^2 - \abs{M_z}\beta^{p-k} \bmod t)$ where
\begin{align*}
    k &= \begin{cases}
        0 & \text{if } \abs{M_x} > \abs{M_z},\\
        1 & \text{otherwise,}
    \end{cases}\\
    M_x &= \frac{x}{\ulp(x)},\\
    M_z &= \frac{z}{\ulp(z)}.
\end{align*}
Let $r \in \reals^*$ be a root of $f$. If $\fl$ is equitable, $r \geq M_x$, and either $\beta^{p-1} < \abs{M_z} \leq 2\beta^{p-1}$ or $\beta = 2$, then $\PlausRU{Z}(x) \leq \ceil{r} \ulp(x)$.
\end{theorem}
\begin{proof}
Suppose the conditions hold and let $I = (\fl^{-1}[Z] - z)/\ulp(z)$ and $M = \PlausRU{Z}(x) / \ulp(x)$. We divide the proof into two major cases based on the value of $M_z$, and then handle the different values of $k$ within those cases.

Suppose $\beta^{p-1} < \abs{M_z} \leq 2 \beta^{p-1}$. Then, since $\fl$ is equitable, by definition we have
\begin{equation*}
\diam \fl^{-1}[Z] \geq \diam \fl^{-1}[\Set{z}] \geq \beta^{Q(z)} \geq \ulp(z),
\end{equation*}
and therefore,
\begin{align*}
    \diam I \beta^{p-k}
    &= \beta^{p-k} \diam \frac{\fl^{-1}[Z]-z}{\ulp(z)}\\
    &= \frac{\beta^{p-k}}{\ulp(z)} \diam \fl^{-1}[Z]\\
    &\geq \beta^{p-k}.
\end{align*}

If $k = 0$, then $\abs{M_z} < \abs{M_x}$, so \Cref{lem:conditional-wide-z-implies-plausible} implies that $x$ is plausible for $Z$, and thus
\begin{equation*}
\PlausRU{Z}(x) = x = M_x \ulp(x) \leq r \ulp(x) \leq \ceil{r} \ulp(x).
\end{equation*}

Suppose $k = 1$ instead, and let $a = \beta^{p-k}$ and $b = \abs{M_z}$. Then $b \leq 2a = 2\beta^{p-1}$ by assumption and $\diam I \beta^{p-k} \geq a$. We now consider each sign of $M_z$:
\begin{itemize}
    \item Suppose $M_z > 0$. Then $b = M_z$, and hence by \Cref{lem:rounded-root-solves-equitable}, we have either $-(ab \bmod \hat{r}) = -(M_z\beta^{p-k} \bmod \hat{r}) \in I \beta^{p-k}$ or $(-ab \bmod \hat{r}) = (-M_z\beta^{p-k} \bmod \hat{r}) \in I \beta^{p-k}$ for some $\hat{r} \in \Set{\floor{r}, \ceil{r}}$ and hence $M \leq \hat{r} \leq \ceil{r}$ by \Cref{thm:plaus-ru-in-terms-of-integer-mod}.
    \item Suppose $M_z < 0$. Then $b = -M_z$ and by \Cref{lem:rounded-root-solves-equitable} we have either $(ab \bmod \hat{r}) = (-M_z\beta^{p-k} \bmod \hat{r}) \in I \beta^{p-k}$ or $-(-ab \bmod \hat{r}) = -(M_z\beta^{p-k} \bmod \hat{r}) \in I \beta^{p-k}$ for some $\hat{r} \in \Set{\floor{r}, \ceil{r}}$ and hence $M \leq \hat{r} \leq \ceil{r}$ by \Cref{thm:plaus-ru-in-terms-of-integer-mod}.
\end{itemize}
Therefore $M \leq \ceil{r}$ in both cases, and multiplying by $\ulp(x)$, we obtain $\PlausRU{Z}(x) \leq \ceil{r} \ulp(x)$.

Finally, suppose $\beta = 2$ and $\abs{M_z} = \beta^{p-1}$ instead. If $k = 1$, then $\abs{M_x} = \abs{M_z}$, so $x$ is trivially plausible for $Z$ and hence $\PlausRU{Z}(x) = x \leq \ceil{r} \ulp(x)$ immediately. Suppose $k = 0$ instead, and let $a = \abs{M_z}$ and $b = \beta^{p-k}$. Then, since $\beta = 2$, we have $b = 2a$. Since $\fl$ is equitable,
\begin{equation*}
\diam \fl^{-1}[Z] \geq \diam \fl^{-1}[\Set{z}] \geq \beta^{Q(z) - 1} \geq \ulp(z)/\beta,
\end{equation*}
and therefore,
\begin{align*}
    \diam I \beta^{p-k}
    &= \beta^p \diam \frac{\fl^{-1}[Z]-z}{\ulp(z)}\\
    &= \frac{\beta^{p}}{\ulp(z)} \diam \fl^{-1}[Z]\\
    &\geq \beta^{p-1}\\
    &= a.
\end{align*}
We now complete the proof using \Cref{lem:rounded-root-solves-equitable} and \Cref{thm:plaus-ru-in-terms-of-integer-mod}. By the same argument as earlier, we have $M \leq \hat{r} \leq \ceil{r}$ for some $\hat{r} \in \Set{\floor{r}, \ceil{r}}$, and thus $\PlausRU{Z}(x) \leq \ceil{r} \ulp(x)$, as desired.
\end{proof}

\begin{remark}

If $r$ is the least root of $f$ greater than $M_x$, then $f$ is nonzero over $(M_x, r)$. Thus, according to \Cref{lem:quadratic-restriction-correspondence}, the restrictions of $M \mapsto -(M_z\beta^{p-k} \bmod M)$ and $M \mapsto (-M_z \beta^{p-k} \bmod M)$ to $(M_x, r) \cap \integers$ are quadratics. Hence, if neither $x$ nor $\floor{r}\ulp(x)$ are plausible for $Z$, then we necessarily have $\PlausRU{Z}(x) = \ceil{r}\ulp(x)$. Otherwise, $\PlausRU{Z}(x) \leq \floor{r} \ulp(x)$, so we can compute $\PlausRU{Z}(x)$ by solving the quadratics from \Cref{lem:quadratic-restriction-correspondence} to find the earliest point where they lie in the interval $(\fl^{-1}[Z] - z)/\ulp(z)$.
\end{remark}

As discussed above, combining the bound given by \Cref{thm:root-to-solution} with \Cref{lem:quadratic-restriction-correspondence} allows us to solve \Cref{prob:main} for normal numbers in binary floating-point arithmetic. We next turn to assembling our results into an algorithm.

\section{Algorithms}\label{sec:algorithms}
\Cref{thm:classical-intersect-feasible} states that the classical division-based algorithm can be used to produce optimal bounds for floating-point arithmetic if we can devise some way of finding the feasible floating-point numbers nearest any given infeasible number. If the bounds and their quotients are within the range of the normal numbers, \Cref{thm:reduction-to-plausible} tells us that we can do this by instead finding the nearest plausible numbers. Finally, for equitable rounding functions in base 2 (and other bases under certain conditions), \Cref{thm:root-to-solution} allows us to efficiently find the next plausible number. Putting these together, we can derive an algorithm for efficiently computing optimal bounds on the factors of a binary floating-point multiplication. \Cref{alg:next-feasible-2,alg:next-plausible-2,alg:next-divisor-in-bounds,alg:NextQuadraticModLinearRootFloor,alg:NextQuadraticPointWithinBounds,alg:QuadraticRootsFloor} form a complete pseudocode implementation of the results of this paper. They are annotated with the properties they satisfy and justifications for them. The proofs of correctness for \Cref{alg:NextQuadraticModLinearRootFloor,alg:NextQuadraticPointWithinBounds,alg:QuadraticRootsFloor} require some additional minor results which are included in \Cref{sec:additional-proofs}. We thus obtain the following:

\begin{theorem}
\Cref{alg:next-feasible-2,alg:next-plausible-2,alg:next-divisor-in-bounds,alg:NextQuadraticModLinearRootFloor,alg:NextQuadraticPointWithinBounds,alg:QuadraticRootsFloor} are correct. \Cref{alg:next-divisor-in-bounds,alg:NextQuadraticModLinearRootFloor,alg:NextQuadraticPointWithinBounds,alg:QuadraticRootsFloor} can be computed in $O(1)$ arithmetic operations. \Cref{alg:next-feasible-2,alg:next-plausible-2} can be computed in $O(1)$ arithmetic operations and computations of the preimage of a floating-point interval under $\fl$.
\end{theorem}
\begin{remark}
As in \Cref{lem:f-time-complexity}, for typical choices of $\fl$, these algorithms have time complexity $O(M(p) + \log(\emax - \emin + 1))$, where $M(p)$ is the time complexity of multiplying two $p$-digit integers. We again do not take the base, precision or exponent range as constants in our analysis.
\end{remark}

According to \Cref{thm:classical-intersect-feasible}, given the preconditions hold, applying \Cref{alg:next-feasible-2} to the bounds produced by the traditional algorithm gives optimal bounds on the factors. This takes $O(1)$ arithmetic operations and computations of $\fl^{-1}$, which is much less than the exponential worst case we proposed in \Cref{conj:f-worst-case-complexity} for just iterating the classical algorithm.

Although the preconditions on \Cref{alg:next-feasible-2} almost always hold when $\beta = 2$, they must still be checked. Otherwise, even though the algorithm always returns either a feasible number or $+\infty$, the value returned may not in fact be the next feasible number (or, in the case of $+\infty$ being infeasible, the problem instance may still be satisfiable).

In the event that some or all of the inputs are subnormal or we are working in a higher base, we still have some options for using these results. By \Cref{lem:plausible-is-lower-bound}, the next plausible number is always a lower bound on the next feasible number, and \Cref{alg:next-plausible-2} works correctly for subnormal input, since there is no such distinction over the unbounded floating-point numbers. Furthermore, it can be shown that \Cref{alg:next-plausible-2} returns a lower bound on the next plausible number even in higher bases. However, we make no claims to the efficiency under such conditions of iterating \Cref{alg:next-plausible-2} to find the next feasible number.

Now, as we demonstrated in \Cref{ex:classical-is-suboptimal}, iterating the classical algorithm is insufficient to find optimal bounds on solution sets in the general case. That is, given a problem instance $(X, Y, Z)$ where $X, Y, Z \subseteq \overline{\floats}$ are floating-point intervals, iterating $F$ until a fixed point is reached may not produce optimal bounds on $Z$. We proposed solving this by using binary search to find the greatest and least values in $Z$ for which the instance is satisfiable. (As a reminder, an instance is satisfiable if and only if its optimal bounds are nonempty.) Naively, this search takes $O(p + \log(\emax - \emin + 1))$ satisfiability queries. However, we can do better. \Cref{lem:very-wide-z-implies-feasible} says that if we have at least $\beta$ values in $Z$, then almost any value is feasible for $Z$. Thus, in most cases, we only need to consider first and last $\beta$ values of $Z$ to obtain optimal intervals. When $\beta = 2$, this means that we typically only need to trim at most one value from each endpoint of $Z$. Note that this situation only occurs when the problem instance is satisfiable---otherwise, our optimal bounds on $X$ and $Y$ will be empty, and thus so will the optimal bounds on $Z$.

\begin{algorithm}
\caption{Finding the next feasible number.}\label{alg:next-feasible-2}
\begin{algorithmic}[1]
\Require{$\fl$ equitable; normal floating-point number $x$; floating-point interval $Z$; $z \in Z$ such that $\beta^\emin \leq \abs{z/x} \leq \max \floats$}; either $\beta = 2$ or $1 < \abs{z/\ufp(z)} \leq 2$
\Ensure{returns $\inf \Feas(Z) \cap [x, +\infty]$}
\Function{NextFeasible}{$x, z, Z$}
    \If{$x \in \Feas(Z)$} \Comment{using \Cref{lem:equiv-for-feasible}}
        \State \Return{$x$}
    \EndIf
    \State $x_b \coloneqq \text{\Call{NextPlausible}{$x, z, Z$}}$ \Comment{by \Cref{thm:reduction-to-plausible}, if a solution exists, it is either $x_b$ or $x_c$}
    \If{$x_b \in \Feas(Z)$} \Comment{check that $x_b \in \overline{\floats}$ and then use \Cref{lem:equiv-for-feasible}}
        \State \Return{$x_b$}
    \EndIf
    \State $x_c \coloneqq \text{\Call{NextPlausible}{$\abs{x}, z, Z$}}$
    \If{$x_c \in \Feas(Z)$}
        \State \Return{$x_c$}
    \EndIf
    \State \Return{$+\infty$} \Comment{$\Feas(Z) \cap [x, +\infty]$ is empty by \Cref{thm:reduction-to-plausible}}
\EndFunction
\end{algorithmic}
\end{algorithm}

\begin{algorithm}
\caption{NextPlausible}\label{alg:next-plausible-2}
\begin{algorithmic}[1]
\Require{$\fl$ equitable; $x \in \floats^*_\infty$; floating-point interval $Z$; nonzero and finite $z \in Z$; either $\beta = 2$ or $1 < \abs{z/\ufp(z)} \leq 2$}
\Ensure{returns $\PlausRU{Z}(x)$}
\Function{NextPlausible}{$x, z, Z$}
    \State $M_x \coloneqq x/\ulp(x)$
    \State $M_z \coloneqq z/\ulp(z)$
    \State $I \coloneqq (\fl^{-1}[Z]-z)/\ulp(z)$
    \If{$\abs{M_x} > \abs{M_z}$}
        \State $a \coloneqq M_z$
        \State $b \coloneqq \beta^p$
        \State $I' \coloneqq I \beta^p$
    \Else
        \State $a \coloneqq \beta^{p-1}$
        \State $b \coloneqq M_z$
        \State $I' \coloneqq I \beta^{p-1}$
    \EndIf
    \State $M \coloneqq \text{\Call{NextDivisorInBounds}{$a, b, M_x, I'$}}$ \Comment{by \Cref{thm:plaus-ru-in-terms-of-integer-mod,thm:root-to-solution}}
    \State \Return{$M \ulp(x)$}
\EndFunction
\end{algorithmic}
\end{algorithm}

\begin{algorithm}
\caption{NextDivisorInBounds}\label{alg:next-divisor-in-bounds}
\begin{algorithmic}[1]
\Require{integers $a, b, n$ such that $0 < \abs{a} \leq \abs{n} \leq \abs{b} \leq \abs{2a}$; interval $I$ containing zero such that $\diam I \geq \abs{a}$}
\Ensure{returns $\min{\Set{m \in \integers \given m \geq n \land ((-ab \bmod m) \in I \lor -(ab \bmod m) \in I)}}$}
\Function{NextDivisorInBounds}{$a, b, n, I$}
    \If{$(-ab \bmod n) \in I \lor -(ab \bmod n) \in I$}
        \State \Return{$n$}
    \EndIf

    \State $\underline{r} \coloneqq \text{\Call{NextQuadraticModLinearRootFloor}{$-1, -\abs{ab}, n$}}$
    
    \If{$(-ab \bmod \underline{r}) \notin I \land -(ab \bmod \underline{r}) \notin I$}
        \State \Return{$\underline{r} + 1$} \Comment{by \Cref{lem:rounded-root-solves-equitable}, since there are no solutions in $[n, \underline{r}]$}
    \EndIf
    
    \State $q \coloneqq (-n^2 - \abs{ab})/n$ \Comment{solution is in $[n, \underline{r}]$; find minimum using \Cref{lem:quadratic-restriction-correspondence}}
    
    \If{$ab > 0$}
        \State $c \coloneqq \text{\Call{NextQuadraticPointWithinBounds}{$-1, -\floor{q}, -ab, n, I$}}$
        \State $d \coloneqq \text{\Call{NextQuadraticPointWithinBounds}{$-1, -\ceil{q}, -ab, n, I$}}$
    \Else
        \State $c \coloneqq \text{\Call{NextQuadraticPointWithinBounds}{$1, \ceil{q}, -ab, n, I$}}$
        \State $d \coloneqq \text{\Call{NextQuadraticPointWithinBounds}{$1, \floor{q}, -ab, n, I$}}$
    \EndIf
    \State \Return{$\min{\Set{c, d}}$}
\EndFunction
\end{algorithmic}
\end{algorithm}

\begin{algorithm}
\caption{NextQuadraticModLinearRootFloor}\label{alg:NextQuadraticModLinearRootFloor}
\begin{algorithmic}[1]
\Require{nonzero integers $a, c, n$}
\Ensure{returns $\min{\Set{\floor{x} \given x \in [n, +\infty) \land (ax^2 + c \bmod x) = 0}}$}
\Function{NextQuadraticModLinearRootFloor}{$a, c, n$}
\State $q \coloneqq (an^2 + c)/n$
\State $R_1 \coloneqq \text{\Call{QuadraticRootsFloor}{$a, -\floor{q}, c$}}$
\State $R_2 \coloneqq \text{\Call{QuadraticRootsFloor}{$a, -\ceil{q}, c$}}$
\State \Return{$\min{\Set{m \in R_1 \cup R_2 \given m \geq n}}$} \Comment{by \Cref{lem:nearest-quotient-integer-values}}
\EndFunction
\end{algorithmic}
\end{algorithm}

\begin{algorithm}
\caption{NextQuadraticPointWithinBounds}\label{alg:NextQuadraticPointWithinBounds}
\begin{algorithmic}[1]
\Require{integers $a, b, c, n$ where $a \neq 0$; interval $I$}
\Ensure{returns $\inf{\Set{m \in \integers \given m \geq n \land am^2 + bm + c \in I}}$}
\Function{NextQuadraticPointWithinBounds}{$a, b, c, n, I$}
    \If{$an^2 + bn + c \in I$}
        \State \Return{n}
    \EndIf
    
    \State $I' \coloneqq I \cap \integers$
    \If{$I' = \emptyset$}
        \State \Return{$+\infty$}
    \EndIf
    
    \State $R_1 \coloneqq \text{\Call{QuadraticRootsFloor}{$a, b, c - \max I'$}}$
    \State $R_2 \coloneqq \text{\Call{QuadraticRootsFloor}{$a, b, c - \min I'$}}$
    \State $\underline{R} \coloneqq R_1 \cup R_2$
    \State $\overline{R} \coloneqq \Set{r + 1 \given r  \in \underline{R}}$ \Comment{solutions are ceilings of roots per \Cref{lem:satisfying-integers-are-ceils-of-roots}}
    \State \Return{$\min{\Set{m \in \underline{R} \cup \overline{R} \given m \geq n \land am^2 + bm + c \in I} \cup \Set{+\infty}}$}
\EndFunction
\end{algorithmic}
\end{algorithm}

\begin{algorithm}
\caption{QuadraticRootsFloor}\label{alg:QuadraticRootsFloor}
\begin{algorithmic}[1]
\Require{integers $a, b, c$ where $a \neq 0$}
\Ensure{returns $\Set{\floor{x} \given x \in \reals \land ax^2 + bx + c = 0}$}
\Function{QuadraticRootsFloor}{$a, b, c$}
\If{$a < 0$}
    \State $a \coloneqq -a$
    \State $b \coloneqq -b$
    \State $c \coloneqq -c$
\EndIf
\State $\Delta \coloneqq b^2 - 4ac$
\If{$\Delta < 0$}
    \State \Return{$\emptyset$}
\EndIf
\State $r_1 \coloneqq \floor{\frac{-b + \floor{-\sqrt{\Delta}}}{2a}}$ \Comment{by applying \Cref{lem:floor-real-div-integer} to the quadratic formula}
\State $r_2 \coloneqq \floor{\frac{-b + \floor{\sqrt{\Delta}}}{2a}}$
\State \Return{$\Set{r_1, r_2}$}
\EndFunction
\end{algorithmic}
\end{algorithm}

\section{Conclusion}\label{sec:conclusion}
In this paper, we have rigorously presented the first step towards an efficient solution to the problem of computing optimal interval bounds on the variables of the floating-point constraint $x \otimes y = z$. The algorithms presented here are sound, complete, and efficient for binary floating-point arithmetic involving normal numbers. This has immediate implications for floating-point decision procedures based on interval constraint solving.
We hope that the techniques developed here will be useful in future study of floating-point theory.

As suggested earlier, a number of questions remain open in this topic:
\begin{itemize}
    \item What is the optimal time complexity for computing feasible numbers in general (i.e.\ in all bases and with subnormals)? Can we asymptotically outperform brute force search? Does an efficient algorithm exist for this task?
    \item Although the algorithms in this paper are relatively brief, the work to develop them is much longer. Is there are shorter derivation of this solution?
    \item The results of this paper rely on a mixture of integer and floating-point reasoning. Can these results be stated naturally using only floating-point numbers? If so, is there a version of this algorithm which uses only floating-point arithmetic?
    \item Separately, \Cref{sec:optimizing-remainders} contains results which may be useful for solving mod constraints beyond the scope of this paper. Can they be applied more broadly?
\end{itemize}

\backmatter

\bmhead{Acknowledgments}
This research was supported by an Australian Government Research Training Program (RTP) Scholarship.

\begin{appendices}
\crefalias{section}{appendix}

\section{Additional proofs} \label{sec:additional-proofs}

This section contains some additional results used to show the correctness of \Cref{alg:NextQuadraticModLinearRootFloor,alg:NextQuadraticPointWithinBounds,alg:QuadraticRootsFloor}. There is also a proof of a minor property on the digital expansions of floating-point quotients.
\begin{lemma}\label{lem:floor-real-div-integer}
Let $x \in \reals$ and $n \in \integers$. If $n > 0$, then $\floor{x/n} = \floor{\floor{x}/n}$.
\end{lemma}
\begin{proof}
Suppose $n > 0$ and let $\delta_1 = x - \floor{x}$ and $\delta_2 = (\floor{x} \bmod n)$. Since $n$ is a positive integer and $\floor{x}$ is an integer, we have $0 \leq \delta_2 \leq n - 1$. Hence $0 \leq \delta_1 + \delta_2 < n$ and thus $\floor{(\delta_1 + \delta_2)/n} = 0$. By the definition of mod, we also have $(\floor{x} - \delta_2)/n = \floor{\floor{x}/n}$ and thus
\begin{align*}
\floor{x/n}
&= \floor*{\frac{\floor{x} + \delta_1 + \delta_2 - \delta_2}{n}}\\
&= \floor{\floor{\floor{x}/n} + (\delta_1 + \delta_2)/n}\\
&= \floor{\floor{x}/n} + \floor{(\delta_1 + \delta_2)/n}\\
&= \floor{\floor{x}/n}.
\end{align*}
\end{proof}

\begin{lemma}\label{lem:satisfying-integers-are-ceils-of-roots}
Let $f$ be a continuous real function and let $n \in \integers$ and $a, b \in \reals$. Let $M = \Set{m \in \integers \given m \geq n \land f(m) \in [a, b]}$. If $M$ is nonempty, then either $\min M = n$ or $\min M = \ceil{x}$ for some $x \in \reals$ such that either $f(x) = a$ or $f(x) = b$.
\end{lemma}
\begin{proof}
If $f(n) \in [a, b]$, the result follows trivially. Suppose instead that $M$ is nonempty and $m > n$ where $m = \min M$. Then $a \leq f(m) \leq b$ and either $f(m - 1) < a$ or $f(m - 1) > b$ by assumption. Suppose $f(m - 1) < a$. Then, by the intermediate value theorem, there is some $x \in (m - 1, m]$ such that $f(x) = a$. Since $\ceil{x} = m$, the result follows. Suppose $f(m - 1) > b$ instead. Then, similarly, there is some $x \in (m - 1, m]$ such that $f(x) = b$. Since $\ceil{x} = m$ again, we are finished.
\end{proof}

\begin{lemma}\label{lem:quotient-limit}
Let $a, c \in \reals^*$ and let $q \from \reals^* \to \reals$ be a function such that $q(t) = (at^2 + c)/t$. For all $x \in \reals^*$, there exists some $y \in [x, +\infty)$ such that $q$ is continuous over $[x, y]$ and $q(y) \in \Set{\floor{q(x)}, \ceil{q(x)}}$.
\end{lemma}
\begin{proof}
In the following, note that $q$ is continuous over its domain and $q(t) = at + c/t$ for all $t \in \reals^*$. We proceed by cases.

Let $x \in \reals^*$ and suppose $x$ is positive. Suppose $a$ is also positive. Then,
\begin{equation*}
\lim_{t \to \infty} q(t) = \lim_{t \to \infty} at + \lim_{t \to \infty} c/t = \infty + 0 = +\infty,
\end{equation*}
and thus $q$ grows without bound. Since it is also continuous over $(0, +\infty)$, it therefore attains $\ceil{q(x)}$ at some $y \in [x, +\infty)$, and hence the result follows.

Suppose $a$ is negative instead. Then $\lim_{t \to \infty} q(t) = -\infty$ similarly, and hence $q$ has no lower bound. Since it is continuous over $(0, +\infty)$, we thus have $q(y) = \floor{q(x)}$ for some $y \in [x, +\infty)$, and hence the result follows.

Now, suppose $x$ is negative instead. Suppose $c$ is positive. Then,
\begin{equation*}
\lim_{t \to 0^-} q(t) = \lim_{t \to 0^-} at + \lim_{t \to 0^-} c/t = 0 - \infty = -\infty,
\end{equation*}
and therefore $q$ has no lower bound over $(-\infty, 0)$. Since it is also continuous over $(-\infty, 0)$, we thus have $q(y) = \floor{q(x)}$ for some $y \in [x, 0)$, as desired.

Suppose $c$ is negative instead. Then $\lim_{t \to 0^-} q(t) = +\infty$, and hence $q$ has no upper bound over $(-\infty, 0)$. Since $q$ is continuous over $(-\infty, 0)$, it follows that $q(y) = \ceil{q(x)}$ for some $y \in [x, 0)$. Therefore the result holds in all cases, and we are done.
\end{proof}

\begin{lemma}\label{lem:nearest-quotient-integer-values}
Let $a, c \in \reals^*$ and let $q \from \reals^* \to \reals$ be a function such that $q(t) = (at^2 + c)/t$. Let $x \in \reals^*$ and $R = \Set{t \in [x, +\infty) \given q(t) \in \integers}$. Then $R$ has a minimum, and either $q(r) = \floor{q(x)}$ or $q(r) = \ceil{q(x)}$ where $r = \min R$.
\end{lemma}
\begin{proof}
Let $S$ be a set such that
\begin{equation*}
S = \Set{t \in [x, +\infty) \given q(t) \in \Set{\floor{q(x)}, \ceil{q(x)})}}.
\end{equation*}
Clearly, $S \subseteq R$. For all $t \in \reals^*$ and $u \in \reals$, we have $q(t) = u$ if and only if $at^2 - ut + c = 0$. Therefore, for any particular $u$, there are at most 2 choices of $t$ that satisfy the equation, and hence $S$ has at most 4 elements.
By \Cref{lem:quotient-limit}, there is some $y \in [x, +\infty)$ such that $q$ is continuous over $[x, y]$ and $q(y) \in \Set{\floor{q(x)}, \ceil{q(x)})}$, and so $y \in S$. Therefore $S$ is nonempty and finite, and hence has a minimum.

Now, let $s = \min S$. Since $S \subseteq R$, we thus have $s \in R$. Let $t \in R \cap [x, s]$. Then $q(t)$ is an integer by definition, and hence either $q(t) \geq \ceil{q(x)}$ or $q(t) \leq \floor{q(x)}$. Therefore, since $q$ is continuous over $[x, y]$ and thus also continuous over $[x, t]$, the intermediate value theorem implies that either $q(u) = \ceil{q(x)}$ or $q(u) = \floor{q(x)}$ for some $u \in [x, t]$. Therefore $u \in S$, and hence $s \leq u$ by definition. Since $u \leq t \leq s$, we thus have $s = t = u$ and hence $R \cap [x, s] = \Set{s}$, so $\min R = s$. By the definition of $S$, we have either $q(s) = \floor{q(x)}$ or $q(s) = \ceil{q(x)}$, so we are done.
\end{proof}
\begin{corollary}
Let $a, c \in \reals^*$ and let $q \from \reals^* \to \reals$ be a function such that $q(t) = (at^2 + c)/t$. For all $x \in \reals^*$,
\begin{equation*}
\min{\Set{t \in [x, +\infty) \given (at^2 + c \bmod t) = 0}} = \min{\Set{t \in [x, +\infty) \given q(t) \in \Set{\floor{q(x)}, \ceil{q(x)}}}}.
\end{equation*}
\end{corollary}

\begin{lemma}\label{lem:quotients-in-prime-bases}
For all $x, y \in \floats^*_\infty$, if $\beta$ is prime, then either $x/y \in \floats^*_\infty$ or $x/y$ has no finite representation in base $\beta$.
\end{lemma}
\begin{proof}
Let $x, y \in \floats^*_\infty$ and suppose $\beta$ is prime and $x/y$ has a finite representation in base $\beta$. We shall prove that $x/y \in \floats^*_\infty$. Let $M_x = x/\ulp(x)$ and $M_y = y/\ulp(y)$. Then $M_x/M_y$ has a finite base-$\beta$ representation, and thus $M_x/M_y = n/\beta^k$ for some $n, k \in \integers$. If $k \leq 0$, then
\begin{equation*}
\abs{n} = \abs{M_x/M_y} \beta^k \leq \abs{M_x/M_y} \leq \abs{M_x} < \beta^p.  
\end{equation*}
Suppose $k > 0$ instead, and suppose without loss of generality that $\beta$ does not divide $n$ and that $M_x/M_y$ is in lowest terms. Then, since $n = M_x\beta^k/M_y$ is an integer and $M_x$ has no common factors with $M_y$, it follows that $\beta^k / M_y$ is an integer. Therefore, since $\beta$ is prime and $M_y$ is an integer, $\abs{M_y}$ is a power of $\beta$. However, since $\beta$ is not a factor of $n$, it follows that $\abs{M_y} = \beta^k$, and thus $\abs{n} = \abs{M_x} < \beta^p$ again. Therefore $M_x/M_y = n/\beta^k \in \floats^*_\infty$ by definition in all cases, and so
$x/y \in \floats^*_\infty$.
\end{proof}

\section{Glossary}\label{sec:glossary}
\subsection{List of terms}
\begin{description}
\item[diameter] the greatest distance between any pair of a set's elements
\item[exponent] the integer $e$ in $\pm d_1.d_2\ldots d_p \times \beta^e$
\item[equitable] see \Cref{def:equitable}; roughly, a rounding function under which no floating-point number has an especially small preimage; alternatively, a rounding function which does not allocate too many reals to any floating-point number
\item[faithful] a rounding function is faithful if it always rounds to either the nearest number above or below (i.e.\ does not jump over floating-point numbers)
\item[feasible] $x$ is feasible for $Z$ if $x$ is a floating-point factor of some $z \in Z$
\item[floating-point factor] $x$ is a floating-point factor of $z$ if $x \otimes y = z$ for some floating-point $y$
\item[floating-point interval] a set of consecutive floating-point numbers
\item[floating-point number] see \Cref{def:floats}; roughly, a number of the form $\pm d_1.d_2\ldots d_p \times \beta^e$
\item[integral significand] the significand multiplied by $\beta^{p - 1}$, making it an integer
\item[normal] a floating-point number which can be written without leading zeros with an exponent between $\emin$ and $\emax$; equivalently, a floating-point number with magnitude no less than $\beta^\emin$
\item[plausible] the unbounded floating-point numbers $x$ and $y$ are plausible for $Z$ if $\fl(xy) \in Z$
\item[quantum] the place value of a 1 in the last place of a floating-point number
\item[quantum exponent] the exponent of the quantum (the quantum is always a power of the base)
\item[rounding function] a function from $\overline{\reals}$ to $\overline{\floats}$
\item[significand] the $\pm d_1.d_2\ldots d_p$ part of $\pm d_1.d_2\ldots d_p \times \beta^e$
\item[subnormal] nonzero, but smaller than any normal number
\item[unit in first place] the place value of a 1 in the first digit of a number written in base-$\beta$ scientific notation; equally, $\beta^e$ where $e$ is the exponent of the number
\item[unit in last place] the place value of a 1 in the last digit of a number written in base-$\beta$ precision-$p$ scientific notation; equally, $\beta^{e-p+1}$ where $e$ is the exponent of the number
\end{description}

\subsection{List of symbols}
\begin{trivlist}
\item \begin{tabularx}{\textwidth}{l X}
$\beta$ & base (alt.\ radix) \\
$p$ & precision\\
$m$ & significand\\
$M$ & integral significand\\
$e$ & exponent\\
$q$ & quantum exponent: satisfies $q = e - p + 1$\\
$E(x)$ & exponent of $x$\\
$Q(x)$ & quantum exponent of $x$\\
$\otimes$ & rounded multiplication: $x \otimes y = \fl(xy)$\\
$\fl$ & any nondecreasing and faithful rounding function \\
$\RD(x)$ & $x$ rounded downward (i.e.\ to the nearest number below) \\
$\RU(x)$ & $x$ rounded upward (i.e.\ to nearest number above) \\
$\RN(x)$ & $x$ rounded to the nearest number; no rule is specified for breaking rounding ties \\
$\fl^{-1}[Z]$ & preimage of $Z$ under rounding; the set of extended real numbers that round to some $z \in Z$\\
$\Feas(Z)$ & set of all floating-point numbers feasible for $Z$\\
$\Plaus(Z)$ & set of all unbounded floating-point numbers plausible for $Z$ \\
$\PlausRU{Z}(x)$ & the least $x' \in \Plaus(Z)$ greater than or equal to $x$\\
$\ufp(x)$ & unit in first place of $x$\\
$\ulp(x)$ & unit in last place of $x$\\
$F(X, Y, Z)$ & classical interval-based narrowing of $X, Y, Z$; see \Cref{def:classical-iteration}\\
$\diam X$ & diameter of the set $X$ \\
$(x \bmod y)$ & remainder of floored division of $x$ by $y$\\
$\integers$ & integers\\
$\integers^*$ & nonzero integers\\
$\reals$ & real numbers \\
$\reals^*$ & nonzero real numbers \\
$\overline{\reals}$ & real numbers with $+\infty$ and $-\infty$ \\
$\overlinestar{\reals}$ & nonzero real numbers with $+\infty$ and $-\infty$ \\
$\floats^*$ & nonzero finite floating-point numbers\\
$\floats$ & finite floating-point numbers \\
$\overline{\floats}$ & finite floating-point numbers with $+\infty$ and $-\infty$\\
$\floats^*_\infty$ & nonzero finite unbounded floating-point numbers\\
$\floats_\infty$ & finite unbounded floating-point numbers \\
$\overlineinf{\floats}$ & finite unbounded floating-point numbers with $+\infty$ and $-\infty$
\end{tabularx}
\end{trivlist}
\end{appendices}

\bibliography{Floating_point}


\begin{thebibliography}{30}
\ifx \bisbn   \undefined \def \bisbn  #1{ISBN #1}\fi
\ifx \binits  \undefined \def \binits#1{#1}\fi
\ifx \bauthor  \undefined \def \bauthor#1{#1}\fi
\ifx \batitle  \undefined \def \batitle#1{#1}\fi
\ifx \bjtitle  \undefined \def \bjtitle#1{#1}\fi
\ifx \bvolume  \undefined \def \bvolume#1{\textbf{#1}}\fi
\ifx \byear  \undefined \def \byear#1{#1}\fi
\ifx \bissue  \undefined \def \bissue#1{#1}\fi
\ifx \bfpage  \undefined \def \bfpage#1{#1}\fi
\ifx \blpage  \undefined \def \blpage #1{#1}\fi
\ifx \burl  \undefined \def \burl#1{\textsf{#1}}\fi
\ifx \doiurl  \undefined \def \doiurl#1{\url{https://doi.org/#1}}\fi
\ifx \betal  \undefined \def \betal{\textit{et al.}}\fi
\ifx \binstitute  \undefined \def \binstitute#1{#1}\fi
\ifx \binstitutionaled  \undefined \def \binstitutionaled#1{#1}\fi
\ifx \bctitle  \undefined \def \bctitle#1{#1}\fi
\ifx \beditor  \undefined \def \beditor#1{#1}\fi
\ifx \bpublisher  \undefined \def \bpublisher#1{#1}\fi
\ifx \bbtitle  \undefined \def \bbtitle#1{#1}\fi
\ifx \bedition  \undefined \def \bedition#1{#1}\fi
\ifx \bseriesno  \undefined \def \bseriesno#1{#1}\fi
\ifx \blocation  \undefined \def \blocation#1{#1}\fi
\ifx \bsertitle  \undefined \def \bsertitle#1{#1}\fi
\ifx \bsnm \undefined \def \bsnm#1{#1}\fi
\ifx \bsuffix \undefined \def \bsuffix#1{#1}\fi
\ifx \bparticle \undefined \def \bparticle#1{#1}\fi
\ifx \barticle \undefined \def \barticle#1{#1}\fi
\bibcommenthead
\ifx \bconfdate \undefined \def \bconfdate #1{#1}\fi
\ifx \botherref \undefined \def \botherref #1{#1}\fi
\ifx \url \undefined \def \url#1{\textsf{#1}}\fi
\ifx \bchapter \undefined \def \bchapter#1{#1}\fi
\ifx \bbook \undefined \def \bbook#1{#1}\fi
\ifx \bcomment \undefined \def \bcomment#1{#1}\fi
\ifx \oauthor \undefined \def \oauthor#1{#1}\fi
\ifx \citeauthoryear \undefined \def \citeauthoryear#1{#1}\fi
\ifx \endbibitem  \undefined \def \endbibitem {}\fi
\ifx \bconflocation  \undefined \def \bconflocation#1{#1}\fi
\ifx \arxivurl  \undefined \def \arxivurl#1{\textsf{#1}}\fi
\csname PreBibitemsHook\endcsname

\bibitem{michel2002}
\begin{bchapter}
\bauthor{\bsnm{Michel}, \binits{C.}}:
\bctitle{Exact projection functions for floating point number constraints}.
In: \bbtitle{Seventh {{International Symposium}} on {{Artificial Intelligence}}
  and {{Mathematics}}}.
\bsertitle{{{AI}}\&{{M}} 15-2002},
\bconflocation{{Fort Lauderdale, FL, USA}}
(\byear{2002})
\end{bchapter}
\endbibitem

\bibitem{ziv2003}
\begin{bchapter}
\bauthor{\bsnm{Ziv}, \binits{A.}},
\bauthor{\bsnm{Aharoni}, \binits{M.}},
\bauthor{\bsnm{Asaf}, \binits{S.}}:
\bctitle{Solving range constraints for binary floating-point instructions}.
In: \bbtitle{Proceedings of the 16th {{IEEE Symposium}} on {{Computer
  Arithmetic}}}.
\bsertitle{{{ARITH}}'03},
pp. \bfpage{158}--\blpage{164}.
\bpublisher{{IEEE Computer Society}},
\blocation{{Los Alamitos, CA, USA}}
(\byear{2003}).
\doiurl{10.1109/ARITH.2003.1207674}
\end{bchapter}
\endbibitem

\bibitem{bagnara2016}
\begin{barticle}
\bauthor{\bsnm{Bagnara}, \binits{R.}},
\bauthor{\bsnm{Carlier}, \binits{M.}},
\bauthor{\bsnm{Gori}, \binits{R.}},
\bauthor{\bsnm{Gotlieb}, \binits{A.}}:
\batitle{Exploiting binary floating-point representations for constraint
  propagation}.
\bjtitle{INFORMS Journal on Computing}
\bvolume{28}(\bissue{1}),
\bfpage{31}--\blpage{46}
(\byear{2016}).
\doiurl{10.1287/ijoc.2015.0663}
\end{barticle}
\endbibitem

\bibitem{gallois-wong2020}
\begin{barticle}
\bauthor{\bsnm{{Gallois-Wong}}, \binits{D.}},
\bauthor{\bsnm{Boldo}, \binits{S.}},
\bauthor{\bsnm{Cuoq}, \binits{P.}}:
\batitle{Optimal inverse projection of floating-point addition}.
\bjtitle{Numerical Algorithms}
\bvolume{83}(\bissue{3}),
\bfpage{957}--\blpage{986}
(\byear{2020}).
\doiurl{10.1007/s11075-019-00711-z}
\end{barticle}
\endbibitem

\bibitem{andrlon2019}
\begin{bchapter}
\bauthor{\bsnm{Andrlon}, \binits{M.}},
\bauthor{\bsnm{Schachte}, \binits{P.}},
\bauthor{\bsnm{Søndergaard}, \binits{H.}},
\bauthor{\bsnm{Stuckey}, \binits{P.J.}}:
\bctitle{Optimal bounds for floating-point addition in constant time}.
In: \bbtitle{2019 {{IEEE}} 26th {{Symposium}} on {{Computer Arithmetic}}
  ({{ARITH}})},
pp. \bfpage{159}--\blpage{166}
(\byear{2019}).
\doiurl{10.1109/ARITH.2019.00038}
\end{bchapter}
\endbibitem

\bibitem{brain2015}
\begin{bchapter}
\bauthor{\bsnm{Brain}, \binits{M.}},
\bauthor{\bsnm{Tinelli}, \binits{C.}},
\bauthor{\bsnm{Rümmer}, \binits{P.}},
\bauthor{\bsnm{Wahl}, \binits{T.}}:
\bctitle{An automatable formal semantics for {{IEEE-754}} floating-point
  arithmetic}.
In: \bbtitle{Proceedings of the 22nd {{IEEE Symposium}} on {{Computer
  Arithmetic}}},
pp. \bfpage{160}--\blpage{167}.
\bpublisher{{IEEE Computer Society}},
\blocation{{Los Alamitos, CA, USA}}
(\byear{2015}).
\doiurl{10.1109/ARITH.2015.26}
\end{bchapter}
\endbibitem

\bibitem{boldo2011}
\begin{bchapter}
\bauthor{\bsnm{Boldo}, \binits{S.}},
\bauthor{\bsnm{Melquiond}, \binits{G.}}:
\bctitle{Flocq: a unified library for proving floating-point algorithms in
  {{Coq}}}.
In: \bbtitle{2011 {{IEEE}} 20th {{Symposium}} on {{Computer Arithmetic}}},
pp. \bfpage{243}--\blpage{252}
(\byear{2011}).
\doiurl{10.1109/ARITH.2011.40}
\end{bchapter}
\endbibitem

\bibitem{boldo2017}
\begin{bbook}
\bauthor{\bsnm{Boldo}, \binits{S.}},
\bauthor{\bsnm{Melquiond}, \binits{G.}}:
\bbtitle{Computer {{Arithmetic}} and {{Formal Proofs}}: {{Verifying
  Floating-point Algorithms}} with The {{Coq System}}}.
\bpublisher{{ISTE Press}},
\blocation{{London, United Kingdom}}
(\byear{2017}).
\doiurl{10.1016/C2015-0-01301-6}
\end{bbook}
\endbibitem

\bibitem{harrison2006}
\begin{bchapter}
\bauthor{\bsnm{Harrison}, \binits{J.}}:
\bctitle{Floating-point verification using theorem proving}.
In: \beditor{\bsnm{Bernardo}, \binits{M.}},
\beditor{\bsnm{Cimatti}, \binits{A.}} (eds.)
\bbtitle{Formal {{Methods}} for {{Hardware Verification}}: 6th {{International
  School}} on {{Formal Methods}} for the {{Design}} of {{Computer}},
  {{Communication}}, and {{Software Systems}}, {{SFM}} 2006, {{Bertinoro}},
  {{Italy}}, {{May}} 22-27, 2006, {{Advances Lectures}}}.
\bsertitle{Lecture {{Notes}} in {{Computer Science}}},
vol. \bseriesno{3965},
pp. \bfpage{211}--\blpage{242}.
\bpublisher{{Springer-Verlag Berlin Heidelberg}},
\blocation{{Germany}}
(\byear{2006}).
\doiurl{10.1007/11757283_8}
\end{bchapter}
\endbibitem

\bibitem{harrison1999}
\begin{bchapter}
\bauthor{\bsnm{Harrison}, \binits{J.}}:
\bctitle{A machine-checked theory of floating point arithmetic}.
In: \bbtitle{Theorem {{Proving}} in {{Higher Order Logics}}:12th
  {{International Conference}}, {{TPHOLs}} '99, {{Nice}}, {{France}},
  {{September}} 14-17, 1999, {{Proceedings}}}.
\bsertitle{Lecture {{Notes}} in {{Computer Science}}},
vol. \bseriesno{1690},
pp. \bfpage{113}--\blpage{130}.
\bpublisher{{Springer-Verlag Berlin Heidelberg}},
\blocation{{Germany}}
(\byear{1999}).
\doiurl{10.1007/3-540-48256-3_9}
\end{bchapter}
\endbibitem

\bibitem{yu2013}
\begin{botherref}
\oauthor{\bsnm{Yu}, \binits{L.}}:
A formal model of {{IEEE}} floating point arithmetic.
Archive of Formal Proofs
(2013)
\end{botherref}
\endbibitem

\bibitem{rummer2010}
\begin{bchapter}
\bauthor{\bsnm{Rümmer}, \binits{P.}},
\bauthor{\bsnm{Wahl}, \binits{T.}}:
\bctitle{An {{SMT-LIB}} theory of binary floating-point arithmetic}.
In: \bbtitle{Informal Proceedings of the 8th {{International Workshop}} on
  {{Satisfiability Modulo Theories}} ({{SMT}}) at {{FLoC}}, {{Edinburgh}},
  {{Scotland}}},
p. \bfpage{14}
(\byear{2010})
\end{bchapter}
\endbibitem

\bibitem{miner1995}
\begin{barticle}
\bauthor{\bsnm{Miner}, \binits{P.S.}}:
\batitle{Formal specification of {{IEEE}} floating-point arithmetic using
  {{PVS}}}.
\bjtitle{IFAC Proceedings Volumes}
\bvolume{28}(\bissue{25}),
\bfpage{31}--\blpage{36}
(\byear{1995}).
\doiurl{10.1016/S1474-6670(17)44820-8}
\end{barticle}
\endbibitem

\bibitem{loiseleur1997}
\begin{botherref}
\oauthor{\bsnm{Loiseleur}, \binits{P.}}:
Formalisation en {{Coq}} de la norme {{IEEE}} 754 sur l’arithmétique à
  virgule flottante.
Research report,
{École normale supérieure},
{Paris, France}
(September 1997).
\url{https://web.archive.org/web/20001003100321/http://www.lri.fr/%7Eloisel/rapport-stage-dea.ps.gz}
\end{botherref}
\endbibitem

\bibitem{carreno1995}
\begin{bchapter}
\bauthor{\bsnm{Carreño}, \binits{V.A.}},
\bauthor{\bsnm{Miner}, \binits{P.S.}}:
\bctitle{Specification of the {{IEEE-854}} floating-point standard in {{HOL}}
  and {{PVS}}}.
In: \bbtitle{Higher {{Order Logic Theorem Proving}} and {{Its Applications}}}.
\bsertitle{Category {{B}} proceedings},
\bconflocation{{Aspen Grove, UT}}
(\byear{1995}).
\burl{https://web.archive.org/web/20000919054331/http://lal.cs.byu.edu/lal/hol95/Bprocs/carreno.ps}
\end{bchapter}
\endbibitem

\bibitem{carreno1995a}
\begin{botherref}
\oauthor{\bsnm{Carreño}, \binits{V.A.}}:
Interpretation of {{IEEE-854}} floating-point standard and definition in the
  {{HOL}} system.
{{NASA}} technical memorandum 110189,
{NASA Langley Research Center},
{Hampton, VA}
(September 1995).
\url{https://ntrs.nasa.gov/api/citations/19960008463/downloads/19960008463.pdf}
Accessed 2022-08-03
\end{botherref}
\endbibitem

\bibitem{barrett1989}
\begin{barticle}
\bauthor{\bsnm{Barrett}, \binits{G.}}:
\batitle{Formal methods applied to a floating-point number system}.
\bjtitle{IEEE Transactions on Software Engineering}
\bvolume{15}(\bissue{5}),
\bfpage{611}--\blpage{621}
(\byear{1989}).
\doiurl{10.1109/32.24710}
\end{barticle}
\endbibitem

\bibitem{boldo2006}
\begin{botherref}
\oauthor{\bsnm{Boldo}, \binits{S.}},
\oauthor{\bsnm{Muñoz}, \binits{C.A.}}:
A high-level formalization of floating-point number in {{PVS}}.
{{NASA}} contractor report 2006-214298,
{NASA Langley Research Center},
{Hampton, VA}
(October 2006).
\url{https://ntrs.nasa.gov/citations/20070003560}
Accessed 2022-08-04
\end{botherref}
\endbibitem

\bibitem{harrison1995}
\begin{bchapter}
\bauthor{\bsnm{Harrison}, \binits{J.}}:
\bctitle{Floating point verification in {{HOL}}}.
In: \beditor{\bsnm{Thomas~Schubert}, \binits{E.}},
\beditor{\bsnm{Windley}, \binits{P.J.}},
\beditor{\bsnm{{Alves-Foss}}, \binits{J.}} (eds.)
\bbtitle{Higher {{Order Logic Theorem Proving}} and {{Its Applications}}: 8th
  {{International Workshop}}, {{Aspen Grove}}, {{UT}}, {{USA}}, {{September}}
  11-14, 1995. {{Proceedings}}}.
\bsertitle{Lecture {{Notes}} in {{Computer Science}}},
vol. \bseriesno{971},
pp. \bfpage{186}--\blpage{199}.
\bpublisher{{Springer-Verlag Berlin Heidelberg}},
\blocation{{Germany}}
(\byear{1995}).
\doiurl{10.1007/3-540-60275-5_65}
\end{bchapter}
\endbibitem

\bibitem{jacobsen2015}
\begin{barticle}
\bauthor{\bsnm{Jacobsen}, \binits{C.}},
\bauthor{\bsnm{Solovyev}, \binits{A.}},
\bauthor{\bsnm{Gopalakrishnan}, \binits{G.}}:
\batitle{A parameterized floating-point formalizaton in {{HOL Light}}}.
\bjtitle{Electronic Notes in Theoretical Computer Science}
\bvolume{317},
\bfpage{101}--\blpage{107}
(\byear{2015}).
\doiurl{10.1016/j.entcs.2015.10.010}
\end{barticle}
\endbibitem

\bibitem{jacobi2005}
\begin{barticle}
\bauthor{\bsnm{Jacobi}, \binits{C.}},
\bauthor{\bsnm{Berg}, \binits{C.}}:
\batitle{Formal verification of the {{VAMP}} floating point unit}.
\bjtitle{Formal Methods in System Design}
\bvolume{26}(\bissue{3}),
\bfpage{227}--\blpage{266}
(\byear{2005}).
\doiurl{10.1007/s10703-005-1613-y}
\end{barticle}
\endbibitem

\bibitem{miner1995a}
\begin{botherref}
\oauthor{\bsnm{Miner}, \binits{P.S.}}:
Defining the {{IEEE-854}} floating-point standard in {{PVS}}.
Technical Report NASA-TM-110167
(June 1995).
\url{https://ntrs.nasa.gov/citations/19950023402}
Accessed 2022-08-04
\end{botherref}
\endbibitem

\bibitem{daumas2001}
\begin{bchapter}
\bauthor{\bsnm{Daumas}, \binits{M.}},
\bauthor{\bsnm{Rideau}, \binits{L.}},
\bauthor{\bsnm{Théry}, \binits{L.}}:
\bctitle{A generic library for floating-point numbers and its application to
  exact computing}.
In: \beditor{\bsnm{Boulton}, \binits{R.J.}},
\beditor{\bsnm{Jackson}, \binits{P.B.}} (eds.)
\bbtitle{Theorem {{Proving}} in {{Higher Order Logics}}: 14th {{International
  Conference}}, {{TPHOLs}} 2001, {{Edinburgh}}, {{Scotland}}, {{UK}},
  {{September}} 3-6, 2001. {{Proceedings}}}.
\bsertitle{Lecture {{Notes}} in {{Computer Science}}},
vol. \bseriesno{2152},
pp. \bfpage{169}--\blpage{184}.
\bpublisher{{Springer-Verlag Berlin Heidelberg}},
\blocation{{Germany}}
(\byear{2001}).
\doiurl{10.1007/3-540-44755-5_13}
\end{bchapter}
\endbibitem

\bibitem{melquiond2012}
\begin{barticle}
\bauthor{\bsnm{Melquiond}, \binits{G.}}:
\batitle{Floating-point arithmetic in the {{Coq}} system}.
\bjtitle{Information and Computation}
\bvolume{216},
\bfpage{14}--\blpage{23}
(\byear{2012}).
\doiurl{10.1016/j.ic.2011.09.005}
\end{barticle}
\endbibitem

\bibitem{pan1990}
\begin{bchapter}
\bauthor{\bsnm{Pan}, \binits{J.}},
\bauthor{\bsnm{Levitt}, \binits{K.N.}}:
\bctitle{A formal specification of the {{IEEE}} floating-point standard with
  application to the verification of {{F}}}.
In: \bbtitle{1990 {{Conference Record Twenty-Fourth Asilomar Conference}} on
  {{Signals}}, {{Systems}} and {{Computers}}, 1990.},
vol. \bseriesno{1},
p. \bfpage{505}
(\byear{1990}).
\doiurl{10.1109/ACSSC.1990.523389}
\end{bchapter}
\endbibitem

\bibitem{muller2018}
\begin{bbook}
\bauthor{\bsnm{Muller}, \binits{J.-M.}},
\bauthor{\bsnm{Brunie}, \binits{N.}},
\bauthor{\bparticle{de} \bsnm{Dinechin}, \binits{F.}},
\bauthor{\bsnm{Jeannerod}, \binits{C.-P.}},
\bauthor{\bsnm{Joldes}, \binits{M.}},
\bauthor{\bsnm{Lefèvre}, \binits{V.}},
\bauthor{\bsnm{Melquiond}, \binits{G.}},
\bauthor{\bsnm{Revol}, \binits{N.}},
\bauthor{\bsnm{Torres}, \binits{S.}}:
\bbtitle{Handbook of {{Floating-Point Arithmetic}}},
\bedition{2}nd edn.
\bpublisher{{Birkhäuser}},
\blocation{{Cham, Switzerland}}
(\byear{2018}).
\doiurl{10.1007/978-3-319-76526-6}
\end{bbook}
\endbibitem

\bibitem{goldberg1991}
\begin{barticle}
\bauthor{\bsnm{Goldberg}, \binits{D.}}:
\batitle{What every computer scientist should know about floating-point
  arithmetic}.
\bjtitle{ACM Computing Surveys}
\bvolume{23}(\bissue{1}),
\bfpage{5}--\blpage{48}
(\byear{1991}).
\doiurl{10.1145/103162.103163}
\end{barticle}
\endbibitem

\bibitem{rump2008}
\begin{barticle}
\bauthor{\bsnm{Rump}, \binits{S.}},
\bauthor{\bsnm{Ogita}, \binits{T.}},
\bauthor{\bsnm{Oishi}, \binits{S.}}:
\batitle{Accurate {{Floating-Point Summation Part I}}: {{Faithful Rounding}}}.
\bjtitle{SIAM Journal on Scientific Computing}
\bvolume{31}(\bissue{1}),
\bfpage{189}--\blpage{224}
(\byear{2008}).
\doiurl{10.1137/050645671}
\end{barticle}
\endbibitem

\bibitem{ieee754_2019}
\begin{botherref}
\oauthor{\bsnm{{IEEE Std. 754-2019}}}:
{{IEEE Standard}} for {{Floating-Point Arithmetic}}.
Technical report,
{IEEE}
(July 2019)
\end{botherref}
\endbibitem

\bibitem{brain2019}
\begin{bchapter}
\bauthor{\bsnm{Brain}, \binits{M.}},
\bauthor{\bsnm{Niemetz}, \binits{A.}},
\bauthor{\bsnm{Preiner}, \binits{M.}},
\bauthor{\bsnm{Reynolds}, \binits{A.}},
\bauthor{\bsnm{Barrett}, \binits{C.}},
\bauthor{\bsnm{Tinelli}, \binits{C.}}:
\bctitle{Invertibility conditions for floating-point formulas}.
In: \beditor{\bsnm{Dillig}, \binits{I.}},
\beditor{\bsnm{Tasiran}, \binits{S.}} (eds.)
\bbtitle{Computer {{Aided Verification}}: 31st {{International Conference}},
  {{CAV}} 2019, {{New York City}}, {{NY}}, {{USA}}, {{July}} 15-18, 2019,
  {{Proceedings}}, {{Part II}}}.
\bsertitle{Lecture {{Notes}} in {{Computer Science}}},
vol. \bseriesno{11562},
pp. \bfpage{116}--\blpage{136}.
\bpublisher{{Springer}},
\blocation{{Cham, Switzerland}}
(\byear{2019}).
\doiurl{10.1007/978-3-030-25543-5_8}
\end{bchapter}
\endbibitem

\end{thebibliography}

\end{document}